\documentclass[lettersize,journal]{IEEEtran}
\usepackage{amsmath,amsfonts}
\usepackage{algorithmic}
\usepackage{algorithm}
\usepackage{array}
\usepackage[caption=false,font=normalsize,labelfont=sf,textfont=sf]{subfig}
\usepackage{textcomp}
\usepackage{stfloats}
\usepackage{url}
\usepackage{verbatim}
\usepackage{graphicx}
\usepackage{cite}
\usepackage{rotating}

\usepackage{url}
\usepackage{xcolor}
\usepackage{epsfig}
\usepackage{setspace} 
\usepackage{soul,color}
\usepackage{csquotes}
\usepackage{listings}
\usepackage{tikz}
\usepackage[official]{eurosym}
\usetikzlibrary{positioning}
\usepackage{adjustbox}
\usepackage{array}
\usepackage{multirow}
\usepackage{fancyhdr}

\usepackage{enumitem}
\usepackage{bbding}
\usepackage{graphicx}
\usepackage{array}
\usepackage{booktabs}
\usepackage{balance}
\usepackage{lineno}

\usepackage{tabularx}
\usepackage{booktabs}
\usepackage{color, colortbl}

\usepackage[acronym,toc,shortcuts]{glossaries}

\newacronym{2D}{2D}{two dimensional}
\newacronym{3D}{3D}{three dimensional}
\newacronym{4G}{4G}{Fourth Generation}
\newacronym{5G}{5G}{Fifth Generation}
\newacronym{6G}{6G}{Sixth Generation}
\newacronym{5G PPP}{5G PPP}{5G Infrastructure Public Private Partnership}
\newacronym{3GPP}{3GPP}{3rd Generation Partnership Project}
\newacronym{4D}{4D}{four dimensional}
\newacronym{AAA}{AAA}{Authentication, Authorization and Accounting}
\newacronym{ABS}{ABS}{Aerial Base Station}
\newacronym{ABSF}{ABSF}{Almost-Blank Subframe}
\newacronym{AES}{AES}{Advanced Encryption Standard }
\newacronym{AI}{AI}{Artificial Intelligence}
\newacronym{AMC}{AMC}{Adaptive Modulation and Coding}
\newacronym{AMF}{AMF}{Access and Mobility Management Function}
\newacronym{AP}{AP}{access point}
\newacronym{API}{API}{Application Programming Interface}
\newacronym{APN}{APN}{Access Point Name}
\newacronym{AR}{AR}{Augmented Reality}
\newacronym{AWGN}{AWGN}{additive white Gaussian noise}
\newacronym{BBU}{BBU}{Baseband Unit}
\newacronym{BCN}{BCN}{Blockchain Network}
\newacronym{BE}{BE}{best-effort}
\newacronym{BET}{BET}{Blind Equal Throughput}
\newacronym{BLAST}{BLAST}{Bell Laboratories Layered Space-Time}
\newacronym{BLER}{BLER}{Block Error Rate}
\newacronym{BS}{BS}{base station}
\newacronym{BTP}{BTP}{Backhaul Transport Provider}
\newacronym{BTS}{BTS}{Base Transceiver Station}
\newacronym{CA}{CA}{carrier aggregation}
\newacronym{MBAR}{MBAR}{Mobile Backhaul Aggregation Router}
\newacronym{CAPEX}{CapEx}{capital expenditure}
\newacronym{CDF}{CDF}{Cumulative Distribution Function}
\newacronym{CELL-ID}{CELL-ID}{cell identification ID}
\newacronym{CIO}{CIO}{cell individual offset}
\newacronym{CDN}{CDN}{Content Delivery Network}
\newacronym{CN}{CN}{core network}
\newacronym{CP}{CP}{Control Plane}
\newacronym{CPU}{CPU}{central processing unit}
\newacronym{CoMP}{CoMP}{Coordinated Multipoint}
\newacronym{CoW}{CoW}{Cell on Wheels}
\newacronym{CSR}{CSR}{Cell Site Router}
\newacronym{CQI}{CQI}{Channel Quality Indicator}
\newacronym{C-RAN}{C-RAN}{Cloud RAN}
\newacronym{CS}{CS}{central scheduler}
\newacronym{CSI}{CSI}{channel state information}
\newacronym{CRE}{CRE}{cell range expansion}
\newacronym{D2D}{D2D}{Device-to-Device}
\newacronym{DLT}{DLT}{Distributed Ledger Technology}
\newacronym{DFT}{DFT}{discrete Fourier transform}
\newacronym{DSL}{DSL}{Digital subscriber line}
\newacronym{EARFCN}{EARFCN}{E-UTRA Absolute Radio Frequency Channel Number}
\newacronym{EC}{EC}{European Commission}
\newacronym{e2e}{e2e}{end-to-end}
\newacronym{eICIC}{eICIC}{enhanced inter-cell interference cancellation}
\newacronym{eMBB}{eMBB}{enhanced Mobile Broadband}
\newacronym{eNodeB}{eNodeB}{Evolved Node B}
\newacronym{EPC}{EPC}{Evolved Packet Core}
\newacronym{EPS}{EPS}{Evolved Packet System}
\newacronym{ETSI}{ETSI}{European Telecommunications Standards Institute}
\newacronym{E-UTRAN}{E-UTRAN}{Evolved Universal Terrestrial Radio Access Network}
\newacronym{EV}{EV}{Electric Vehicle}
\newacronym{FANET}{FANET}{Fly Ad Hoc Network}
\newacronym{FDMA}{FDMA}{frequency division multiple access}
\newacronym{FFT}{FFT}{fast Fourier transform}
\newacronym{FTP}{FTP}{File Transfer Protocol}
\newacronym{FU}{FU}{Frame Usage}
\newacronym{GEO}{GEO}{Geostationary Earth Orbit} 
\newacronym{GIS}{GIS}{Geographical Information Systems}
\newacronym{GGSN}{GGSN}{Gateway GPRS Support Node}
\newacronym{GPS}{GPS}{global positioning system}
\newacronym{GRA}{GRA}{Grey relational analysis}
\newacronym{GSM}{GSM}{Global System for Mobile Communications}
\newacronym{GTP}{GTP}{GPRS Tunneling Protocol}
\newacronym{GTP-U}{GTP-U}{GPRS Tunneling Protocol-User Plane}
\newacronym{HAPS}{HAPS}{High Altitude Platform Stations}
\newacronym{HDFS}{HDFS}{Hadoop Distributed File System}
\newacronym{HetNet}{HetNet}{Heterogeneous Network}
\newacronym{HiveQL}{HiveQL}{Hive Query language}
\newacronym{HD}{HD}{High Definition}
\newacronym{HEO}{HEO}{High Earth Orbit}
\newacronym{HO}{HO}{handover}
\newacronym{HARQ}{HARQ}{Hybrid automatic repeat request}
\newacronym{HS-DSCH}{HS-DSCH}{High Speed Downlink Shared Channel}
\newacronym{HSS}{HSS}{Home Subscriber Station}
\newacronym{HTS}{HTS}{High Throughput Satellite}
\newacronym{HTTP}{HTTP}{Hypertext Transfer Protocol}
\newacronym{IAB}{IAB}{Integrated Access and Backhaul}
\newacronym{ICIC}{ICIC}{inter-cell interference cancellation}
\newacronym{ICN}{ICN}{information-centric network}
\newacronym{IEEE}{IEEE}{Institute of Electrical and Electronics Engineers}
\newacronym{IMEI}{IMEI}{International Mobile Station Equipment Identity}
\newacronym{IMSI}{IMSI}{International Mobile Subscriber Identity}
\newacronym{IMS}{IMS}{IP Multimedia Subsystem}
\newacronym{IMT-A}{IMT-A}{International Mobile Telecommunications - Advanced}
\newacronym{ITU}{ITU}{International Telecommunication Union}
\newacronym{IP}{IP}{Internet Protocol}
\newacronym{IPsec}{IPsec}{Internet Protocol Security}
\newacronym{IoT}{IoT}{Internet of Things}
\newacronym{ISAC}{ISAC}{Integrated Sensing and Communication}
\newacronym{JSON}{JSON}{JavaScript Object Notation}
\newacronym{KPI}{KPI}{key performance indicator}
\newacronym{LAC}{LAC}{location area code}
\newacronym{LEO}{LEO}{Low Earth Orbit}
\newacronym{LOS}{LOS}{Line-of-Sight}
\newacronym{LPWAN}{LPWAN}{Low Power Wide Area Network}
\newacronym{LTE}{LTE}{Long Term Evolution}
\newacronym{LTE-A}{LTE-A}{Long Term Evolution Advanced}
\newacronym{mmWave}{mmWave}{millimeter wave}
\newacronym{MAC}{MAC}{Medium Access Control}
\newacronym{MADM}{MADM}{Multiple Attribute Decision Making}
\newacronym{MANET}{MANET}{Mobile Ad Hoc Network}
\newacronym{MBH}{MBH}{Mobile Backhaul}
\newacronym{MCCS}{MCCS}{Mission Critical Communication System}
\newacronym{MCD}{MCD}{Mission Critical Data}
\newacronym{MCS}{MCS}{Modulation Coding Scheme}
\newacronym{MCX}{MCX}{Mission Critical Services}
\newacronym{MCPPT}{MCPPT}{Mission Critical Push-to-Talk}
\newacronym{MCVideo}{MCVideo}{Mission Critical Video}
\newacronym{MDRU}{MDRU}{ovable and Deployable Resource Unit}
\newacronym{MEC}{MEC}{Multi-access Edge Computing}
\newacronym{MEO}{MEO}{Medium Earth Orbit}
\newacronym{MEW}{MEW}{multiplicative exponent weighting}
\newacronym{MIMO}{MIMO}{multiple-input multiple-output}
\newacronym{ML}{ML}{Machine Learning}
\newacronym{MME}{MME}{Mobility Management Entity}
\newacronym{mMTC}{mMTC}{massive Machine Type Communications}
\newacronym{MMF}{MMF}{max-min fairness}
\newacronym{MMSE}{MMSE}{minimum mean square error}
\newacronym{MPLS}{MPLS}{Multiprotocol Label Switching}
\newacronym{MSISDN}{MSISDN}{Mobile Station International Subscriber Directory Number}
\newacronym{MSP}{MSP}{Mobile Service Provider}
\newacronym{MT}{MT}{Maximum Throughput}
\newacronym{NAS}{NAS}{Non Access Stratum}
\newacronym{NE}{NE}{Nash Equilibrium}
\newacronym{NLP}{NLP}{Natural Language Processing}
\newacronym{NOMA}{NOMA}{Non-Orthogonal Multiple Access}
\newacronym{NR}{NR}{New Radio}
\newacronym{NTN}{NTN}{Non-Terrestrial Network}
\newacronym{NFV}{NFV}{Network Functions Virtualization}
\newacronym{NoSQL}{NoSQL}{Not Only SQL}
\newacronym{OAM}{OAM}{Operation, Administration and Management}
\newacronym{OFDM}{OFDM}{orthogonal frequency division multiplexing}
\newacronym{OFDMA}{OFDMA}{orthogonal frequency division multiple access}
\newacronym{ONF}{ONF}{open networking foundation}
\newacronym{ONOS}{ONOS}{Open Network Operating System}
\newacronym{OPEX}{OpEx}{operating expenditure}
\newacronym{OS}{OS}{operating system}
\newacronym{OTT}{OTT}{over-the-top}
\newacronym{OWC}{OWC}{Optical Wireless Communication}
\newacronym{PCI}{PCI}{Physical Cell Identity}
\newacronym{PCRF}{PCRF}{Policy and Charging Rules Function}
\newacronym{PDF}{PDF}{Probability Distribution Function}
\newacronym{PDN}{PDN}{packet data network}
\newacronym{PDCP}{PDCP}{Packet Data Convergence Control}
\newacronym{PDSCH}{PDSCH}{Physical Downlink Shared Channel}
\newacronym{PDU}{PDU}{Protocol Data Unit}
\newacronym{PF}{PF}{Proportional Fair}
\newacronym{PGW}{P-GW}{Packet Data Gateway}
\newacronym{PHY}{PHY}{physical layer}
\newacronym{PoC}{PoC}{Proof-of-Concept}
\newacronym{PPP}{PPP}{{P}oisson point process}
\newacronym{PPDR}{PPDR}{Public Protection and Disaster Relief}
\newacronym{PSNs}{PSNs}{Public Safety Networks}
\newacronym{PTP}{PTP}{Precision Time Protocol}
\newacronym{PV}{PV}{photovoltaic} 
\newacronym{QKD}{QKD}{Quantum Key Distribution}
\newacronym{QoE}{QoE}{quality-of-experience}
\newacronym{QoS}{QoS}{quality-of-service}
\newacronym{QCI}{QCI}{QoS Class Identifier}
\newacronym{PSC}{PSC}{Primary Scrambling Code}
\newacronym{PSD}{PSD}{power spectral density}
\newacronym{RACH}{RACH}{random access channel}
\newacronym{RAN}{RAN}{Radio Access Network}
\newacronym{RAT}{RAT}{Radio Access Technology}
\newacronym{RB}{RB}{Resource Block}
\newacronym{RE}{RE}{range extension}
\newacronym{RES}{RES}{Renewable Energy Sources}
\newacronym{RF}{RF}{radio frequency}
\newacronym{RG}{RG}{rate guarantee}
\newacronym{IRS}{IRS}{Intelligent Reflective Surfaces}
\newacronym{RLC}{RLC}{Radio Link Controller}
\newacronym{RNC}{RNC}{Radio Network Controller}
\newacronym{RR}{RR}{Round Robin}
\newacronym{RRC}{RRC}{Radio Resource Control}
\newacronym{RRH}{RRH}{remote radio head}
\newacronym{RRU}{RRU}{Remote Radio Unit}
\newacronym{RRM}{RRM}{radio resource management}
\newacronym{RSI}{RSI}{RACH Root Sequence Index}
\newacronym{RSS}{RSS}{received signal strength}
\newacronym{RSSI}{RSSI}{received signal strength indicator}
\newacronym{RSRP}{RSRP}{reference signal received power}
\newacronym{RTT}{RTT}{Round Trip Time}
\newacronym{SAC}{SAC}{service area code}
\newacronym{SANET}{SANET}{Sea Ad Hoc Network}
\newacronym{SAW}{SAW}{simple additive weighting}
\newacronym{SC-FDMA}{SC-FDMA}{single carrier frequency division multiple access}
\newacronym{SCN}{SCN}{small cell network}
\newacronym{SCTP}{SCTP}{Stream Control Transmission Protocol}
\newacronym{SDN}{SDN}{Software-Defined Networking}
\newacronym{SDO}{SDO}{Standard Development Organization}
\newacronym{SDMN}{SDMN}{Software Defined Mobile Network}
\newacronym{SDU}{SDU}{Service Data Unit}
\newacronym{SecGW}{SecGW}{Security Gateway}
\newacronym{SGSN}{SGCN}{Serving GPRS Support Node}
\newacronym{SGW}{S-GW}{Serving Gateway}
\newacronym{SHARING}{SHARING}{Self-organized Heterogeneous Advanced RadIo Networks Generation}
\newacronym{SNR}{SNR}{signal-to-noise ratio}
\newacronym{SINR}{SINR}{signal-to-interference-plus-noise ratio}
\newacronym{SISO}{SISO}{single-input single-output}
\newacronym{SSID}{SSID}{Service Set Identification}
\newacronym{ST}{ST}{Standart Multi-User TOPSIS}
\newacronym{STBCs}{STBCs}{space-time block codes}
\newacronym{SVM}{SVM}{Support Vector Machine}
\newacronym{SWIPT}{SWIPT}{Simultaneous Wireless Information and Power Transfer}
\newacronym{SyncE}{SyncE}{Synchronous Ethernet}
\newacronym{TB}{TB}{Transport Block}
\newacronym{TBS}{TBS}{Transport Block Size}
\newacronym{TCP}{TCP}{Transport Control Protocol}
\newacronym{TDMA}{TDMA}{Time Division Multiple Access}
\newacronym{TEID}{TEID}{tunnel endpoint identifier}
\newacronym{TOPSIS}{TOPSIS}{Total Order Preference By Similarity to the Ideal Solution}
\newacronym{TTI}{TTI}{transmission time interval}
\newacronym{UAV}{UAV}{Unmanned Aerial Vehicle}
\newacronym{UAV-BS}{UAV-BS}{Unmanned Aerial Vehicles-Base Station}
\newacronym{UARN}{UARN}{UAV-aided relay network}
\newacronym{UDM}{UDM}{Unified Data Management}
\newacronym{UDP}{UDP}{User Datagram Protocol}
\newacronym{UE}{UE}{user equipment}
\newacronym{UGV}{UGV}{Unmanned Ground Vehicle}
\newacronym{UL}{UL}{Uplink}
\newacronym{UQD}{UQD}{UAV-BS QoS Determination}
\newacronym{UP}{UP}{User Plane}
\newacronym{UPF}{UPF}{User Plane Function}
\newacronym{UMTS}{UMTS}{Universal Mobile Telecommunications Service} 
\newacronym{URLLC}{URLLC}{Ultra-reliable low latency communications}
\newacronym{XR}{XR}{Extended Reality}
\newacronym{VANET}{VANET}{Vehicular Ad Hoc Network}
\newacronym{VoIP}{VoIP}{voice over IP}
\newacronym{VLC}{VLC}{Visible Light Communication}
\newacronym{VPN}{VPN}{virtual private network}
\newacronym{VR}{VR}{Virtual Reality}
\newacronym{VSAT}{VSAT}{Very Small Aperture Terminal}
\newacronym{W-CDMA}{W-CDMA}{Wideband Code Division Multiple Access}
\newacronym{WiFi}{WiFi}{Wireless Fidelity}
\newacronym{Wi-Fi}{Wi-Fi}{Wireless Fidelity}
\newacronym{WiMAX}{WiMAX}{Worldwide Interoperability for Microwave Access}
\newacronym{WLAN}{WLAN}{Wireless Local Area Network}
\newacronym{WMN}{WMN}{Wireless Mesh Network}
\newacronym{WSN}{WSN}{Wireless Sensor Network}
\newacronym{WMC}{WMC}{weighted Markov chain}
\newacronym{ZF}{ZF}{zero-forcing}
\newacronym{MNO}{MNO}{Mobile Network Operator}
\newacronym{SON}{SON}{Self Organizing Network}
\newacronym{ANR}{ANR}{Automatic Neighbor Relation}
\newacronym{MRO}{MRO}{Mobility Robustness Optimizer}
\newacronym{MLB}{MLB}{Mobility Load Balancing}
\newacronym{CQO}{CQO}{cell quality offset}

\newacronym{LoS}{LoS}{Line-of-Sight}
\newacronym{FSO}{FSO}{Free Space Optics}
\newacronym{THz}{THz}{Terahertz}
\newacronym{RIS}{RIS}{Reconfigurable Intelligent Surfaces}

\newacronym{BESS}{BESS}{Battery Energy Storage Systems}
\newacronym{UPS}{UPS}{Uninterrupted Power Supplies}
\newacronym{HES}{HES}{Hydrogen Energy Storage}

\newacronym{LoRa}{LoRa}{Long-Range}

\usepackage[subtle]{savetrees}

\makeatletter
 \patchcmd{\@maketitle}
  {\addvspace{0.5\baselineskip}\egroup}
   {\addvspace{-1\baselineskip}\egroup}
   {}
   {}
 \makeatother

\usepackage{etoolbox}
\let\mybibitem\bibitem
\renewcommand{\bibitem}[1]{%
  \ifstrequal{#1}{STERBENZ20101245}
    {\color{black}\mybibitem{#1}}
     {\color{black}\mybibitem{#1}}}

\begin{document}

\title{Solutions for Sustainable and Resilient Communication Infrastructure in Disaster Relief and Management Scenarios}

\author{Bilal Karaman, Ilhan Basturk, ~\IEEEmembership{Senior Member,~IEEE}, Sezai Taskin, Engin Zeydan, ~\IEEEmembership{Senior Member,~IEEE},\\ Ferdi Kara,~\IEEEmembership{Senior Member,~IEEE}, Esra Aycan Beyazıt,~\IEEEmembership{Member,~IEEE,} Miguel Camelo,\\ Emil Bj\"ornson,~\IEEEmembership{Fellow,~IEEE}, and Halim Yanikomeroglu~\IEEEmembership{Fellow,~IEEE.}  

\thanks{The work of F. Kara and E. Bj\"ornson is supported by the Swedish Foundation for Strategic Research and the SweWIN Vinnova Competence Center. The work of B. Karaman is supported by the Study in Canada Scholarship (SICS) by Global Affairs Canada (GAC).}

\thanks{B. Karaman, I. Basturk and S. Taskin are with Manisa Celal Bayar University, Turkiye. B. Karaman is also with the Department of Systems and Computer Engineering, Carleton University, Ottawa, ON, K1S 5B6 Canada, emails: \{bilal.karaman, ilhan.basturk, sezai.taskin\}@cbu.edu.tr. E. Zeydan is with Centre Tecnològic de Telecomunicacions de Catalunya (CTTC), Barcelona, Spain, 08860, email: ezeydan@cttc.es. F. Kara and E. Bj\"ornson are with the Department of Computer Science, KTH Royal Institute of Technology, Stockholm, Sweden, 16440. F. Kara is also with the Department of Computer Engineering, Zonguldak Bulent Ecevit University, Zonguldak Turkiye, 67100, e-mails: \{ferdi, emilbjo\} @kth.se. E. A. Beyazıt and M. Camelo are with IDLab Research Group, University of Antwerp - IMEC, Belgium, e-mails: \{esra.aycanbeyazit, miguel.camelo\}@imec.be. H. Yanikomeroglu is with the Department of Systems and Computer Engineering, Carleton University, Ottawa, ON, K1S 5B6 Canada, e-mail: halim@sce.carleton.ca.} 
}




\maketitle

\begin{abstract}

As natural disasters become more frequent and severe, ensuring a resilient communications infrastructure is of paramount importance for effective disaster response and recovery.  This disaster-resilient infrastructure should also respond to sustainability goals by providing an energy-efficient and economically feasible network that is accessible to everyone.  To this end,  this paper provides a comprehensive exploration of the technological solutions and strategies necessary to build and maintain resilient communications networks that can withstand and quickly recover from disaster scenarios. The paper starts with a survey of existing literature and related reviews to establish a solid foundation, followed by an overview of the global landscape of disaster communications and power supply management. We then introduce the key enablers of communications and energy resource technologies to support communications infrastructure, examining emerging trends that improve the resilience of these systems. Pre-disaster planning is emphasized as a critical phase where proactive communication and energy supply strategies can significantly mitigate the impact of disasters. We also explore the essential technologies for disaster response, focusing on real-time communications and energy solutions that support rapid deployment and coordination in times of crisis. The paper then presents post-disaster communication and energy management planning for effective rescue and evacuation operations. The main findings derived from the comprehensive survey are also summarized for each disaster phase. This is followed by an analysis of existing vendor products and services as well as standardization efforts and ongoing projects that contribute to the development of resilient infrastructures. A detailed case study of the Turkiye earthquakes is presented to illustrate the practical application of these technologies and strategies. Finally, we address the open issues and challenges in realizing sustainable and resilient communication infrastructures and provide insights into future research directions. By incorporating lessons learned from various disaster scenarios, this paper presents strategic recommendations that enhance the resilience and adaptability of communication systems in the context of disaster relief and management.
\end{abstract}

\begin{IEEEkeywords}
Disaster management, pre-disaster planning, disaster response, earthquake, communication enablers, energy enablers, post-disaster, standardization, use cases.
\end{IEEEkeywords}


\section{Introduction}
\label{introduction}

Major natural \textcolor{black}{hazards} and public safety incidents significantly disrupt communication network infrastructure. In the aftermath of major disasters, such as earthquakes or storms, the primary telecommunication infrastructures and other public infrastructures, such as power sources, are often severely damaged or completely destroyed. This results in the unavailability of cellular networks or \textcolor{black}{I}nternet connectivity. The telecommunications infrastructure is as crucial as other basic life needs like shelter, food, and clean water in disasters. Without a reliable and uninterrupted communication channel, it is very challenging to coordinate rescue operations to find those affected. It is also crucial to organize and provide basic life needs to rescued people.  Therefore, the discontinuities in communications severely restrict the central controlling authority from obtaining timely information, which is critical for ensuring coordination between rescue teams, quickly transmitting vital information, and responding to calls for help about the disaster area \cite{HAZRA202054}. For example, after two significant earthquakes, with magnitudes of 7.7 and 7.6, in Turkiye on February 6, 2023, hundreds of cellular towers were damaged. This resulted in a lack of cellular and Internet connectivity in more than 10 cities \cite{preventionweb2023}. For this reason, it is crucial that critical communication is preserved with an efficient temporary communication infrastructure in the disaster area until the conventional communication infrastructures are restored. This temporary network will help to connect various stakeholders, including volunteers and rescue/relief teams, enabling them to exchange information seamlessly and in a timely manner. This will help to facilitate effective and well-coordinated rescue, relief, and recovery efforts. In this context, \ac{PPDR} agencies have been exploring reliable wireless communication systems to ensure the public safety sector, efficient coordination of first responders and necessary support to the affected regions by reducing the likelihood of casualties and economic damage in the affected areas \cite{MARQUES201549}.  \ac{PPDR} communication systems primarily rely on private (professional) mobile radio (PMR) technologies, which offer a comprehensive range of voice services tailored to the specific needs of \ac{PPDR} systems, such as push-to-talk and call priority. Therefore, their data transmission capabilities are comparatively limited and lag behind current telecommunication technologies. Initial efforts have been aimed at improving communications capabilities for \ac{PPDR} agencies through the introduction of \ac{4G} technologies that go beyond the capabilities of the PMR system. The emergence of \ac{3GPP} Release 15 (originally \ac{5G}) is seen as a crucial standard, encompassing a broader range of functionalities. Although users \textcolor{black}{are expected to have} access to broadband voice, data, and video capabilities, including support for mission-critical services, there are key limitations to interoperability between different technologies that can hinder time-critical emergency management  \cite{Fantacci16}. Robust and secure \ac{PPDR} networks are therefore still required.

Emergency management involves the coordination of various functions to deal with major emergencies, including prevention, preparedness, response, and rehabilitation. In emergencies, coordination between different functions is crucial. Law enforcement focuses on the prevention, investigation, and apprehension of individuals suspected or convicted of criminal offenses. Emergency Medical Services (EMS) provide critical care, transportation, and disaster medicine, involving professionals such as doctors, paramedics, and volunteers. Firefighting deals with extinguishing hazardous fires that threaten people and property. Protection of the environment involves safeguarding ecosystems through the monitoring and intervention of organizations such as forest guards and volunteers. Search and rescue aims to find missing persons and bring them to safety, which is often carried out by organizations like firefighters and EMS. Border security, carried out by the police or specialized guard services, focuses on controlling borders to ensure security and economic well-being. Overall, emergency management centralizes command and control for public safety agencies during emergencies. To realize these functions within \ac{PPDR} services, real-time access to information via a broadband connection is critical. This opens the door to a variety of data-centric, multimedia applications that greatly enhance the capabilities for communication in emergency scenarios.

The white paper \cite{bapon_fakhruddin_2019_3406127} proposes the next generation of disaster data infrastructure to successfully collect, process, and display disaster data in reducing the impact of natural hazards. Data collection plays a crucial role for the proposed solutions in\cite{bapon_fakhruddin_2019_3406127}, and most of the time, this data collection is completed through mobile networks. Therefore, uninterrupted connectivity is required. However, it is likely to be affected by a disaster strike. On the other hand, all these mobile network technologies require energy sources to operate and are also prone to disaster strikes. With the United Nations (UN) aiming to reach net-zero CO$_2$ emissions by 2050, the energy consumption of \ac{RAN} components such as \glspl{BS} and data centers, powered mainly by non-renewable sources, poses a significant challenge to sustainability efforts \cite{UN_sustainable}. Transitioning to green energy models in powering \ac{RAN} infrastructure and energy-efficient solutions is crucial for reducing carbon footprints and ensuring that next-generation communication systems align with global sustainability goals. This is paramount since the information and communication technologies (ICT) sector is expected to account for up to 20 percent of global energy consumption by 2030 \cite{kement2023sustaining}. Hence, green energy technologies and integrated communication architectures should be considered in the development of disaster-resistant communication systems.

\subsection{Related Surveys and Reviews}

In order to emphasize the importance of communication and energy issues in disaster scenarios, many survey papers have been presented. In this section, a brief summary of them will be \textcolor{black}{presented} to understand which gap in the literature will be \textcolor{black}{covered} by our survey paper. In \cite{gomes2016survey}, the network solutions for different phases of disasters have been examined. Approaches for network vulnerability assessment and strategies for enhancing the robustness of an existing network, and solutions for achieving resilient routing, including disaster-aware routing, have been presented.
In \cite{2017Wang}, a hybrid communication network architecture that combined ground, air, and space levels \textcolor{black}{was examined} for emergency communication scenarios, and the challenges of this hybrid network \textcolor{black}{were also discussed} in a detailed survey. In \cite{2018Pervez}, the design choices and the current status of the major wireless-based emergency response systems proposed in the literature were examined, and a comprehensive comparison of the wireless-based emergency response technologies based on various considerations (including bandwidth, range, and throughput) \textcolor{black}{was performed}. In \cite{PozzaNIB2018}, a concept called Networks-In-a-Box (i.e., networks characterized by a low number of physical devices) \textcolor{black}{was presented}. These networks have been designed to provide on-demand connectivity to rescue operators and survivors in after-disaster scenarios or to support soldiers on the battlefield. \ac{MANET} technology, which could be established temporarily in disaster-stricken areas, \textcolor{black}{was studied} for different routing protocols in \cite{2019Jahir}. A systematic review study that focused on the recent technologies for emergency communication systems \textcolor{black}{was also presented} in \cite{2022Coch_WT}. In \cite{2022Debnath}, the widely used communication technologies that were applied for setting up an emergency communication network to mitigate the disaster aftermath \textcolor{black}{were examined}. The authors \textcolor{black}{of} \cite{2022Matracia} presented a detailed survey paper on post-disaster communications. They \textcolor{black}{mentioned} the wireless technologies as well as physical and network layer issues, and proposed a use case scenario in their work. Wireless technologies \textcolor{black}{were classified} into three categories: recovery of terrestrial networks, installation of aerial networks, and use of space/satellite networks. In the physical layer issues, the related works \textcolor{black}{were evaluated} in terms of channel modeling, coverage, capacity, radio resource management, localization, and energy efficiency. Moreover, the existing literature \textcolor{black}{was classified and discussed} in terms of routing, delay-tolerant networks, edge computing, and integrated space-air-ground architecture in the network layer issues. In \cite{wang2023overview}, extensive research into emergency communication technology, including satellite networks, \textit{ad hoc} networks, cellular networks, and wireless private networks, \textcolor{black}{was presented}. The networks used in emergency rescue operations and the future development direction of emergency communication networks \textcolor{black}{were also analyzed}. In \cite{krichen2023managing}, some technological advancements such as remote sensing, satellite imaging, and social media \textcolor{black}{were analyzed} along with their opportunities and challenges for different phases of natural disasters. \ac{IoT} solutions in the field of Early Warning for natural disasters \textcolor{black}{were described} in detail in \cite{esposito2022recent,abdalzaher2023early}. In \cite{erdelj2017help}, it \textcolor{black}{was discussed} how \ac{UAV} could be utilized for various tasks such as assessing the extent of damage, identifying affected areas, and aiding in search and rescue missions during natural disasters. Additionally, the importance of integrating \ac{UAV}s with wireless sensor networks to enhance data collection and communication capabilities in disaster-stricken areas \textcolor{black}{was emphasized}. The importance of \ac{UAV}-based solutions and their possible challenges in disaster management \textcolor{black}{was discussed} in \cite{2020Khan}. A systematic review that focused on the \ac{UAV} path planning problem in emergency situations \textcolor{black}{was also presented} in \cite{2021QADIR}.

\ac{PSNs}, which were crucial for public protection and disaster relief, \textcolor{black}{were reviewed} for different technologies in \cite{2013Baldini_PSN,2018YU,2019Jarwan_PSN_LTE,2020Perez_PSN2020,2021Ali,HildmanDrones2019,hasan2021search}. The regulatory and standardization issues of \ac{PSNs} \textcolor{black}{were reviewed} in \cite{2013Baldini_PSN}. The potentials of \ac{D2D} communications and dynamic wireless networks for \ac{PSNs} \textcolor{black}{were analyzed} in \cite{2018YU}. In addition, the progress of standardization of \ac{D2D} and dynamic wireless networks for public safety communications was investigated. \ac{PSNs} were also examined in detail for \ac{LTE} and \ac{5G} technologies in \cite{2019Jarwan_PSN_LTE} and \cite{2020Perez_PSN2020}\cite{2021Ali}, respectively. In \cite{HildmanDrones2019}, the advantages of \ac{UAV}s in different areas, including public safety, \textcolor{black}{were summarized}. The importance of \ac{IoT} technologies for disaster management systems and public safety communications \textcolor{black}{was highlighted} in \cite{2021Ali,hasan2021search}.

Disaster management issues that have been integrated with novel concepts such as \ac{AI}, \ac{ML}, and Blockchain have also been analyzed in detail in different surveys. In \cite{2018Nunawath}, a systematic review \textcolor{black}{was conducted} on the \ac{AI} applications that analyzed and processed social media big data for efficient disaster management. In \cite{Sun2020AI}, an overview of the current applications of \ac{AI} in disaster management during mitigation, preparedness, response, and recovery phases \textcolor{black}{was provided}. In \cite{2021Chamola}, the \ac{ML} algorithms and how they could be combined with other technologies to address disaster and pandemic management \textcolor{black}{were examined} in detail. A survey on the integration of Blockchain with aerial communications for disaster management \textcolor{black}{was provided} in \cite{kumar2021blockchain}.

The literature \textcolor{black}{discussed} above presents communication solutions for disaster management. However, a sustainable (e.g., supported by green energy resources, accessible by everyone, and economically feasible) communication infrastructure that can withstand the challenges of disasters is also a must \cite{ghorbanian2019communication, imoize20216g}. This infrastructure should be resilient to damage, easy to repair, and energy efficient. It should also be affordable and accessible to all. The potential applications and recommendations of 6G in the \ac{RES} sector \textcolor{black}{were discussed} in \cite{yap2022future}. The paper \cite{jahid2019toward} \textcolor{black}{focused} on an essential energy management approach to improve energy efficiency and reduce fuel consumption of off-grid cellular networks whose \glspl{BS} were powered by hybrid power sources including solar \ac{PV} systems and diesel generators (DG). The paper \cite{hassan2019novel} \textcolor{black}{proposed} a new approach to configure and operate \glspl{BS} to provide ancillary services to the smart grid. Depending on various system parameters of \ac{LTE}, a simulation-based feasibility analysis using the Hybrid Optimization Model for Electric Renewables (HOMER) \textcolor{black}{was conducted} to evaluate the optimal system, energy production, total net present cost (NPC), cost of electricity (COE), and greenhouse gas (GHG) emissions in \cite{hossain2020solar}. \textcolor{black}{Reference \cite{lee2019adaptive}} \textcolor{black}{considered} a cellular system in which \glspl{BS} were fed by both renewable and on-grid energy sources. 

\textcolor{black}{Resilience, in this context, refers to “the ability of a network to provide and maintain an acceptable level of service in the face of disaster-induced faults and challenges to normal operation”~\cite{rak2020guide}. It encompasses a broad set of disciplines aimed at ensuring continued service through redundancy, diversity, disruption tolerance, and survivability mechanisms, especially under conditions such as large-scale failures or correlated disruptions.} \textcolor{black}{To evaluate resilience, both topological and functional metrics are outlined in~\cite{rak2020guide}. Topological metrics assess structural properties such as connectivity, centrality, and the size of the largest connected component. Functional metrics, on the other hand, evaluate service performance, including packet loss, latency, jitter, and user-perceived quality indicators like Mean Opinion Score (MOS) and Peak Signal-to-Noise Ratio (PSNR). These metrics provide a quantitative basis for comparing different disaster-resilient communication designs and are crucial for guiding the development of robust infrastructures.} The authors \textcolor{black}{of} \cite{bhusal2020power} \textcolor{black}{provided} a critical review of existing resilience definitions and metrics, and examined widely used approaches from various organizations and researchers. \textcolor{black}{ Reference \cite{batista2024survey} provided a comprehensive taxonomy and systematic analysis of resilience strategies in information sharing across diverse network environments, highlighting applied techniques such as redundancy and monitoring to address threats from malicious behavior, network disruptions, and performance issues.} The authors \textcolor{black}{of} \cite{wang2022systematic} \textcolor{black}{reviewed} power system resilience in terms of generation, networks, and loads. They also \textcolor{black}{discussed} resilience enhancement strategies involving distributed generation, conventional and renewable generators, energy storage, microgrids, load shifting, and demand response. In \cite{younesi2022trends}, the importance of addressing power system resilience and standard resilience definitions \textcolor{black}{was discussed}. The authors also \textcolor{black}{reviewed} the benefits of smart microgrids for enhancing system resilience and demonstrated their effectiveness through numerical simulation case studies. The authors \textcolor{black}{of} \cite{kaloti2023toward} \textcolor{black}{explored} resilience frameworks and metrics while analyzing damage costs and risks associated with extreme events. Moreover, they \textcolor{black}{examined} case studies on network risk estimation and the effectiveness of resilience improvement techniques for enhancing grid resilience.

\subsection{Comparing This Survey with Related Surveys}

A comparative analysis between our survey and other relevant surveys is presented in Table \ref{vertical_table} where N/D \textcolor{black}{indicates} Not-Defined. We have focused on the three main topics while making this comparison: Disaster Phase (Pre, In, Post), Key Contribution (Energy, Communication), and Network structure (Space, Air, Ground, Sea).

Many devastating disasters (such as the series of earthquakes in Turkiye in 2023) have revealed the importance of considering communications and energy solutions together to build resilient and sustainable infrastructure. In these earthquakes, communication was disrupted even in areas where the communication infrastructure remained functional but could not be used due to power outages \cite{karaman2024enhancing}. The previous literature contains certain studies \cite{erdelj2017help,2017Wang, 2018Nunawath, 2018Pervez, PozzaNIB2018, 2018YU, HildmanDrones2019, 2019Jahir, 2019Jarwan_PSN_LTE, 2020Perez_PSN2020, Sun2020AI, 2021Ali, 2021Chamola, 2021QADIR, 2020Khan, 2022Coch_WT, 2022Debnath, 2022Matracia, wang2023overview} concentrated on communication solutions, while others \cite{bhusal2020power, wang2022systematic, younesi2022trends, kaloti2023toward} concentrated on energy solutions, addressing these aspects separately. Compared to these studies, to the best of our knowledge, our study is the first one to consider communication and energy supply solutions jointly.

From a different point of view, our survey focuses on an integrated network model which covers ground, air, space and sea levels contrary to many existing surveys in the literature \cite{erdelj2017help, 2018Nunawath,2019Jahir, 2020Perez_PSN2020,PozzaNIB2018,2018Pervez,2019Jarwan_PSN_LTE,2021Ali}. This integrated network model is vital for disaster management because it provides a complete view of a situation by combining information from satellites, planes, ground sensors, and sea sources. This system offers real-time data, improving the accuracy of damage assessment and response planning. It also enables smooth communication and coordination among different organizations and stakeholders through a unified data platform. By utilizing various data sources, an integrated network model helps to assess risks, respond to emergencies effectively, and allocate resources efficiently. Overall, this integrated network strengthens disaster resilience and improves the management of both immediate and long-term impacts.

Another feature that sets our article apart from others is that communication and energy technologies are examined in detail for each phase of the disaster individually. In many other publications, this distinction has not been thoroughly considered \cite{2017Wang, 2018Nunawath,  PozzaNIB2018,2019Jahir,2021QADIR,wang2023overview}.

\textcolor{black}{Furthermore, recent disaster scenarios have shown the importance of rapidly deployable and resilient communication systems. For example, Starlink satellite terminals were deployed to restore connectivity in Tonga following the volcanic eruption in 2022, while AT\&T’s Flying Cell on Wings drone-based LTE base stations were deployed in hurricane-hit areas \cite{2022Matracia}. Similarly, FirstNet in the U.S. offers a dedicated public-safety LTE network on Band 14 \cite{firstnet}. While prior studies such as \cite{gomes2016survey}, \cite{PozzaNIB2018}, \cite{2022Debnath}, \cite{wang2023overview}, \cite{ghorbanian2019communication}, and  \cite{yap2022future} have made valuable contributions by analyzing architectural aspects for emergency communication, they often focus on either conceptual frameworks or individual system categories. In contrast, this survey distinguishes itself by presenting a comprehensive, cross-layer analysis of both the academic proposals and the operational solutions deployed by the industry. This survey presents not only a comprehensive taxonomy but also a detailed evaluation of real-world technologies and vendor-supported platforms, with a focus on sustainability, interoperability, deployability, and performance under disaster-related constraints.}

\textcolor{black}{It is worth noting that sustainability in communication infrastructures encompasses a wider spectrum of considerations beyond energy efficiency. These include economic viability, environmental impact, infrastructure reuse, modular deployment, and community inclusiveness. In this survey, we focus primarily on energy-oriented sustainability. We therefore comprehensively evaluate solutions that both mitigate the impact of power-grid failures during disasters and ensure that a significant share of the network’s energy consumption in normal operations is supplied by renewable sources.}

\subsection{Contributions}
Several emerging technologies are being integrated into disaster management to enhance its effectiveness across all stages of the life-cycle. However, existing surveys tend to focus on either communication technologies or energy solutions in isolation, missing the critical integration of these enablers. This paper discusses prospective solutions for sustainable communication infrastructure in disaster relief and management scenarios.

The key issues addressed in this paper are as follows: 
 
\begin{itemize}
    \item  This paper identifies and evaluates emerging key technologies that enhance communication networks during pre-disaster, in-disaster, and post-disaster phases. It specifically examines how these technologies can be effectively deployed to support disaster relief and management, emphasizing their role in maintaining communication continuity. 
    \item  This paper underscores the critical importance of developing robust and resilient communication networks that can withstand the impacts of disasters, particularly in scenarios like the Turkiye earthquakes. It highlights the necessity of integrating energy requirements for sustainability and resiliency, ensuring that communication infrastructures remain operational even in the most challenging conditions. 
    \item This paper explores the main advantages that emerging communication technologies, such as \ac{UAV}s, \ac{HAPS}, mesh networks, and satellite communication, can bring to disaster relief scenarios compared to traditional networks. It discusses how these technologies can enhance the efficiency, speed, and reliability of communication during emergencies, thereby improving overall disaster response efforts. 
    
   \item \textcolor{black}{This paper introduces a comprehensive integrated network architecture that bridges space, air, ground and sea. communication layers. Unlike previous studies that are limited to a single network domain, this work emphasizes the importance of multi-domain coordination to ensure end-to-end communication resilience and situational awareness across heterogeneous platforms during disasters.} 
    
    \item This paper addresses the limitations of both existing and new communication technologies in disaster relief and management scenarios. It evaluates the challenges in implementing these technologies, including energy constraints, deployment difficulties, and the need for continuous innovation to overcome these barriers. 
    \item This paper reviews the progress made in constructing robust communication infrastructure that can support disaster scenarios, with a particular focus on energy sustainability and resiliency. It highlights case studies, including the Turkiye earthquakes, to demonstrate how these infrastructures have been tested and improved in real-world situations, ensuring enhanced preparedness and response capabilities.

    \item \textcolor{black}{This paper presents a cross-layer analytical framework that not only categorizes emerging technologies, but also assesses their interoperability, deployability and energy efficiency under disaster-induced constraints. By combining academic findings with real-world implementations and ongoing standardization efforts, the gap between theoretical models and operational disaster communication systems is bridged.}

\end{itemize}

\begin{sidewaystable*}[htp!]
\scriptsize
\captionsetup{font=sc,  position=above, justification=centering, labelsep=newline,singlelinecheck=true}
\centering
 \caption{Summary of survey literature on disaster response and planning phases in communication and energy domain}
\label{vertical_table}
\centering
\begin{tabular}{|l|l|l|lll|l|ll|llll|}
\hline
\rowcolor{gray!25}
\multirow{1}{*}{\textbf{Refs}} & \multicolumn{1}{c|}{\multirow{1}{*}{\textbf{Year}}} & \multicolumn{1}{c|}{\multirow{1}{*}{\textbf{Description}}} & \multicolumn{3}{c|}{\textbf{Disaster Phase}} & \multicolumn{1}{c|}{\multirow{2}{*}{\textbf{\begin{tabular}[c]{@{}c@{}}Disaster \\ Type\end{tabular}}}} & \multicolumn{2}{c|}{\textbf{Key contributions}} & \multicolumn{4}{c|}{\textbf{Network}} \\ \cline{4-6} \cline{8-13} 
\rowcolor{gray!25}
& \multicolumn{1}{c|}{} & \multicolumn{1}{c|}{} & \multicolumn{1}{c|}{\textbf{Pre}} & \multicolumn{1}{c|}{\textbf{In}} & \textbf{Post} & \multicolumn{1}{c|}{} & \multicolumn{1}{c|}{\textbf{Comm.}} & \multicolumn{1}{c|}{\textbf{Energy}} & \multicolumn{1}{l|}{\textbf{Space}} & \multicolumn{1}{l|}{\textbf{Air}} & \multicolumn{1}{l|}{\textbf{Ground}} & \textbf{Sea} \\ \hline

\multicolumn{1}{|c|}{ Erdelj et al. \cite{erdelj2017help}} & \multicolumn{1}{c|}{2017} & \multicolumn{1}{c|}{\begin{tabular}[c]{@{}c@{}}Discusses  how UAVs can be utilized for \\ various  tasks such as assessing the extent of  damage, \\ identifying affected areas. \end{tabular} }  &  \multicolumn{1}{l|}{\checkmark} & \multicolumn{1}{l|}{\checkmark} & \multicolumn{1}{l|}{\checkmark} & \multicolumn{1}{l|}{\begin{tabular}[c]{@{}c@{}} All natural and man made disasters \\ are examined in  a categorized fashion. \end{tabular} } & \multicolumn{1}{l|}{\checkmark} & \multicolumn{1}{l|}{x}  & \multicolumn{1}{l|}{x} & \multicolumn{1}{l|}{\checkmark} & \multicolumn{1}{l|}{\checkmark} & \multicolumn{1}{l|}{x} \\ \hline

\multicolumn{1}{|c|}{ Wang et al. \cite{2017Wang}} & \multicolumn{1}{c|}{2017} & \multicolumn{1}{c|}{\begin{tabular}[c]{@{}c@{}}Focus on the  hybrid Satellite-Aerial-Terrestrial \\ Networks in
Emergency Scenarios. \end{tabular} }  &  \multicolumn{1}{l|}{N/D} & \multicolumn{1}{l|}{N/D} & \multicolumn{1}{l|}{N/D} & \multicolumn{1}{l|}{\begin{tabular}[c]{@{}c@{}} Natural disasters not for a specific one. \end{tabular} } & \multicolumn{1}{l|}{\checkmark} & \multicolumn{1}{l|}{x}  & \multicolumn{1}{l|}{\checkmark} & \multicolumn{1}{l|}{\checkmark} & \multicolumn{1}{l|}{\checkmark} & \multicolumn{1}{l|}{x} \\ \hline

 \multicolumn{1}{|c|}{ \begin{tabular}[c]{@{}c@{}} Nunawath and \\ Goodwin \cite{2018Nunawath}.
 \end{tabular}   } & \multicolumn{1}{c|}{2018} & \multicolumn{1}{c|}{\begin{tabular}[c]{@{}c@{}}Review on application of  AI on  social media big data \\ for efficient disaster management.
 \end{tabular} }  &  \multicolumn{1}{l|}{x} & \multicolumn{1}{l|}{\checkmark} & \multicolumn{1}{l|}{x} & \multicolumn{1}{l|}{\begin{tabular}[c]{@{}c@{}} Any natural or man  made disasters. \end{tabular} } & \multicolumn{1}{l|}{\checkmark} & \multicolumn{1}{l|}{x}  & \multicolumn{1}{l|}{x} & \multicolumn{1}{l|}{x} & \multicolumn{1}{l|}{\checkmark} & \multicolumn{1}{l|}{x} \\ \hline

\multicolumn{1}{|c|}{ Pervez et al. \cite{2018Pervez} } & \multicolumn{1}{c|}{2018} & \multicolumn{1}{c|}{\begin{tabular}[c]{@{}c@{}}Survey of the design choices and the current status of \\ the major  wireless-based emergency response systems.
 \end{tabular} }  &  \multicolumn{1}{l|}{x} & \multicolumn{1}{l|}{\checkmark} & \multicolumn{1}{l|}{\checkmark} & \multicolumn{1}{l|}{\begin{tabular}[c]{@{}c@{}} Natural disaster, terrorism  event, \\ battlefield scenario, \ and patient  \\ monitoring. \end{tabular} } & \multicolumn{1}{l|}{\checkmark} & \multicolumn{1}{l|}{x}  & \multicolumn{1}{l|}{x} & \multicolumn{1}{l|}{x} & \multicolumn{1}{l|}{\checkmark} & \multicolumn{1}{l|}{x} \\ \hline

  \multicolumn{1}{|c|}{ Pozza et al. \cite{PozzaNIB2018}  } & \multicolumn{1}{c|}{2018} & \multicolumn{1}{c|}{\begin{tabular}[c]{@{}c@{}}Networks-In-a-Box to provide on-demand connectivity \\
to rescue operators and survivors.
 \end{tabular} }  &  \multicolumn{1}{l|}{x} & \multicolumn{1}{l|}{x} & \multicolumn{1}{l|}{\checkmark} & \multicolumn{1}{l|}{\begin{tabular}[c]{@{}c@{}}Any natural or man made disasters. \end{tabular} } & \multicolumn{1}{l|}{\checkmark} & \multicolumn{1}{l|}{x}  & \multicolumn{1}{l|}{\checkmark} & \multicolumn{1}{l|}{x} & \multicolumn{1}{l|}{\checkmark} & \multicolumn{1}{l|}{x} \\ \hline

\multicolumn{1}{|c|}{ Yu et al. \cite{2018YU} } & \multicolumn{1}{c|}{2018} & \multicolumn{1}{c|}{\begin{tabular}[c]{@{}c@{}} Public safety communication is addressed.
 \end{tabular} }  &  \multicolumn{1}{l|}{\checkmark} & \multicolumn{1}{l|}{\checkmark} & \multicolumn{1}{l|}{\checkmark} & \multicolumn{1}{l|}{\begin{tabular}[c]{@{}c@{}} Any natural or man made disasters. \end{tabular} } & \multicolumn{1}{l|}{\checkmark} & \multicolumn{1}{l|}{x}  & \multicolumn{1}{l|}{x} & \multicolumn{1}{l|}{\checkmark} & \multicolumn{1}{l|}{\checkmark} & \multicolumn{1}{l|}{x} \\ \hline

\multicolumn{1}{|c|}{ \begin{tabular}[c]{@{}c@{}} Hildman and \\ Kovacs \cite{HildmanDrones2019}.
 \end{tabular}    } & \multicolumn{1}{c|}{2019} & \multicolumn{1}{c|}{\begin{tabular}[c]{@{}c@{}} Advantages of UAV in different \\ areas including the public safety.
 \end{tabular} }  &  \multicolumn{1}{l|}{\checkmark} & \multicolumn{1}{l|}{\checkmark} & \multicolumn{1}{l|}{\checkmark} & \multicolumn{1}{l|}{\begin{tabular}[c]{@{}c@{}} Any natural or man  made disasters. \end{tabular} } & \multicolumn{1}{l|}{\checkmark} & \multicolumn{1}{l|}{x}  & \multicolumn{1}{l|}{x} & \multicolumn{1}{l|}{\checkmark} & \multicolumn{1}{l|}{x} & \multicolumn{1}{l|}{x} \\ \hline

\multicolumn{1}{|c|}{ Jahir et al. \cite{2019Jahir}  } & \multicolumn{1}{c|}{2019} & \multicolumn{1}{c|}{\begin{tabular}[c]{@{}c@{}}  Advantages, disadvantages and performance of the routing protocols \\ used in the disaster area networks.
 \end{tabular} }  &  \multicolumn{1}{l|}{x} & \multicolumn{1}{l|}{\checkmark} & \multicolumn{1}{l|}{x} & \multicolumn{1}{l|}{\begin{tabular}[c]{@{}c@{}} Any natural or man made disasters. \end{tabular} } & \multicolumn{1}{l|}{\checkmark} & \multicolumn{1}{l|}{x}  & \multicolumn{1}{l|}{x} & \multicolumn{1}{l|}{x} & \multicolumn{1}{l|}{\checkmark} & \multicolumn{1}{l|}{x} \\ \hline

\multicolumn{1}{|c|}{ Jarwan et al. \cite{2019Jarwan_PSN_LTE} } & \multicolumn{1}{c|}{2019} & \multicolumn{1}{c|}{\begin{tabular}[c]{@{}c@{}}  LTE based \ac{PSNs}
 \end{tabular} }  &  \multicolumn{1}{l|}{\checkmark} & \multicolumn{1}{l|}{\checkmark} & \multicolumn{1}{l|}{\checkmark} & \multicolumn{1}{l|}{\begin{tabular}[c]{@{}c@{}} Any natural or man made disasters. \end{tabular} } & \multicolumn{1}{l|}{\checkmark} & \multicolumn{1}{l|}{x}  & \multicolumn{1}{l|}{x} & \multicolumn{1}{l|}{x} & \multicolumn{1}{l|}{\checkmark} & \multicolumn{1}{l|}{x} \\ \hline

\multicolumn{1}{|c|}{ Perez et al. \cite{2020Perez_PSN2020} } & \multicolumn{1}{c|}{2020} & \multicolumn{1}{c|}{\begin{tabular}[c]{@{}c@{}}  Reviews \ac{PSNs} based on 5G.
 \end{tabular} }  &  \multicolumn{1}{l|}{\checkmark} & \multicolumn{1}{l|}{\checkmark} & \multicolumn{1}{l|}{x} & \multicolumn{1}{l|}{\begin{tabular}[c]{@{}c@{}} Any natural or man made disasters. \end{tabular} } & \multicolumn{1}{l|}{\checkmark} & \multicolumn{1}{l|}{x}  & \multicolumn{1}{l|}{x} & \multicolumn{1}{l|}{x} & \multicolumn{1}{l|}{\checkmark} & \multicolumn{1}{l|}{x} \\ \hline

\multicolumn{1}{|c|}{ Sun et al. \cite{Sun2020AI} } & \multicolumn{1}{c|}{2020} & \multicolumn{1}{c|}{\begin{tabular}[c]{@{}c@{}}  An overview of current applications of AI in disaster management.
 \end{tabular} }  &  \multicolumn{1}{l|}{\checkmark} & \multicolumn{1}{l|}{\checkmark} & \multicolumn{1}{l|}{\checkmark} & \multicolumn{1}{l|}{\begin{tabular}[c]{@{}c@{}} Any natural or man  made disasters. \end{tabular} } & \multicolumn{1}{l|}{\checkmark} & \multicolumn{1}{l|}{x}  & \multicolumn{1}{l|}{N/D} & \multicolumn{1}{l|}{N/D} & \multicolumn{1}{l|}{N/D} & \multicolumn{1}{l|}{N/D} \\ \hline

\multicolumn{1}{|c|}{ Bhusal et al. \cite{bhusal2020power} } & \multicolumn{1}{c|}{2020} & \multicolumn{1}{c|}{\begin{tabular}[c]{@{}c@{}}  Reviews current power system resilience  metrics \\ and evaluation methods, and   discusses future directions for developing \\ standardized  definitions, metrics, and enhancement strategies.
 \end{tabular} }  &  \multicolumn{1}{l|}{\checkmark} & \multicolumn{1}{l|}{\checkmark} & \multicolumn{1}{l|}{\checkmark} & \multicolumn{1}{l|}{\begin{tabular}[c]{@{}c@{}} Any natural or man made disasters. \end{tabular} } & \multicolumn{1}{l|}{x} & \multicolumn{1}{l|}{\checkmark}  & \multicolumn{1}{l|}{x} & \multicolumn{1}{l|}{x} & \multicolumn{1}{l|}{\checkmark} & \multicolumn{1}{l|}{x} \\ \hline

\multicolumn{1}{|c|}{ Ali et al. \cite{2021Ali} } & \multicolumn{1}{c|}{2021} & \multicolumn{1}{l|}{\begin{tabular}[c]{@{}c@{}}  Discusses the suitability of potential \\ 5G wireless communication technologies in disaster.
 \end{tabular} }  &  \multicolumn{1}{l|}{\checkmark} & \multicolumn{1}{l|}{\checkmark} & \multicolumn{1}{l|}{\checkmark} & \multicolumn{1}{l|}{\begin{tabular}[c]{@{}c@{}} Any natural disasters. \end{tabular} } & \multicolumn{1}{l|}{\checkmark} & \multicolumn{1}{l|}{x}  & \multicolumn{1}{l|}{x} & \multicolumn{1}{l|}{\checkmark} & \multicolumn{1}{l|}{\checkmark} & \multicolumn{1}{l|}{x} \\ \hline

\multicolumn{1}{|c|}{Chamola et al. \cite{2021Chamola}} & \multicolumn{1}{c|}{2021} & \multicolumn{1}{c|}{\begin{tabular}[c]{@{}c@{}} ML to address disaster and pandemic management.
 \end{tabular} }  &  \multicolumn{1}{l|}{\checkmark} & \multicolumn{1}{l|}{x} & \multicolumn{1}{l|}{\checkmark} & \multicolumn{1}{l|}{\begin{tabular}[c]{@{}c@{}} Any natural  disasters. \end{tabular} } & \multicolumn{1}{l|}{\checkmark} & \multicolumn{1}{l|}{x}  & \multicolumn{1}{l|}{N/D} & \multicolumn{1}{l|}{N/D} & \multicolumn{1}{l|}{N/D} & \multicolumn{1}{l|}{x} \\ \hline

\multicolumn{1}{|c|}{Qadir et al. \cite{2021QADIR}} & \multicolumn{1}{c|}{2021} & \multicolumn{1}{c|}{\begin{tabular}[c]{@{}c@{}}  Path planning algorithms for UAVs \\  for disaster management.
 \end{tabular} }  &  \multicolumn{1}{l|}{x} & \multicolumn{1}{l|}{x} & \multicolumn{1}{l|}{\checkmark} & \multicolumn{1}{l|}{\begin{tabular}[c]{@{}c@{}} Flood, earthquake and bush fire. \end{tabular} } & \multicolumn{1}{l|}{\checkmark} & \multicolumn{1}{l|}{x}  & \multicolumn{1}{l|}{x} & \multicolumn{1}{l|}{\checkmark} & \multicolumn{1}{l|}{\checkmark} & \multicolumn{1}{l|}{x} \\ \hline

\multicolumn{1}{|c|}{ Wang et al. \cite{wang2022systematic} } & \multicolumn{1}{c|}{2022} & \multicolumn{1}{c|}{\begin{tabular}[c]{@{}c@{}}  Presents an analysis of power systems \\ in terms of impact analysis, impact quantification, \\ and resilience improvement against disasters.
 \end{tabular} }  &  \multicolumn{1}{l|}{\checkmark} & \multicolumn{1}{l|}{\checkmark} & \multicolumn{1}{l|}{\checkmark} & \multicolumn{1}{l|}{\begin{tabular}[c]{@{}c@{}} Any natural or man made disasters. \end{tabular} } & \multicolumn{1}{l|}{x} & \multicolumn{1}{l|}{\checkmark}  & \multicolumn{1}{l|}{x} & \multicolumn{1}{l|}{x} & \multicolumn{1}{l|}{\checkmark} & \multicolumn{1}{l|}{x} \\ \hline

\multicolumn{1}{|c|}{ Younesi et al. \cite{younesi2022trends} } & \multicolumn{1}{c|}{2022} & \multicolumn{1}{c|}{\begin{tabular}[c]{@{}c@{}}  Discusses methods to evaluate and enhance \\ modern power system resilience, comparing common \\  metrics for both long-term and \\ short-term resilience assessment.
 \end{tabular} }  &  \multicolumn{1}{l|}{\checkmark} & \multicolumn{1}{l|}{\checkmark} & \multicolumn{1}{l|}{\checkmark} & \multicolumn{1}{l|}{\begin{tabular}[c]{@{}c@{}} Any natural or man made disasters. \end{tabular} } & \multicolumn{1}{l|}{x} & \multicolumn{1}{l|}{\checkmark}  & \multicolumn{1}{l|}{x} & \multicolumn{1}{l|}{x} & \multicolumn{1}{l|}{\checkmark} & \multicolumn{1}{l|}{x} \\ \hline
 
\multicolumn{1}{|c|}{ Khan et al. \cite{2020Khan}  } & \multicolumn{1}{c|}{2022} & \multicolumn{1}{c|}{\begin{tabular}[c]{@{}c@{}} Surveys how UAVs can be helpful in \\ any disaster scenario
 \end{tabular} }  &  \multicolumn{1}{l|}{\checkmark} & \multicolumn{1}{l|}{\checkmark} & \multicolumn{1}{l|}{\checkmark} & \multicolumn{1}{l|}{\begin{tabular}[c]{@{}c@{}} Any natural or man made disasters. \end{tabular} } & \multicolumn{1}{l|}{\checkmark} & \multicolumn{1}{l|}{x}  & \multicolumn{1}{l|}{x} & \multicolumn{1}{l|}{\checkmark} & \multicolumn{1}{l|}{\checkmark} & \multicolumn{1}{l|}{x} \\ \hline

\multicolumn{1}{|c|}{Coch et al. \cite{2022Coch_WT}} & \multicolumn{1}{c|}{2022} & \multicolumn{1}{c|}{\begin{tabular}[c]{@{}c@{}}  Communication technologies with their challenges \\ in emergency situations.
 \end{tabular} }  &  \multicolumn{1}{l|}{\checkmark} & \multicolumn{1}{l|}{x} & \multicolumn{1}{l|}{\checkmark} & \multicolumn{1}{l|}{\begin{tabular}[c]{@{}c@{}} Any natural or man made disasters. \end{tabular} } & \multicolumn{1}{l|}{\checkmark} & \multicolumn{1}{l|}{x}  & \multicolumn{1}{l|}{\checkmark} & \multicolumn{1}{l|}{\checkmark} & \multicolumn{1}{l|}{\checkmark} & \multicolumn{1}{l|}{x} \\ \hline

\multicolumn{1}{|c|}{Debnath et al. \cite{2022Debnath} } & \multicolumn{1}{c|}{2022} & \multicolumn{1}{c|}{\begin{tabular}[c]{@{}c@{}}  Communication
technologies applied for setting up \\ an emergency communication.
 \end{tabular} }  &  \multicolumn{1}{l|}{\checkmark} & \multicolumn{1}{l|}{\checkmark} & \multicolumn{1}{l|}{\checkmark} & \multicolumn{1}{l|}{\begin{tabular}[c]{@{}c@{}} Any natural or man made disasters. \end{tabular} } & \multicolumn{1}{l|}{\checkmark} & \multicolumn{1}{l|}{x}  & \multicolumn{1}{l|}{x} & \multicolumn{1}{l|}{\checkmark} & \multicolumn{1}{l|}{\checkmark} & \multicolumn{1}{l|}{x} \\ \hline

\multicolumn{1}{|c|}{Matracia et al. \cite{2022Matracia}  } & \multicolumn{1}{c|}{2022} & \multicolumn{1}{c|}{\begin{tabular}[c]{@{}c@{}}  Technical issues related to \\ post-disaster networks.
 \end{tabular} }  &  \multicolumn{1}{l|}{x} & \multicolumn{1}{l|}{\checkmark} & \multicolumn{1}{l|}{\checkmark} & \multicolumn{1}{l|}{\begin{tabular}[c]{@{}c@{}} Any natural disasters. \end{tabular} } & \multicolumn{1}{l|}{\checkmark} & \multicolumn{1}{l|}{x}  & \multicolumn{1}{l|}{\checkmark} & \multicolumn{1}{l|}{\checkmark} & \multicolumn{1}{l|}{\checkmark} & \multicolumn{1}{l|}{x} \\ \hline

\multicolumn{1}{|c|}{ Kaloti et al. \cite{kaloti2023toward} } & \multicolumn{1}{c|}{2023} & \multicolumn{1}{c|}{\begin{tabular}[c]{@{}c@{}}  Analyzes various definitions, frameworks, \\ and metrics for resilience, followed by an \\ assessment of damage costs and risks from extreme events.
 \end{tabular} }  &  \multicolumn{1}{l|}{\checkmark} & \multicolumn{1}{l|}{\checkmark} & \multicolumn{1}{l|}{\checkmark} & \multicolumn{1}{l|}{\begin{tabular}[c]{@{}c@{}} Any natural or man made disasters. \end{tabular} } & \multicolumn{1}{l|}{x} & \multicolumn{1}{l|}{\checkmark}  & \multicolumn{1}{l|}{x} & \multicolumn{1}{l|}{x} & \multicolumn{1}{l|}{\checkmark} & \multicolumn{1}{l|}{x} \\ \hline

\multicolumn{1}{|c|}{Wang et al. \cite{wang2023overview} } & \multicolumn{1}{c|}{2023} & \multicolumn{1}{c|}{\begin{tabular}[c]{@{}c@{}}  Emergency communication technology,
including  satellite, \\ 	\textit{ad hoc} ,  cellular, and wireless  private networks.
 \end{tabular} }  &  \multicolumn{1}{l|}{x} & \multicolumn{1}{l|}{\checkmark} & \multicolumn{1}{l|}{\checkmark} & \multicolumn{1}{l|}{\begin{tabular}[c]{@{}c@{}} Any natural  and man made disasters. \end{tabular} } & \multicolumn{1}{l|}{\checkmark} & \multicolumn{1}{l|}{x}  & \multicolumn{1}{l|}{\checkmark} & \multicolumn{1}{l|}{\checkmark} & \multicolumn{1}{l|}{\checkmark} & \multicolumn{1}{l|}{\checkmark} \\ \hline

\multicolumn{1}{|c|}{\textbf{Our paper}}  & 2025 &  \multicolumn{1}{c|}{\begin{tabular}[c]{@{}c@{}}  Consider energy and  communication enablers\\ for all phases of  disaster response and planning \\  Specialized use case for Turkiye earthquake
\end{tabular}} & \multicolumn{1}{l|}{\checkmark} & \multicolumn{1}{l|}{\checkmark} &  \multicolumn{1}{l|}{\checkmark} & \multicolumn{1}{l|}{\begin{tabular}[c]{@{}c@{}} Earthquake specific disasters. \end{tabular} }  & \multicolumn{1}{l|}{\checkmark} &  \multicolumn{1}{l|}{\checkmark} & \multicolumn{1}{l|}{\checkmark} & \multicolumn{1}{l|}{\checkmark} & \multicolumn{1}{l|}{\checkmark} &  \multicolumn{1}{l|}{\checkmark} \\ \hline
\end{tabular}
\label{summary}
\end{sidewaystable*}

\subsection{Organization}

\begin{figure*}[htp!]
\centering
\includegraphics[width=.99\linewidth]{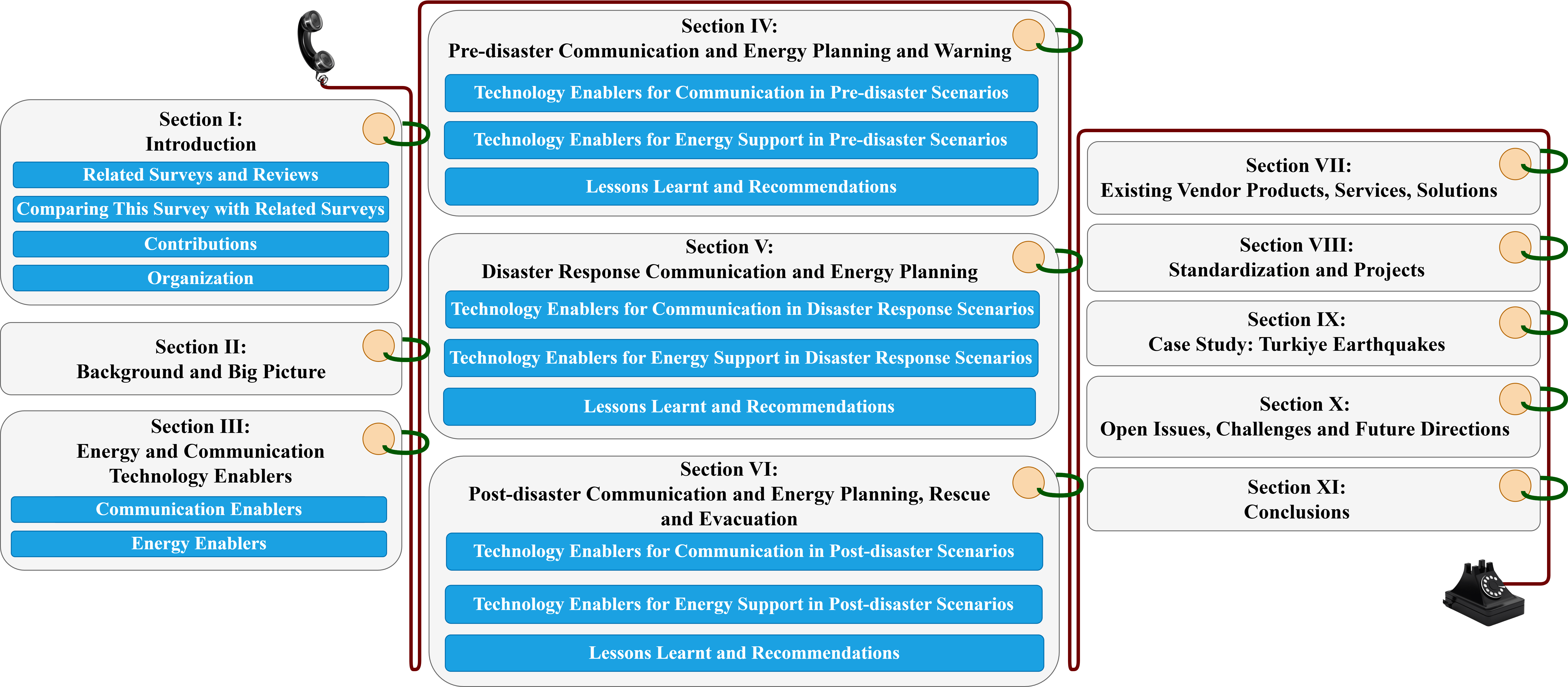}
\caption{Schematic representation of the paper's structure.}
\label{survey_outline}
\end{figure*}

The structure of this paper is schematically \textcolor{black}{presented} in Fig. \ref{survey_outline}. Section \ref{background} provides a comprehensive background and global perspective on the current state of disaster communication and energy management. Section \ref{EnergyCommunicationEnablers} discusses the key enablers of communication and energy technologies, focusing on emerging trends that enhance resilience. Section \ref{predisaster} addresses pre-disaster communication and energy planning, emphasizing the importance of proactive strategies for mitigating disaster impacts. Section \ref{indisaster} provides a detailed exploration of technology enablers during the disaster response phase, including real-time communication and energy solutions that support rapid deployment and coordination.  Section \ref{afterdisaster}  covers post-disaster communication planning, focusing on rescue and evacuation efforts, and the technologies that support these critical operations. Section \ref{vendors} examines existing vendor products and services. Section \ref{standadization} provides standardization efforts, and ongoing projects that contribute to building resilient infrastructures. A case study on the Turkiye earthquakes is presented to illustrate the practical application of the discussed technologies and strategies in Section \ref{case_study}. Section \ref{issues} of the paper discusses open issues, challenges and future directions for the realization of sustainable and resilient communication infrastructures, Finally, Section \ref{conclusion} gives conclusions of the paper.

\section{Background and \textcolor{black}{Big} Picture}
\label{background}

The UN's disaster management life-cycle \cite{UN_disasterPhases} consists of pre-disaster (mitigation and preparedness), response, and post-disaster recovery activities.\footnote{\textcolor{black}{To cope with the effects of natural hazards, the disaster management plans for wireless infrastructure may include different phases. For instance, in \cite{STERBENZ20101245}, the authors propose a disaster management plan where the diagnose and refine phases aim to prepare the wireless infrastructure 
better for a new disaster following lessons learnt from the recent one. However, in this paper, we follow the UN's structure since it is widely accepted.}} Each phase plays a critical role in mitigating the effects of disasters and facilitating effective recovery. A detailed explanation of each phase is given as follows.

\textbf{1) Pre-Disaster Phase:} The pre-disaster phase focuses on activities aimed at mitigating the effects of disasters and preparing for their occurrence \cite{WHO_preparedness}. This phase includes:
\begin{itemize}[leftmargin=*]
    \item \textit{Mitigation:} The mitigation involves identifying and implementing measures to reduce the vulnerability of communities and infrastructure to disasters. These may include land use planning, building codes and regulations, infrastructure improvements, and public awareness campaigns.
    \item \textit{Preparedness:} Preparedness activities involve the development of plans, procedures, and resources to enable an effective response to disasters. These efforts include creating emergency response plans, conducting simulation exercises, establishing robust communication networks, and training emergency personnel. \textcolor{black}{Disasters can be broadly  classified into predictable and unpredictable events \cite{rak2020guide}, and accordingly, the scope and focus of preparedness largely depend on the predictability of the disaster. For predictable disasters such as hurricanes, floods and some wildfires, preparedness measures can be more targeted and time-sensitive. Forecasting systems, early warning mechanisms and evacuation protocols can be activated in advance, enabling the strategic deployment of temporary infrastructure (e.g. mobile LTE stations, backup power units ) and pre-positioning of emergency supplies. In contrast, there is little to no warning in the event of unpredictable disasters such as earthquakes, industrial accidents or tsunamis triggered by seismic events. Therefore, preparedness for these events tends to focus on long-term strategies, including building resilient and redundant communications and energy systems, enforcing strict security regulations, and ensuring that emergency response teams and communities are trained to act quickly and autonomously in uncertain conditions.}
\end{itemize}

The goal of the pre-disaster phase is to minimize potential damage and loss of life by being proactive and well-prepared when disaster strikes.

\textbf{2) Response Phase:} The response phase takes place during and immediately after a disaster. It focuses on immediate actions taken to save lives, protect property, and meet the basic needs of affected people \cite{WHO_response}. Key elements of the response phase include:
\begin{itemize}[leftmargin=*]
    \item \textit{Emergency warning and communications:} Rapid and accurate communications are critical during the response phase. This includes issuing early warnings, activating emergency alert systems, and establishing communication channels for emergency responders and affected communities.
    \item \textit{Search and Rescue:} Search and rescue operations are carried out to find and rescue people who may be trapped or in immediate danger. This includes coordination between emergency teams with specialized equipment and techniques to locate and rescue survivors.
    \item \textit{Emergency Shelter and Relief:} Providing emergency shelter, food, water, medical assistance, and other essentials to affected people is an important aspect of the response phase. Emergency shelters and distribution centers will be set up to meet the immediate needs of people affected by the disaster.
\end{itemize}

\textbf{3) Post-Disaster Recovery Phase:}
The post-disaster recovery phase focuses on the recovery and reconstruction of disaster-affected communities. The post-disaster recovery phase aims to build stronger and more resilient communities and to incorporate lessons learned from the disaster to reduce future vulnerabilities. This phase includes:
\begin{itemize}[leftmargin=*]
    \item \textit{Damage assessment:} Assessing the extent of damage to infrastructure, buildings, and the environment is essential to planning the reconstruction process. This includes conducting structural assessments, evaluating damage to critical facilities, and identifying areas that require immediate attention.
    \item \textit{Infrastructure restoration:} Repairing and rebuilding damaged infrastructure such as roads, bridges, utilities, and communications networks is an important part of the reconstruction phase. This will ensure the restoration of essential services and facilitate the return to normalcy.
    \item \textit{Rehabilitation of communities:} The recovery phase also includes helping affected communities rebuild their lives. This includes psycho-social support, facilitating access to medical care and education, and assisting in the restoration of livelihoods.
\end{itemize}

Overall, the disaster management life-cycle described above encompasses a comprehensive approach to addressing the challenges posed by disasters. By focusing on mitigation, preparedness, response, and recovery, agencies and communities can work together to minimize the impact of disasters and improve resilience.

\section{Energy and Communication Technology Enablers}
\label{EnergyCommunicationEnablers}

\begin{figure*}[htp!]
\centering
\includegraphics[width=0.780\linewidth]{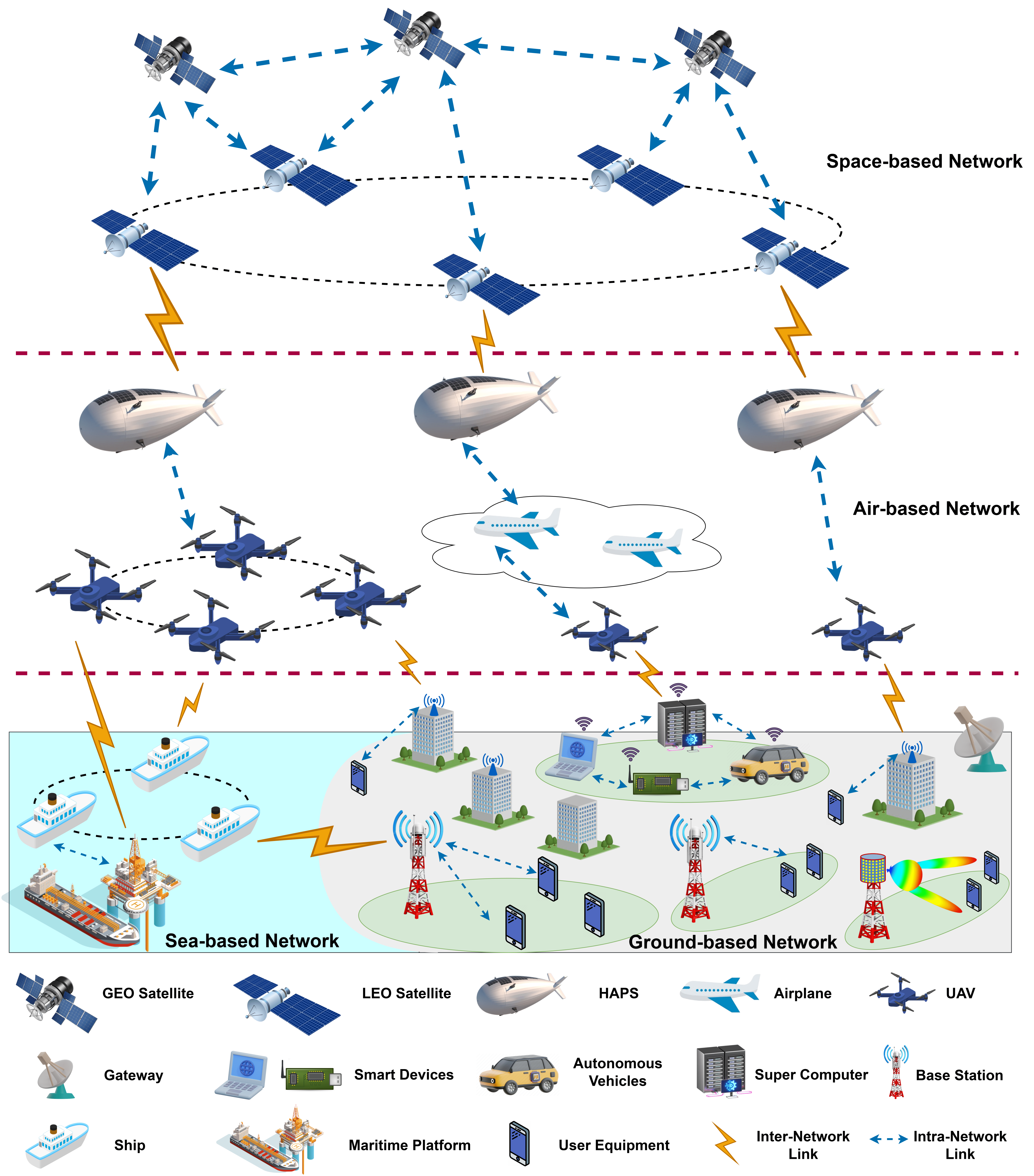}
\caption{System topology of the integrated space-air-ground-sea emergency communication network.}
\label{Integrated_Networks}
\vspace{-.5cm}
\end{figure*}

The current terrestrial mobile network infrastructure is inherently vulnerable to disasters and is prone to failure or interruption during such events. Therefore, a resilient mobile network structure is needed to cope with natural \textcolor{black}{hazards}, which are becoming more frequent due to climate change. On the other hand, it should also meet the UN sustainability goals, at least in some sense by becoming cost and energy effective. From the operational point of view, this infrastructure should also be scalable and easy to deploy.
Infrastructure redundancy is one way to ensure sustained operation even when some infrastructure layers are severely damaged.
\textcolor{black}{Figure} \ref{Integrated_Networks} shows such as a multi-layered communications infrastructure consisting of integrated space-based, air-based, sea-based, and ground-based networks, all interconnected to create a comprehensive and resilient communications system, especially in disaster scenarios where the traditional infrastructure may be compromised. The space-based network comprises \ac{GEO}, \ac{MEO} and \ac{LEO} satellites that form a mesh network for global communications coverage. These satellites are interconnected via cross-network links and communicate with the air-based networks via dedicated downlinks. The air-based network is represented by the \ac{HAPS} and \glspl{UAV}, which act as intermediary points between the space-based \ac{HAPS} and the ground-based infrastructure. \ac{HAPS} stationed in the stratosphere connect to \glspl{UAV} and aircrafts, forming a dynamic network layer that provides communication services over large areas, including disaster-affected areas. The sea-based network consists of maritime platforms and ships equipped with communication capabilities. These platforms are connected to both the air-based network and ground stations, facilitating seamless communication over oceans and large water bodies. The ground-based network consists of various elements such as \glspl{BS}, gateways, autonomous vehicles, and smart devices. These components are interconnected via in-network connections that facilitate communication within the ground network. The figure also shows that ground-based networks are connected to other layers via inter-network links. This ensures that all network layers work together to provide continuous and reliable communication in various environments.

\begin{figure*}[htp!]
\centering
\includegraphics[width=.99\linewidth]{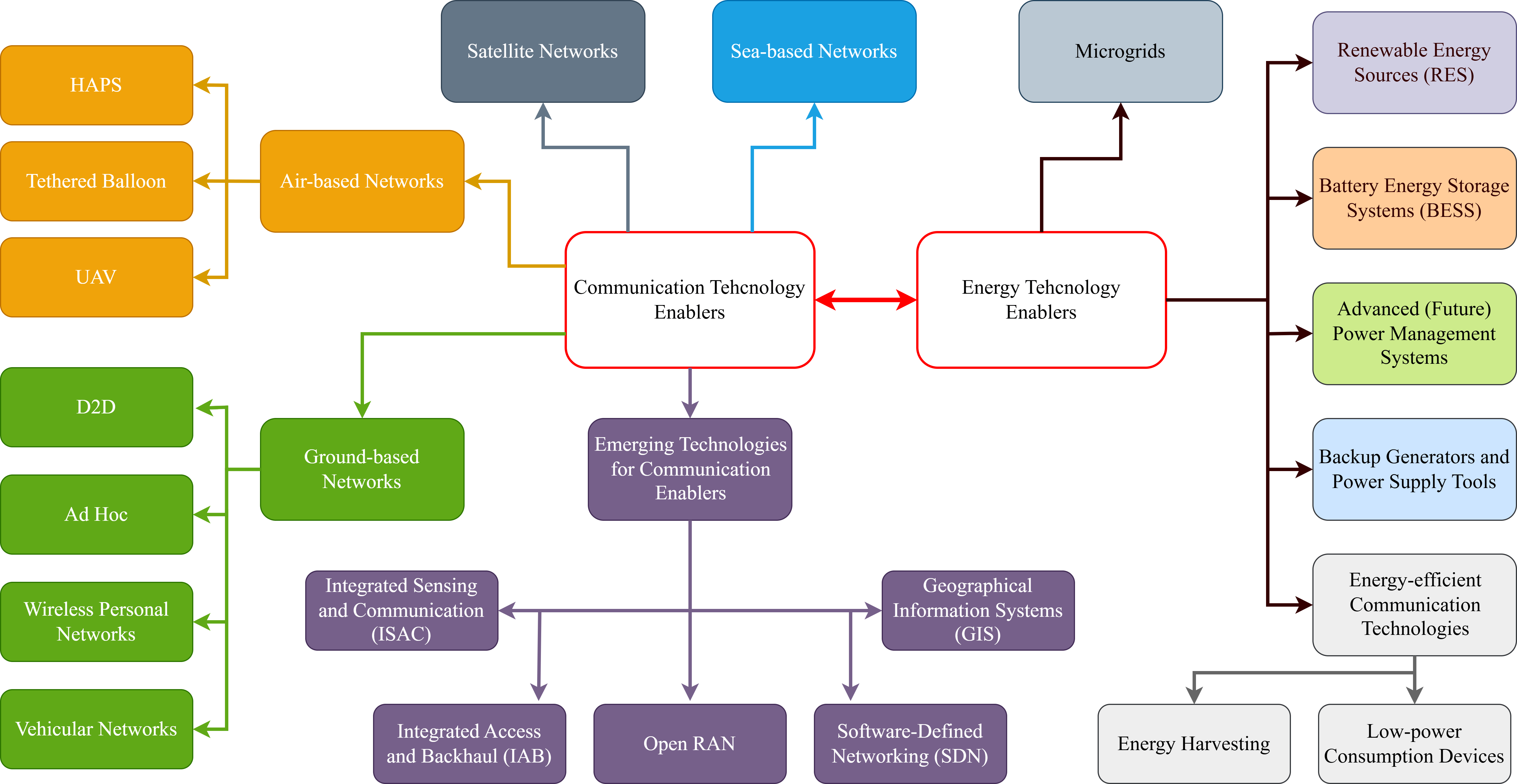}
\caption{\textcolor{black}{Taxonomy diagram for the communication and energy technology enablers for disaster response in the literature.}}
\label{comm_ene_taxonomy}
\end{figure*}

\textcolor{black}{In Section III.A, communication technology enablers are introduced, while Section III.B presents the energy technology enablers for disaster response communication, for which a related taxonomy diagram is provided in Fig.~\ref{comm_ene_taxonomy}.} 

\subsection{Communication Enablers}

\begin{table*}[ht]
\footnotesize
\centering
\caption{\textcolor{black}{Summary of Communication Technologies for Disaster Recovery}}
\begin{tabularx}{\textwidth}
{|p{2.0cm}|p{3.0cm}|p{3.0cm}|p{2.5cm}|X|p{2cm}|}
\hline
\rowcolor{gray!25}
\textbf{Communication Technology} & \textbf{Key Features} & \textbf{Use Case in Disaster Recovery} & \textbf{Advantages} & \textbf{Limitations} & \textbf{References} \\
\hline
Satellite Networks & Global coverage, independent of terrestrial infrastructure, remote sensing & Early warning, communication in remote/damaged areas, real-time data sharing & Wide-area coverage, resilience, remote monitoring & Weather-sensitive, latency, requires specialized equipment & \cite{casoni2015integration,lwin2019city,saeed2021point, rajput2020impact} \\
\hline
HAPS (High Altitude Platform Stations) & Stationary, solar-powered, wide coverage, stratospheric deployment & Backhaul replacement, direct coverage for affected zones & Stable LoS, quick deployment, less weather-sensitive (S-band) & Deployment cost, logistics & \cite{karaman2024enhancing, kurt2021vision} \\
\hline
Tethered Balloons & Low-cost, scalable, wireless payloads & Rapid deployment of temporary networks & Inexpensive, flexible, already field-tested & Limited coverage, affected by weather, tethering logistics & \cite{alsamhi2018disaster} \\
\hline
UAVs (Drones) & Aerial monitoring, temporary network relay, thermal imaging & Search and rescue, network extension, real-time data collection & Flexible, mobile, real-time imagery & Limited flight time, packet loss, energy constraints & \cite{Deepak2019Over,Panda2019UAV ,WuFSObasedDrone2020, Wang2021DisasterReliefWireless} \\
\hline
Ground-Based Networks & Fixed infrastructure, includes BSs, smart devices, vehicular networks & Primary communication layer, integration with other layers & High capacity, currently deployed technology & High vulnerability to physical damage & \cite{matracia2024reliability} \\
\hline
D2D \& Ad Hoc Networks & Infrastructure-free, decentralized & Emergency messaging, coordination, survivor tracking & Robust, immediate deployment & Short range, interference, scalability & \cite{AliD2D2018, 2018SaxenaD2D, Ramakrishnan2022, hossain2020smartdr}, \cite{raza2020establishing} \\
\hline
ISAC (Integrated Sensing and Communication) & Sensing + communication capabilities, real-time analytics & Damage assessment, survivor detection, situational awareness & Actionable data, real-time updates, life detection & Technology maturity, integration complexity & \cite{gonzalez2025six, zhu2024enabling, alnoman2024emerging} \\
\hline
IAB (Integrated Access and Backhaul) & Shared spectrum for access and backhaul & Backup connectivity, rapid network extension & Efficient spectrum use, easy deployment & Performance trade-offs under heavy load & \cite{madapatha2020integrated} \\
\hline
Open RAN & Vendor-agnostic modular design & Flexible component replacement, quick reconfiguration & Cost-effective, flexible upgrades & Interoperability issues, standardization in progress & \cite{firouzi20225g} \\
\hline
SDN (Software Defined Networking) & Centralized control, dynamic resource allocation & Post-failure re-routing, resilient transport layers & Efficient management, resilience improvements & Dependency on controllers, complex deployment & \cite{Kaleem2019UAV, machuca2016technology} \\
\hline
GIS (Geographic Information Systems) & Geospatial data analysis and visualization & Risk mapping, evacuation planning, resource allocation & Real-time maps, public communication, situational awareness & Data accuracy and timeliness, requires integration with sensors & \cite{van2002remote} \\
\hline
\end{tabularx}
\label{tab:summary_comm}
\end{table*}

\textbf{Satellite networks} play a crucial role in disaster response communication planning, particularly in the context of earthquakes, due to their wide coverage, resilience, and ability to provide connectivity in geographically remote or damaged areas. They can operate independently of terrestrial networks, providing a reliable and resilient communication infrastructure and support for emergency response efforts in remote or hard-to-reach areas \cite{casoni2015integration,lwin2019city,saeed2021point}.  They can be used to establish a temporary communication infrastructure in disaster-prone areas. For example, satellite-based mobile networks can be deployed to provide connectivity to emergency responders and local communities in the aftermath of a disaster, when terrestrial networks may be damaged or non-functional. Satellite networks offer wide-area coverage, allowing communication to reach even the most remote or inaccessible regions affected by an earthquake. They can provide connectivity to disaster response teams, emergency personnel, and affected communities, facilitating real-time communication, coordination, and information exchange over large geographical areas. Rescue teams, government agencies, and relief organizations can use satellite connectivity to share vital information, coordinate resources, and efficiently manage rescue and relief operations.  Satellites equipped with remote sensing capabilities can provide valuable data for disaster response planning and decision-making. They can capture high-resolution imagery, monitor ground movements, and create accurate maps of affected areas. This information aids in identifying critical infrastructure damage, assessing the extent of the disaster, and guiding resource allocation and deployment.

Regarding early warning systems, satellite imagery and remote sensing technologies can be used to monitor natural hazards enabling early warnings and better preparedness. In the pre-disaster phase, satellite networks can be used to establish communication channels between emergency responders, local communities, and other stakeholders. These networks can provide connectivity in remote or hard-to-reach areas, where terrestrial networks may not be available or may be unreliable. Satellite networks can also be used to support the exchange of critical information, such as maps, images, and video feeds between emergency responders and command centers \cite{rajput2020impact}. They can also be used to support warning and dissemination of critical information to local communities. For example, satellite-based warning systems can be used to alert people in disaster-prone areas of impending natural \textcolor{black}{hazards} or emergencies. These warning systems can be used to deliver text messages, voice messages, or other forms of communication to people in remote or hard-to-reach areas. \textcolor{black}{These messages can be delivered to the people either by using direct access to mobile handheld devices or using the satellite-based systems as relays to convey this information to terrestrial systems in a large footprint.} Another use of satellite networks in pre-disaster communication planning is to support situational awareness and information sharing among emergency responders. Satellite networks can provide real-time information on the extent of damage, the location of survivors, and other critical information that can be used to support emergency response efforts.

\textcolor{black}{Although satellite communication offers significant benefits in extending coverage and maintaining connectivity in disaster-affected areas, satellite links are often more susceptible to rain and other atmospheric conditions. Commonly used SatCom protocols frequently face limitations in latency, bandwidth, and hardware availability. As noted in~\cite{rsshydro2023}, while satellite-based systems can bridge gaps in terrestrial connectivity, they are vulnerable to weather-induced disruptions and typically require specialized ground terminals, which can delay rapid deployment. However, it is important to note that different types of satellite links exhibit varying degrees of robustness against such environmental factors, depending on their frequency bands and system design. For instance, traditional RF satellite links operate in L-, C-, X-, Ku-, Ka-, and V-bands, with higher frequencies providing greater bandwidth but being more susceptible to atmospheric attenuation. Particularly, L- and C-bands are more resistant to rain that makes them suitable for emergency response coordination but requires specific satellite phones and VSAT terminals while Ku- and Ka-bands provide higher data rates but are prone to rain attenuation. On the other hand, FSO communication uses laser beams for high-capacity optical communication between satellites and ground stations. However, it is highly vulnerable to atmospheric conditions like dense fog, heavy rain, and dust storms, which severely degrade signal quality due to scattering and absorption. It works optimally in clear-sky conditions, making it more suitable for inter-satellite communication rather than direct-to-ground disaster response. To this end, hybrid RF-Optical satellite links combine RF and optical communication to utilize both strengths while mitigating their weaknesses. Such adaptive systems switch between RF and optical links based on real-time atmospheric conditions.}

\textcolor{black}{In the context of satellite protocols, their integration with terrestrial systems is further hindered by incompatibilities in signaling standards, lack of unified control protocols, and insufficient real-time synchronization, particularly during fast-evolving disaster scenarios. These limitations highlight the need for hybrid architectures that enable dynamic interoperability between satellite and terrestrial communication layers. To better understand the opportunities and constraints in building such hybrid systems, examining the dominant satellite communication protocols currently employed in disaster relief scenarios is essential.}

\textcolor{black}{\paragraph{Dominant Satellite Communication Protocols in Disaster Relief Scenarios}
Satellite communication systems play a crucial role in maintaining connectivity in disaster scenarios, especially when terrestrial networks are compromised. Two main categories of protocols are commonly used in such contexts \cite{kagai2024rapidly}:}

\textcolor{black}{\textit{Consultative Committee for Space Data Systems (CCSDS) Protocol Suite:} The CCSDS (Consultative Committee for Space Data Systems) has developed a set of internationally standardized protocols that are commonly used in space and disaster resilient communication systems. These protocols support robust and interoperable communication across heterogeneous systems. Notable examples include the Telemetry Channel Coding and Telecommand protocols for low bit rate reliable command and control, the CCSDS File Delivery Protocol (CFDP) for efficient store-and-forward file transmission; the Space Link Extension (SLE) protocol to support interoperability of remote ground stations, and the Proximity 1 protocol for short-range inter-satellite or ground spacecraft communication. These standards offer high reliability and are well suited for critical and infrastructure-independent communication tasks in emergency response operations \cite{ccsdsinterop2021}.}

\textcolor{black}{\textit{Commercial Satellite Protocols:} In addition to standardized CCSDS protocols, proprietary protocols are used in various commercial satellite systems. Iridium satellites use Short Burst Data (SBD) and voice services over closed transport and network layers. Inmarsat employs protocols like Broadband Global Area Network (BGAN) and IsatPhone, optimized for mobile voice and data communication in remote environments. Other satellite providers such as Globalstar and Thuraya rely on proprietary codecs such as QCELP and AMBE for low-bitrate voice encoding. These systems are designed for rapid deployment, lightweight infrastructure requirements, and global reach, making them suitable for fast-response communication during disasters.}

\textcolor{black}{\paragraph{Integration Challenges with Terrestrial Protocol} Despite the complementary benefits of satellite systems, their seamless integration with terrestrial communication protocols remains a challenge due to several factors listed below~\cite{satcom_disaster2020}.}

\begin{itemize}
    \item \textcolor{black}{ \textit{Protocol Incompatibility:} Many satellite systems use specialized protocol stacks, such as the CCSDS suite, which differ from the TCP/IP-based stacks used in terrestrial networks. This discrepancy makes direct interoperability difficult and requires protocol converters or gateways.}
    
    \item \textcolor{black}{ \textit{Latency and QoS mismatch:} GEO satellites, in particular, lead to significant latency, e.g. over 500 ms roundtrip time, which affects the performance of real-time services and TCP-based congestion control mechanisms that are sensitive to delays.}
    
    \item \textcolor{black}{\textit{Lack of interoperability standards:} Many commercial satellite networks work with proprietary technologies that are not fully adapted to terrestrial standards such as 3GPP, which leads to interworking failures in hybrid setups.}
    
    \item \textcolor{black}{\textit{Complex deployment requirements:} Satellite communications often require specialized equipment such as large dish antennas or high-power modems that are not readily compatible with the lightweight terrestrial infrastructure used in emergency scenarios.}
    
    \item \textcolor{black}{\textit{Security and regulatory differences:} Satellite and terrestrial systems may adopt different security frameworks and operate under distinct regulatory regimes, which complicates the secure and lawful exchange of data in joint operations.}

    \item \textcolor{black}{\textit{Dynamic Topology and Routing Conflicts:} In mobile or dynamic disaster environments, terrestrial networks rely on fast topology updates, such as via OSPF or BGP, which are not directly compatible with static or slower satellite routing architectures, leading to delays or loss in convergence.}
\end{itemize}

\textcolor{black}{\paragraph{Emerging Interoperability Efforts.}
To address these limitations, ongoing efforts in the research and standardization communities are focused on the development of integrated satellite-terrestrial network architectures. These include the incorporation of satellite systems into the 3GPP NTN framework, enabling unified radio access and signaling procedures. Additionally, SDRs and cognitive network technologies are being explored to enable dynamic protocol adaptation across platforms. Ultra-low-bitrate codecs such as Codec2 and open-source communication stacks are also gaining traction as lightweight, interoperable solutions for use in bandwidth-constrained emergency environments. These emerging approaches aim to close the interoperability gap and foster seamless collaboration between satellite and terrestrial networks in disaster relief scenarios \cite{rapidsat2023}.}

\textbf{Air-based networks} incorporate innovative technologies like \ac{HAPS}, tethered balloons, and \ac{UAV}s \textcolor{black}{\cite{karaman2024enhancing}, \cite{kurt2021vision}, \cite{sheng2021space, liu2018space, hong2020space}}.

\textit{HAPS}, in particular, holds significant potential in enhancing the resilience of integrated space-air-ground-sea networks during and after natural disasters such as earthquakes \textcolor{black}{\cite{karaman2024enhancing}}. \ac{HAPS} offers several advantages: its expansive surface area enables almost self-sufficient energy generation through solar panels. A single \ac{HAPS} can serve as a substitute for multiple damaged terrestrial \ac{BS}, providing extensive coverage. Essentially, a single HAPS can function as a multi-sector (directional) BS, establishing direct \ac{LoS} connections for outdoor users, compensating for distance-related path loss or scattering losses without the need for additional intermediate devices like relays, unlike satellites such as Starlink's \ac{LEO} constellation, which have limited LoS windows. Unlike satellites, \ac{HAPS} remain stationary, ensuring stable and continuous service without orbital movements, thus facilitating rapid deployment, especially crucial in the immediate aftermath of earthquakes when ground access to affected areas is challenging. HAPS serves as a vital alternative not only for \ac{RAN} infrastructure but also for backhaul. It can efficiently redirect backhaul traffic to unaffected areas via \ac{FSO} or \ac{THz} communication links with remote HAPS or ground stations \textcolor{black}{\cite{kurt2021vision}}. Additionally, during earthquakes, where fiber optic lines may sustain damage, rendering some BS inactive, HAPS remain physically unaffected by weather events. As mentioned in \cite{STRATXX}, HAPS is expected to function in all stratospheric climate conditions. Moreover, according to the \ac{ITU}’s decision at the World Radiocommunication Conference 2023 (WRC-23), the sub-2.7 GHz frequency band (S-band) has been allocated to High Altitude International Mobile Telecommunications Base Stations (HIBS) \cite{wrcReport}. Compared to the Ka and Ku bands used in satellites, the S-band is relatively less affected by weather events. In this way, uninterrupted service delivery can be ensured.

\textit{Tethered balloon-based emergency network} system uses balloons to deploy wireless communications networks in disaster areas. They can be used for a wide range of communications services including voice, data, and video. They are comparatively inexpensive and can be scaled to meet the needs of different disaster scenarios. In \cite{alsamhi2018disaster}, a fully wireless communications solution relying on a Tethered Balloon that can be deployed immediately, reliably, and easily before or during, and even after the disaster has been proposed. Tethered balloons have already been used in a number of disaster scenarios, including  Puerto Rico and Peru after disasters in 2017 and 2019, respectively \footnote{Online: https://www.itu.int/en/mediacentre/backgrounders/Pages/emergency-telecommunications.aspx, Available: September 2023}. 

\textit{\ac{UAV}} usage for aerial monitoring and communication transmission in earthquake zones is an innovative approach that has gained popularity in recent years. They can be used to establish temporary communication infrastructure in disaster-prone areas \cite{Deepak2019Over,Panda2019UAV ,WuFSObasedDrone2020, Wang2021DisasterReliefWireless}. For example, \glspl{UAV} equipped with wireless communication equipment can be deployed to create \textit{ad hoc} networks or to extend the range of existing networks, providing connectivity to emergency responders and local communities \cite{Matracia2023SG}.  \glspl{UAV} can also be equipped with thermal imaging cameras to locate people in need of assistance and identify areas of high temperature, which may indicate gas leaks or other hazards. \glspl{UAV} can also be used to support situational awareness and information sharing among emergency responders. For aerial assessment and monitoring, drones equipped with cameras or other sensors can provide real-time information on the extent of damage \cite{ali2021real}, rapid aerial assessment of disaster-affected areas, the location of survivors, and other critical information that can be used to support emergency response and search and rescue efforts. One of the most important applications of \glspl{UAV} in disaster relief is their ability to provide high-resolution aerial imagery that can be used to assess damage to infrastructure, identify areas requiring search and rescue, and plan evacuation routes. On the other hand, drones may have difficulty with massive packet loss and handling enormous traffic when used for post-disaster reconnaissance services. In areas where traditional communication networks are disrupted, \glspl{UAV} equipped with communication equipment can serve as a temporary solution to act as communication relays to establish communication between first responders and affected communities. This can be particularly useful in mountainous or remote areas where cell towers and other communications infrastructure may be damaged. In order to benefit from \ac{UAV}s with full efficiency, problems such as energy supply and trajectory design must be overcome \cite{Zhao2019UAV,Feng2020SWIPT,Niu2021EEmax,Sherman2021UAVcharge}.

\textbf{Ground-based networks} comprise diverse elements such as cellular networks, terrestrial Internet (which utilizes physical cables like fiber optics or copper wires for data transmission over long distances), and mobile 	\textit{ad hoc} networks. Unlike air-based networks, ground-based network topology tends to be more fixed and less mobile. During natural disasters like hurricanes, earthquakes, or floods, the ground-based communication infrastructure faces significant risks of severe damage or complete destruction \cite{matracia2024reliability}. 

\textit{\ac{D2D}} technology, offering a decentralized and resilient communication, is an effective approach to address this challenge for ground-based networks in disaster response scenarios \cite{AliD2D2018, 2018SaxenaD2D, Ramakrishnan2022, hossain2020smartdr}. It is mostly useful in scenarios where the use of grid power is inaccessible, and the use of batteries to power is not feasible. It can be used to enhance communication capabilities and improve coordination among responders and affected individuals. In the aftermath of an earthquake, when cellular networks may be congested or disrupted, \ac{D2D} communication can provide a reliable means of communication between responders, allowing them to share critical information, coordinate efforts, and exchange updates on the situation even in the absence of infrastructure. 

\textit{Ad hoc networks} are formed by a group of mobile devices that communicate with each other directly, without the need for a centralized infrastructure such as cellular towers or \ac{Wi-Fi} access points (e.g., \ac{WMN}). \ac{D2D}-based 	\textit{ad hoc} networks can support the dissemination of emergency alerts, location sharing, resource coordination, and survivor tracking \cite{raza2020establishing}. During search and rescue operations, rescuers equipped with \ac{D2D}-capable devices can establish direct communication with survivors, allowing them to gather information about their location, status, and any immediate needs. 

\textit{Wireless personal networks,} such as Bluetooth, \ac{LoRa}, and \ac{Wi-Fi} technologies \cite{HAZRA202054}, can be used to establish ad-hoc networks that can operate independently of cellular networks or other traditional communication infrastructure. These networks can be established using personal devices that are equipped with wireless communication capabilities.

\textit{Vehicular networks} can be used in a variety of ways in disaster response scenarios to improve the speed, efficiency, and effectiveness of emergency response when traditional communications and transportation infrastructure is severely damaged or disrupted. First, they can be used to establish communications between responders and affected areas, especially in areas where traditional communications infrastructure is damaged or overwhelmed. Second, they can facilitate the transport of relief supplies, medical personnel, and other resources to affected areas. Third, \textcolor{black}{vehicular} networks can be used to collect data on road conditions, infrastructure damage, and other information that can be used to create maps to help responders navigate affected areas. Fourth, \textcolor{black}{vehicular} networks can be used to coordinate search and rescue efforts, especially in hard-to-reach areas or in areas where traditional search and rescue methods are impractical. Finally, they can be used to coordinate emergency response efforts among various agencies and organizations to ensure that resources are used effectively and efficiently.

\textbf{Sea-based networks} have crucial importance when conventional communication infrastructure in coastal regions collapses in a disaster situation. Within this network structure, various components, such as ships and unmanned surface vehicles, can enhance communication utilizing UAVs, HAPS, and satellite technologies, depending on their respective positions \textcolor{black}{\cite{guo2021survey}}. They expand the coverage and resilience, offering alternative routes, and mitigating the risk of network congestion and failures. Moreover, they facilitate interoperability among diverse network types (e.g., naval vessels, aircraft, and ground-based command and control centers) through standard or customizable protocols, facilitating seamless communication across different services and agencies.

\textbf{Emerging Technologies for Communication Enablers:} \textit{Integrated Sensing and Communication (ISAC)} technology offers significant utility in earthquake disaster relief by combining multiple sensing technologies with communication systems to provide real-time, actionable data on ground conditions \textcolor{black}{\cite{gonzalez2025six}}.  First, ISAC uses a variety of sensing methods such as seismic sensors, radar, LiDAR (Light Detection and Ranging), and wireless radio frequency sensing to accelerate damage assessment post-earthquake \textcolor{black}{\cite{zhu2024enabling}}. These sensors can be used to monitor structural integrity, ground movement and changes in the environment and transmit this data immediately via communication systems. This rapid transfer of information helps to prioritize recovery efforts and locate areas that require urgent action. Second, ISAC can help to locate survivors trapped under debris by using thermal imaging, sound/vibration sensors, and RF-based motion detectors to detect movement, signs of life, and body heat. This vital data is immediately relayed to rescue teams to speed up search and rescue operations. Third, ISAC provides real-time updates on the location of survivors, the severity of damage, and the progress of relief efforts \textcolor{black}{\cite{alnoman2024emerging}}. All of this is made possible through a combination of wireless sensor networks, drone surveillance, and ground-based radar, enabling informed decision-making and a coordinated response. Finally, ISAC supports seamless communication between response teams by enabling the exchange of information in real-time over secure communication links, optimizing resource allocation, and streamlining relief efforts.

\textit{\ac{IAB}} technology can be used in pre-disaster scenarios to enhance the resilience of communication infrastructure and ensure that connectivity remains available during and after a disaster. \ac{IAB} combines the functionalities of wireless backhaul and access networks, enabling wireless networks to use the same frequency band for both access and backhaul \cite{madapatha2020integrated}. This allows for easier deployment of small cells and other wireless access points, as well as more efficient use of available spectrum. In a pre-disaster scenario, \ac{IAB} can be used to provide enhanced connectivity to critical infrastructure and services such as hospitals, emergency services, and government offices. The technology can also be used to establish a network of small cells and wireless access points in high-risk areas, providing redundancy and backup connectivity in case of an emergency.

\textit{\ac{AI}}-based solutions can also be used to address several problems for pre-disaster communications. For risk assessment and predictive analytics, \ac{AI}-driven models can analyze historical data, climate patterns, and other variables to predict and assess disaster risks in specific regions. \textcolor{black}{Reference} \cite{Sun2020AI} provides an overview of current applications of AI in disaster management in the four phases: Mitigation, Preparedness, Response, and Recovery.

\textit{Open \ac{RAN}} is an emergent technology 
that builds on open interfaces so that a telecom operator can build a network by mixing equipment from multiple vendors. The vision is to create more cost-effective networks and new innovations through enhanced competition. 
The modular nature can be useful in disaster situations because a broken component can be rapidly replaced by a new such component from any vendor instead of waiting for the original vendor to deliver an exact copy of the broken component. This makes telecom networks compliant with Open RAN more well-suited to maintain connectivity and facilitate emergency response and recovery operations. A methodology for deploying and optimizing FL tasks in O-RAN to deliver distributed intelligence for 5G applications is studied in \cite{firouzi20225g}.

\textit{\ac{SDN}} offers a vital solution for disaster situations, as it can enhance network control, communication efficiency, and overall stability \cite{Kaleem2019UAV}. A survey of disaster resilient \ac{SDN} networks is presented in \cite{machuca2016technology}. A hierarchical, failure- and disaster-resilient Transport Software-Defined Network (T-SDN) control plane is designed, and a heuristic for post-failure switch-controller reassignment is proposed in \cite{lourencco2018robust}. The results indicate this proposed scheme can achieve much higher disaster and failure resiliency, at the cost of slightly larger network-resource utilization.

\textit{\ac{GIS}} technologies provide powerful tools for visualizing, analyzing, and interpreting data related to disaster risks, vulnerabilities, and resources \cite{van2002remote}. \ac{GIS} helps identify vulnerable areas by analyzing geographic and demographic data, enabling better planning and preparedness. During disasters, \ac{GIS} integrates real-time data like weather reports and satellite imagery to provide up-to-date situational awareness. \ac{GIS} optimizes resource allocation by identifying the most efficient routes and locations for aid distribution. By analyzing transportation networks and population density, \ac{GIS} helps design effective evacuation routes and plans. After disasters, \ac{GIS} assesses damage and informs recovery efforts by mapping affected areas and identifying critical infrastructure. \ac{GIS} tools also enhance public communication by providing clear visual information on risks, safety measures, and recovery efforts.

\textcolor{black}{Table \ref{tab:summary_comm} provides a summary of the communication technologies discussed above. Moreover, a detailed comparison of different communication technologies based on several \glspl{KPI} have been given for a more objective evaluation of communication technologies used in disaster response in Table \ref{tab:disaster_comm}. The table provides a comparative analysis of different communication technologies used in disaster response and evaluates them based on \glspl{KPI} such as robustness, latency, deployment and operational costs, energy requirements, mobility, modalities/functionality, interoperability and resilience. Satellites and HAPS stand out for their robustness and comprehensive coverage and are therefore suitable for large-scale, remote disaster scenarios. Technologies such as cellular and WLAN are ideal for urban and semi-urban environments as they offer low latency and high mobility, although their resilience can be compromised during large-scale crises. While D2D communication offers very low latency and high mobility, it is limited in terms of interoperability and scope of application. LoRaWAN is highly energy efficient and suitable for IoT applications for tracking and monitoring, but has higher latency and less functionality. UAVs, which offer high mobility and features such as data relay and video streaming, are ideal for search and damage assessment, but their operating costs and energy consumption are moderate. Overall, the table illustrates the trade-offs between cost, mobility and functionality and enables the selection of the most suitable technology based on the specific needs of disaster response.}


\begin{sidewaystable} 
    \centering
     \caption{\textcolor{black}{Comparison of Communication Enablers in Disaster Relief Applications}}
    \renewcommand{\arraystretch}{1.3} 
    \setlength{\tabcolsep}{6pt} 
    \small 
    
    \resizebox{\textwidth}{!}{ 
    \begin{tabular}{|l|c|c|c|c|c|c|c|}
        \hline
        \rowcolor{gray!25}
        \multirow{1}{*}{\textbf{KPIs}} & \multicolumn{7}{c|}{\textbf{Technology}} \\ 
        \cline{2-8}
        \rowcolor{gray!25} 
        & \textbf{Satellite} & \textbf{Cellular (4G/5G)} & \textbf{WLAN (Wi-Fi, Mesh)} & \textbf{D2D (Device-to-Device)} & \textbf{UAV} & \textbf{LoRaWAN} & \textbf{HAPS} \\
        \hline
        Robustness & High & Medium & Medium & Medium & Medium & Medium & High \\
        \hline
        Latency & High & Low & Low & Very Low & Low & High & Low \\
        \hline
        Deployment \& Operational Cost & High & Medium & Low & Low & Medium & Low & Medium \\
        \hline
        Application Domain & Remote \& large-scale & Urban \& semi-urban & Localized zones & Proximity-based communication & Search \& damage assessment & IoT, tracking & Wide-area response \\
        \hline
        Energy Requirements & High & Medium & Low & Very Low & Medium & Very Low & Medium \\
        \hline
        Mobility & Low & High & High & Very High & Very High & High & High \\
        \hline
        Modalities/Functionality & Voice, Data, GPS & Voice, Video, Data & Data, Voice & Direct device messaging, IoT & Data relay, Video & Data, Text & Voice, Video, Data, IoT \\
        \hline
        Interoperability & Medium & High & High & Low & Medium & Medium & Medium \\
        \hline
        Resiliency & High & Low & Low & Medium & Medium & Medium & High \\
        \hline
        
    \end{tabular}
    } 
   
    \label{tab:disaster_comm}
\end{sidewaystable}

\subsection{Energy Enablers}

\textcolor{black}{While disaster communication infrastructures are typically designed as short-term and ad-hoc systems, energy efficiency remains a critical consideration, particularly in off-grid or resource-constrained settings where refueling or recharging is logistically challenging \cite{qu2023environmentally}. In mountainous regions affected by earthquakes, or during prolonged wildfire operations, power delivery may be severely delayed for several days. Under such conditions, energy-efficient designs not only extend the operational lifespan of off-grid power systems in temporary or remote deployments but also reduce hardware complexity and thermal load, thereby enhancing deployment flexibility and system reliability \cite{cheng2024trace}. Moreover, energy-aware solutions, such as renewable energy integration and energy-harvesting modules, can ensure communication continuity when conventional energy supplies are unavailable \cite{2022Matracia, saif2021efficient, unal2024enhancement}. Therefore, although energy efficiency may appear secondary in short-lived deployments, it can serve as a key operational enabler, especially during the critical initial 72-hour period following a disaster, when logistical disruptions and operational autonomy are the primary concerns.}

\begin{table*}[ht]
\footnotesize
\centering
\caption{\textcolor{black}{Summary of Energy Technologies for Disaster Communication}}
\begin{tabularx}{\textwidth}{|p{2.0cm}|p{3.0cm}|p{3.0cm}|p{2.5cm}|X|p{2cm}|}
\hline
\rowcolor{gray!25}
\textbf{Energy Technology} & \textbf{Key Features} & \textbf{Use Case in Disaster Recovery} & \textbf{Advantages} & \textbf{Limitations} & \textbf{References} \\
\hline
Microgrids & Distributed/local energy, combines RES, autonomous or grid-connected & Backup power for terrestrial communication infrastructure in remote/disconnected areas & Flexible, scalable, high reliability, resilient against grid failure & High initial cost, complex integration & \cite{nadeem2023distributed, khalid2024smart} \\
\hline
Renewable Energy Sources (RES) & Solar, wind, biomass, hydro; clean and decentralized power & Primary or backup power for BSs and remote communication systems & Sustainable, low carbon emissions, off-grid capable & Intermittency, weather dependency & \cite{wan2023pre, ko2023renewable} \\
\hline
Battery Energy Storage Systems (BESS) & High-capacity batteries for energy storage from grid or RES & Backup power for BSs, data centers, mobile communication devices & Reliable backup, peak load support, supports off-grid use & Battery lifespan, high cost & \cite{zhao2023grid, rana2023applications} \\
\hline
Advanced Power Management Systems / Smart Grids & Digital monitoring, load balancing, automated fault isolation & Continuous power for critical communication infrastructure & Efficient, resilient, real-time monitoring and control & Infrastructure upgrade needed, cybersecurity risks & \cite{khalid2024smart, ali2023optimal} \\
\hline
Backup Generators and UPS & Diesel/petrol generators, battery-based uninterrupted supply & Immediate and long-term backup power for communication nodes & Reliable, proven technology, ensures continuous operation & Fuel dependency, emissions & \cite{cabrera2023energy} \\
\hline
Energy-Efficient Communication Technologies & Low-power devices, energy harvesting, optimized protocols & Sustain critical communication services under limited energy availability & Reduced power consumption, cost-effective, enables longer operation & Lower processing power, limited range/bandwidth & \cite{pandey2024uav, messaoudi2024ugv} \\
\hline
\end{tabularx}
\label{tab:summary_ene}
\end{table*}

\textit{Microgrids:} Microgrids are designed to enhance the reliability, resilience, and efficiency of electrical energy distribution, particularly in remote areas. They can operate autonomously or in parallel with the main grid, ensuring uninterrupted electricity generation during failures or outages. The energy grid is expected to shift from centralized energy production to a more distributed and localized model in the near future \cite{nadeem2023distributed}. Microgrid components include renewable energy sources (RES) such as solar panels and wind turbines, distributed generation units, energy storage systems, \glspl{EV}, and the utility grid. By combining multiple energy sources (e.g., solar, wind, diesel generators) with decentralized energy solutions, microgrids enhance the reliability and flexibility of the grid in response to disasters\cite{khalid2024smart}. Additionally, powering terrestrial communication systems with a microgrid structure can significantly reduce energy-related issues during disasters.

\textit{Renewable Energy Sources (RES):} Given the increasing global concern over climate change and the imperative to reduce energy costs, especially in off-grid areas, there is growing interest in leveraging \ac{RES} \cite{wan2023pre}. \ac{RES} also play a crucial role in improving energy sustainability and resilience within microgrids. Solar, wind, biomass, and hydropower are examples of \ac{RES} that provide clean, sustainable, and decentralized power generation. In terrestrial networks, \glspl{BS} have become focal points for integrating \ac{RES}, such as solar and wind power \cite{ko2023renewable}. Adopting renewable energy by terrestrial network components offers numerous opportunities for energy conservation, sustainability, and resilience against disasters.

\textit{\ac{BESS}:} A \ac{BESS} is designed to store energy from renewable sources and utility grids in high-capacity battery systems \cite{zhao2023grid}. This stored energy can support the system during peak energy demand on the main grid and meet energy needs during power outages \cite{rana2023applications}. A \ac{BESS} can play a critical role in communication networks by providing backup power during outages, ensuring the continuous operation of essential communication infrastructure. The integration of communication systems and energy storage system solutions has significant potential for various applications, including mobile \glspl{BS}, data centers, emergency services communication, railway communication systems, and maritime communication.

\textit{Advanced (Future) Power Management Systems:} 
Reliable electrical power is essential for the functioning of modern infrastructure, powering homes, businesses, \glspl{BS}, and more. A smart grid is an advanced power grid system that incorporates digital technology to enhance the efficiency, reliability, and sustainability of electrical energy production, distribution, and consumption. By integrating distributed energy resources such as solar panels, wind turbines, and BESSs, smart grids ensure continuous power supply to critical communication infrastructure, even during widespread outages \cite{khalid2024smart}. Smart grids can automatically detect and isolate faults, minimizing downtime and maintaining service continuity. They also help conserve energy and balance loads through demand response strategies, which are crucial during emergencies when energy resources may be limited \cite{ali2023optimal}. By implementing demand response strategies, energy usage can be adjusted based on supply conditions, ensuring that critical communication infrastructure remains powered.

\textit{Backup Generators and Power Supply Tools:} Backup generators and power supply tools, such as \ac{UPS}, are essential for ensuring communication systems remain operational during disasters. These systems provide a crucial layer of redundancy by supplying power when the main source is disrupted, thus preventing critical communication infrastructure from going offline. \ac{UPS} systems provide immediate, short-term power during an outage, allowing for seamless transitions and protecting sensitive equipment from damage caused by sudden power loss or fluctuations. On the other hand, despite requiring fuel replenishment to operate for extended periods, backup generators offer longer-term power solutions, ensuring the sustained operation of communication networks during prolonged outages. These tools are vital for maintaining emergency communication services, supporting coordination efforts, and ensuring the continuous flow of information, which is crucial for disaster response and recovery. The integration of reliable backup power systems significantly enhances the robustness and dependability of communication networks, enabling them to function effectively in adverse conditions \cite{cabrera2023energy}. 

\textit{Energy-efficient Communication Technologies:} Energy-efficient technologies, such as low-power communication devices and energy-efficient network equipment, are crucial for ensuring the sustainability and resilience of modern communication systems. Low-power communication devices help extend battery life, which is critical in situations where energy is limited, such as during disasters and in remote or off-grid areas with restricted or intermittent power supplies. Similarly, energy-efficient network equipment improves the overall efficiency of the communications infrastructure, allowing more data to be transmitted using less energy and enabling the use of \ac{RES} that produce less power than traditional sources. This saves costs and contributes to sustainability by reducing the carbon footprint of terrestrial communication networks. Furthermore, in disaster scenarios, energy-efficient technologies can help maintain critical communication services for more extended periods using backup power sources, thus improving the resilience of communication systems. Approaches such as energy harvesting can significantly reduce the energy consumption constraints of mobile devices during disasters \cite{pandey2024uav}. It is also important for battery-powered systems such as \glspl{UAV}, which support terrestrial communications and extend coverage during natural disasters, to operate efficiently and meet their energy requirements \cite{messaoudi2024ugv}. 

\textcolor{black}{Enhancing power system resilience to ensure continuous communication during and after disasters is an important aspect. Power system resiliency refers to the ability of electrical grids to withstand and recover from disruptive events. Promising approaches in this context include the deployment of smart grids with distributed renewable energy-based microgrids, the integration of BESS, the use of electric vehicle batteries through Vehicle-to-Load (V2L) applications and the use of seismic isolation components to protect critical power system equipments, such as transformers, from strong ground shaking.} \textcolor{black}{To consolidate the findings from the discussion on energy enablers, Table \ref{tab:summary_ene} was introduced, which provides a structured summary of the different energy solutions, including their key features, use cases, benefits, limitations and applicability in disaster scenarios.}

\textcolor{black}{Finally, we conclude this section with Table \ref{tab:vertical_table2}, which provides a broader perspective by mapping existing studies according to the disaster phase they address and the energy and communication technologies used. Overall, the tables presented in this section provide a comprehensive basis for understanding the interplay of energy and communication technologies in improving disaster resilience.}

\begin{table*}[htp!]
\footnotesize
\captionsetup{font=sc,  position=above, justification=centering, labelsep=newline,singlelinecheck=true}
\centering

\caption{\textcolor{black}{Comparison of Existing Literature Across Disaster Phases and Communication-Energy Technologies}}
\centering
\begin{tabular}{|l|l|l|lll|lll|}
\hline
\rowcolor{gray!25}
\multirow{1}{*}{\textbf{Refs}} & \multicolumn{1}{c|}{\multirow{1}{*}{\textbf{Year}}} & \multicolumn{1}{c|}{\multirow{1}{*}{\textbf{Description}}} & \multicolumn{3}{c|}{\textbf{Disaster Phase}} 
&\multicolumn{3}{c|}{\textbf{Network and Technologies}} \\ \cline{4-6} \cline{7-9} 
\rowcolor{gray!25} & \multicolumn{1}{c|}{} & \multicolumn{1}{c|}{} & \multicolumn{1}{c|}{\textbf{Pre}} & \multicolumn{1}{c|}{\textbf{In}} & \textbf{Post} 
 & \multicolumn{1}{c|}{\textbf{Ground}} & \multicolumn{1}{c|}{\textbf{Air}} & \multicolumn{1}{c|}{\textbf{Space}} \\\hline

\multicolumn{1}{|c|}{ \cite{casoni2015integration}} & \multicolumn{1}{c|}{2015} & \multicolumn{1}{c|}{\begin{tabular}[c]{@{}c@{}}A hybrid network design that combines LTE and satellite technologies \\ is suggested to ensure seamless connectivity, broader coverage, and \\reliable performance during disaster situations\end{tabular} }
& \multicolumn{1}{l|}{x} 
& \multicolumn{1}{l|}{\checkmark}  
& \multicolumn{1}{l|}{\checkmark}
& \multicolumn{1}{c|}{LTE} 
& \multicolumn{1}{c|}{x} 
& \multicolumn{1}{c|}{\ac{MEO}}   
\\ \hline

\multicolumn{1}{|c|}{ \cite{AliD2D2018}} & \multicolumn{1}{c|}{2018} & \multicolumn{1}{c|}{\begin{tabular}[c]{@{}c@{}}An energy harvesting based \ac{D2D} clustering model \\ is proposed for emergency and disaster situations. \end{tabular} }
& \multicolumn{1}{l|}{x} 
& \multicolumn{1}{l|}{\checkmark}  
& \multicolumn{1}{l|}{\checkmark}
& \multicolumn{1}{c|}{Cellular, D2D} 
& \multicolumn{1}{c|}{x} 
& \multicolumn{1}{c|}{x}   \\ \hline

\multicolumn{1}{|c|}{ \cite{2018SaxenaD2D}} & \multicolumn{1}{c|}{2018} & \multicolumn{1}{c|}{\begin{tabular}[c]{@{}c@{}}A cooperative \ac{D2D} protocol is presented to both prolong \\ the average battery lifespan and guarantee fair connection \\ opportunities for stuck users in the disaster zone. \end{tabular} }
& \multicolumn{1}{l|}{x} 
& \multicolumn{1}{l|}{\checkmark}  
& \multicolumn{1}{l|}{x}
& \multicolumn{1}{c|}{D2D} 
& \multicolumn{1}{c|}{x} 
& \multicolumn{1}{c|}{x}   \\ \hline

\multicolumn{1}{|c|}{ \cite{Deepak2019Over}} & \multicolumn{1}{c|}{2019} & \multicolumn{1}{c|}{\begin{tabular}[c]{@{}c@{}} The potential solutions such as D2D communication, \\ drone-supported communication, mobile 	\textit{ad hoc} networks, \\  and Internet-of-Things (IoT) are explored for post disaster networks. \end{tabular} }
& \multicolumn{1}{l|}{x}   
& \multicolumn{1}{l|}{x}
& \multicolumn{1}{l|}{\checkmark}
& \multicolumn{1}{c|}{\begin{tabular}[c]{@{}c@{}} LTE, D2D\\MANET, IoT
 \end{tabular} } 
& \multicolumn{1}{c|}{UAV} 
& \multicolumn{1}{c|}{x}   \\ \hline

\multicolumn{1}{|c|}{ \cite{Kaleem2019UAV}} & \multicolumn{1}{c|}{2019} & \multicolumn{1}{c|}{\begin{tabular}[c]{@{}c@{}} A disaster-resilient architecture for public safety LTE \\ that integrates the advantages of software-defined networks and \\ UAV cloudlets is proposed. \end{tabular} }
& \multicolumn{1}{l|}{x}   
& \multicolumn{1}{l|}{\checkmark}
& \multicolumn{1}{l|}{x}
& \multicolumn{1}{c|}{LTE, SDN} 
& \multicolumn{1}{c|}{UAV} 
& \multicolumn{1}{c|}{x}   \\ \hline

\multicolumn{1}{|c|}{ \cite{Panda2019UAV}} & \multicolumn{1}{c|}{2019} & \multicolumn{1}{c|}{\begin{tabular}[c]{@{}c@{}} A UAV-assisted \ac{Wi-Fi} based emergency network model \\ is designed for post-disaster management. \end{tabular} }
& \multicolumn{1}{l|}{x}   
& \multicolumn{1}{l|}{x}
& \multicolumn{1}{l|}{\checkmark}
& \multicolumn{1}{c|}{Wi-Fi} 
& \multicolumn{1}{c|}{UAV} 
& \multicolumn{1}{c|}{x}   \\ \hline

\multicolumn{1}{|c|}{ \cite{Zhao2019UAV}} & \multicolumn{1}{c|}{2019} & \multicolumn{1}{c|}{\begin{tabular}[c]{@{}c@{}} A joint UAV trajectory and scheduling optimization has been \\  studied for emergency networks in disaster. \end{tabular} }
& \multicolumn{1}{l|}{x}   
& \multicolumn{1}{l|}{\checkmark}
& \multicolumn{1}{l|}{\checkmark}
& \multicolumn{1}{c|}{Cellular, D2D} 
& \multicolumn{1}{c|}{UAV} 
& \multicolumn{1}{c|}{x}   \\ \hline

\multicolumn{1}{|c|}{ \cite{Feng2020SWIPT}} & \multicolumn{1}{c|}{2020} & \multicolumn{1}{c|}{\begin{tabular}[c]{@{}c@{}} A framework has been devised for emergency communications, \\ employing UAV-enabled Simultaneous Wireless Information \\ and Power Transfer (SWIPT) for IoT networks. \end{tabular} }
& \multicolumn{1}{l|}{\checkmark}
& \multicolumn{1}{l|}{\checkmark}
& \multicolumn{1}{l|}{x}   
& \multicolumn{1}{c|}{IoT} 
& \multicolumn{1}{c|}{UAV} 
& \multicolumn{1}{c|}{x}   \\ \hline

\multicolumn{1}{|c|}{ \cite{HAZRA202054}} & \multicolumn{1}{c|}{2020} & \multicolumn{1}{c|}{\begin{tabular}[c]{@{}c@{}} Network resource deployment problem is studied in disaster area. \end{tabular} }
& \multicolumn{1}{l|}{x}
& \multicolumn{1}{l|}{x}  
& \multicolumn{1}{l|}{\checkmark}
& \multicolumn{1}{c|}{\begin{tabular}[c]{@{}c@{}} Bluetooth, \ac{Wi-Fi},  \\ \ac{LoRa}
 \end{tabular} } 
& \multicolumn{1}{c|}{x} 
& \multicolumn{1}{c|}{x}   \\ \hline

\multicolumn{1}{|c|}{ \cite{hossain2020smartdr}} & \multicolumn{1}{c|}{2020} & \multicolumn{1}{c|}{\begin{tabular}[c]{@{}c@{}} A method based on D2D technology that use smartphones called \\  as Smartphone Assisted Disaster Recovery (SmartDR) \\ for post-disaster communication is proposed. \end{tabular} }
& \multicolumn{1}{l|}{x}
& \multicolumn{1}{l|}{x}  
& \multicolumn{1}{l|}{\checkmark}
& \multicolumn{1}{c|}{D2D} 
& \multicolumn{1}{c|}{x} 
& \multicolumn{1}{c|}{x}   \\ \hline

\multicolumn{1}{|c|}{ \cite{WuFSObasedDrone2020}} & \multicolumn{1}{c|}{2020} & \multicolumn{1}{c|}{\begin{tabular}[c]{@{}c@{}} The design of a mobile access network, supported by UAVs \\ employing \ac{FSO}, is introduced to \\ rapidly reinstate communication in disaster-stricken areas and  \\ reduce network latency between mobile users in those regions and  \\ operational \glspl{BS} located elsewhere. \end{tabular} }
& \multicolumn{1}{l|}{x}
& \multicolumn{1}{l|}{x}  
& \multicolumn{1}{l|}{\checkmark}
& \multicolumn{1}{c|}{Cellular} 
& \multicolumn{1}{c|}{UAV} 
& \multicolumn{1}{c|}{x}   \\ \hline

\multicolumn{1}{|c|}{ \cite{jahid2020techno}} & \multicolumn{1}{c|}{2020} & \multicolumn{1}{c|}{\begin{tabular}[c]{@{}c@{}} Investigates the potential of using RES, diesel \\ generators, and hybrid PV/electric grid, to power LTE base \\ stations with a focus on technical, economic, and environmental factors. \end{tabular} }
& \multicolumn{1}{l|}{\checkmark}
& \multicolumn{1}{l|}{x}  
& \multicolumn{1}{l|}{x}
& \multicolumn{1}{c|}{ \begin{tabular}[c]{@{}c@{}} RES, \\ Diesel generators
 \end{tabular} } 
& \multicolumn{1}{c|}{x} 
& \multicolumn{1}{c|}{x}   \\ \hline

\multicolumn{1}{|c|}{ \cite{Niu2021EEmax}} & \multicolumn{1}{c|}{2021} & \multicolumn{1}{c|}{\begin{tabular}[c]{@{}c@{}} Energy-efficiency maximization problem for UAV \\ emergency network where a UAV works as an \\ aerial \ac{BS} is studied.
 \end{tabular} }
& \multicolumn{1}{l|}{x}
& \multicolumn{1}{l|}{x}  
& \multicolumn{1}{l|}{\checkmark}
& \multicolumn{1}{c|}{Cellular} 
& \multicolumn{1}{c|}{UAV} 
& \multicolumn{1}{c|}{x}   \\ \hline

\multicolumn{1}{|c|}{ \cite{Sherman2021UAVcharge}} & \multicolumn{1}{c|}{2021} & \multicolumn{1}{c|}{\begin{tabular}[c]{@{}c@{}} A temporary UAV-assisted mobile network model in which \\ renewable energy powered charging stations are used \\to solve the energy problem of UAVs. Moreover, a reinforcement\\ Q-Learning-based charging scheduling  algorithm is \\ proposed to improve the life-span UAV \glspl{BS}.
 \end{tabular} }
& \multicolumn{1}{l|}{x}
& \multicolumn{1}{l|}{x}  
& \multicolumn{1}{l|}{\checkmark}
& \multicolumn{1}{c|}{Cellular} 
& \multicolumn{1}{c|}{UAV} 
& \multicolumn{1}{c|}{x}   \\ \hline

\multicolumn{1}{|c|}{ \cite{Wang2021DisasterReliefWireless}} & \multicolumn{1}{c|}{2021} & \multicolumn{1}{c|}{\begin{tabular}[c]{@{}c@{}} Blockchain and machine learning techniques are integrated \\ to UAV-assisted disaster relief networks \\ to achieve secure and efficient data delivery.
 \end{tabular} }
& \multicolumn{1}{l|}{x}
& \multicolumn{1}{l|}{x}
& \multicolumn{1}{l|}{\checkmark}
& \multicolumn{1}{c|}{Cellular} 
& \multicolumn{1}{c|}{UAV} 
& \multicolumn{1}{c|}{x}   \\ \hline

\multicolumn{1}{|c|}{ \cite{kurt2021vision}} & \multicolumn{1}{c|}{2021} & \multicolumn{1}{c|}{\begin{tabular}[c]{@{}c@{}} Presents a vision and framework for HAPS networks, \\ including use cases and the integration of emerging \\ wireless technologies.
 \end{tabular} }
& \multicolumn{1}{l|}{\checkmark}
& \multicolumn{1}{l|}{\checkmark}
& \multicolumn{1}{l|}{\checkmark}
& \multicolumn{1}{c|}{x} 
& \multicolumn{1}{c|}{HAPS, PV} 
& \multicolumn{1}{c|}{x}   \\ \hline

\multicolumn{1}{|c|}{ \cite{Ramakrishnan2022}} & \multicolumn{1}{c|}{2022} & \multicolumn{1}{c|}{\begin{tabular}[c]{@{}c@{}} A resilient architecture that enables first responder \\ communication by using D2D technology is proposed \\ for challenging situations caused by a disaster.
 \end{tabular} }
& \multicolumn{1}{l|}{x}
& \multicolumn{1}{l|}{\checkmark}
& \multicolumn{1}{l|}{\checkmark}
& \multicolumn{1}{c|}{D2D} 
& \multicolumn{1}{c|}{x} 
& \multicolumn{1}{c|}{x}   \\ \hline

\multicolumn{1}{|c|}{ \cite{Matracia2023SG}} & \multicolumn{1}{c|}{2023} & \multicolumn{1}{c|}{\begin{tabular}[c]{@{}c@{}} Stochastic geometry based performance analysis of \\ UAV-aided post-disaster cellular networks is performed.
 \end{tabular} }
& \multicolumn{1}{l|}{x}
& \multicolumn{1}{l|}{x}
& \multicolumn{1}{l|}{\checkmark}
& \multicolumn{1}{c|}{Cellular} 
& \multicolumn{1}{c|}{\begin{tabular}[c]{@{}c@{}} Ad Hoc \\ UAV
 \end{tabular} } 
& \multicolumn{1}{c|}{x}   \\ \hline

\multicolumn{1}{|c|}{ \cite{israr2023emission}} & \multicolumn{1}{c|}{2023} & \multicolumn{1}{c|}{\begin{tabular}[c]{@{}c@{}} Explores a cost-effective, low-carbon energy solution \\ for small-cell mobile networks using renewable \\ energy and energy storage facilities.
 \end{tabular} }
& \multicolumn{1}{l|}{\checkmark}
& \multicolumn{1}{l|}{x}
& \multicolumn{1}{l|}{x}
& \multicolumn{1}{c|}{RES, BESS} 
& \multicolumn{1}{c|}{x} 
& \multicolumn{1}{c|}{x}   \\ \hline

\multicolumn{1}{|c|}{ \cite{bahri2023economic}} & \multicolumn{1}{c|}{2023} & \multicolumn{1}{c|}{\begin{tabular}[c]{@{}c@{}} Evaluates the various energy sources for a mobile BS \\ in an isolated nanogrid, focusing on economic and environmental \\ aspects and addressing uncertainties such as mobile BS traffic \\ rates and PV generation, using optimization methods.
 \end{tabular} }
& \multicolumn{1}{l|}{\checkmark}
& \multicolumn{1}{l|}{x}
& \multicolumn{1}{l|}{x}
& \multicolumn{1}{c|}{ \begin{tabular}[c]{@{}c@{}} PV, HES, EV, \\ Diesel generators
 \end{tabular}} 
& \multicolumn{1}{c|}{x} 
& \multicolumn{1}{c|}{x}   \\ \hline

\multicolumn{1}{|c|}{ \cite{ghosh2023uav}} & \multicolumn{1}{c|}{2023} & \multicolumn{1}{c|}{\begin{tabular}[c]{@{}c@{}} The energy efficiency maximization problem in a UAV-assisted, \\ energy harvesting-enabled, 
NOMA-based D2D network \\ scenario for disaster management has been investigated.
 \end{tabular} }
& \multicolumn{1}{l|}{x}
& \multicolumn{1}{l|}{\checkmark}
& \multicolumn{1}{l|}{\checkmark}
& \multicolumn{1}{c|}{ \begin{tabular}[c]{@{}c@{}} Wireless Power \\ Transfer, \\ Energy Harvesting
 \end{tabular} } 
& \multicolumn{1}{c|}{x} 
& \multicolumn{1}{c|}{x}   \\ \hline

\multicolumn{1}{|c|}{ \cite{deevela2024review}} & \multicolumn{1}{c|}{2024} & \multicolumn{1}{c|}{\begin{tabular}[c]{@{}c@{}} Provides a comprehensive analysis of various \\ renewable energy-based hybrid energy solutions for tower-type BSs.
 \end{tabular} }
& \multicolumn{1}{l|}{\checkmark}
& \multicolumn{1}{l|}{\checkmark}
& \multicolumn{1}{l|}{x}
& \multicolumn{1}{c|}{ \begin{tabular}[c]{@{}c@{}} RES, BESS, \\ Diesel generators, \\  Fuel cells
 \end{tabular} } 
& \multicolumn{1}{c|}{x} 
& \multicolumn{1}{c|}{x}   \\ \hline

\multicolumn{1}{|c|}{ \cite{matracia2024reliability}} & \multicolumn{1}{c|}{2024} & \multicolumn{1}{c|}{\begin{tabular}[c]{@{}c@{}} An interference mitigation strategy based \\ on BS silencing is proposed for emergency scenarios. 
 \end{tabular} }
& \multicolumn{1}{l|}{x}
& \multicolumn{1}{l|}{x}
& \multicolumn{1}{l|}{\checkmark}
& \multicolumn{1}{c|}{Cellular} 
& \multicolumn{1}{c|}{x} 
& \multicolumn{1}{c|}{x}   \\ \hline

\end{tabular}
\label{tab:vertical_table2}
\end{table*}

\section{Pre-Disaster Communication and Energy Planning and Warning}
\label{predisaster}

We will now consider approaches that can be taken before a disaster occurs to save-guard the network, both from a communication and energy perspective.

\subsection{Technology Enablers for Communication in Pre-disaster Scenarios}

Pre-disaster communication planning and warning systems include various communications technologies and protocols. FEMA\footnote{Online: https://www.fema.gov/about/offices/field-operations/disaster-emergency-communications, Available: April 2023.} provides disaster emergency communications through geographically dispersed mobile emergency response support detachments and a pre-positioned fleet of mobile communications office vehicles. The common alerting protocol is used to enhance messaging in disaster early warning communication systems in \cite{christian2022communicating}. Present emerging disruptive technologies and communication protocols are employed internationally for early warning and communication systems \cite{banzal2022disaster}. Information and communication tools such as the \textcolor{black}{I}nternet, \ac{GIS}, remote sensing, and satellite communication are used for disaster preparedness, warning, and forecasting \cite{mohan2020review}. Remote sensing is a technology that enables the collection of environmental data from a distance, typically through the use of satellites or aircraft. It can be beneficial in monitoring and tracking natural disasters like hurricanes, wildfires, and floods and it is also valuable in creating maps and models that illustrate the damage caused by natural disasters, helping decision-makers prioritize response efforts and allocate resources more effectively. It requires a proper tool like \ac{GIS} to handle the large amount of data obtained by remote sensing. Thus, in \cite{van2002remote}, remote sensing and GIS technologies \textcolor{black}{were discussed} for different types of natural disasters for warning and monitoring duties. Radar and satellite imaging technologies are also vital tools for monitoring and managing natural disasters, as they provide critical data for the surface and atmosphere of the Earth \cite{handwerger2022generating}\cite{villagran2023improved}. The authors \textcolor{black}{of} \cite{omar2020research} \textcolor{black}{demonstrated} the research design for mobile-based decision support of the Flood Early Warning System (FEWS).

\textit{Ad hoc} networks can be used for pre-disaster communication planning and warning by providing a flexible and dynamic communication infrastructure that can be quickly deployed in disaster-prone areas \cite{ahmed2019disaster}.  In the pre-disaster phase, 	\textit{ad hoc} networks can be used to create communication channels between emergency responders, local communities, and other stakeholders. These networks can be established using mobile devices and are quick and easy to set up in an emergency. 	\textit{Ad hoc} networks can also be used to alert and disseminate important information to local communities. For example, community members can receive alerts and warnings about impending natural disasters or emergencies via their mobile devices. Another use of 	\textit{ad hoc} networks in pre-disaster communications planning is to support situational awareness and information sharing among emergency responders. 	\textit{Ad hoc} networks can facilitate the exchange of real-time information such as maps, images and videos between emergency responders and command centers. In \cite{reina2015survey}, different \textit{ad hoc} communication paradigms such as \ac{MANET}, \ac{VANET}, \ac{WMN}, TETRA, etc. \textcolor{black}{were discussed} in detail for disaster management scenarios.

\glspl{UAV}, generally known as drones, can be a valuable tool in pre-disaster communication planning and warning, providing a versatile platform for gathering and disseminating critical information in disaster-prone areas and supporting emergency response efforts \cite{wu2021research}. One use of \glspl{UAV} in pre-disaster communication planning is to conduct aerial surveys of disaster-prone areas to identify potential hazards and vulnerabilities \cite{alawad2023unmanned}. These surveys can be used to generate detailed maps and models of the terrain, as well as to identify infrastructure, buildings, and other critical assets that may be at risk during a disaster.  Another use of \glspl{UAV} in pre-disaster communication planning is to support warnings and disseminate important information to local communities. \glspl{UAV} can be equipped with loudspeakers or other audio-visual devices to broadcast warnings and alerts to people in disaster-prone areas. \glspl{UAV} can also be used to deliver essential supplies such as first aid kits or food and water to people in remote or hard-to-reach areas. In \cite{Feng2020SWIPT}, a \ac{UAV} \textcolor{black}{was used} not only as a data gathering tool from the \ac{IoT} devices but also as a platform for energy transfer to these energy-limited IoT devices, which \textcolor{black}{were distributed} in a disaster area.

\ac{IoT} offers significant potential for the detection and management of natural \textcolor{black}{hazards}. \ac{IoT} sensors can be deployed to collect data to monitor environmental factors such as temperature, humidity, air quality, and seismic activity, providing early warnings of potential natural \textcolor{black}{hazards} \cite{ali2015disaster}\cite{awadallah2019internet}. This data can be used to detect anomalies that may indicate the presence of natural hazards, such as earthquakes  \cite{abdalzaher2023early}\cite{abdalzaher2023toward}. The data can also be used to identify vulnerabilities and risks in critical infrastructure, such as bridges, dams, and power plants, to support situational awareness and hazard identification. \textcolor{black}{ \glspl{WSN}} can be used for environmental monitoring where \ac{IoT} devices and sensors can collect data on environmental conditions in real-time, which is critical for early warning systems and situational awareness \cite{russell2018agile}. \ac{IoT} and \ac{WSN}s can be used for pre-disaster communication planning and warning by providing real-time data and analytics that can be used to support situational awareness, hazard identification, and early warning systems, particularly for critical infrastructures such as gas and oil refinery \cite{salameh2020end}. \ac{IoT} networks can also be used to support situational awareness and information sharing among emergency responders. IoT-based sensors can provide real-time information on the extent of damage, the location of survivors, and other critical information that can be used to support emergency response efforts. Distributing many \ac{IoT} devices, which are usually heterogeneous, with their own communication protocol and/or data formats, in large-scale disasters may face some challenges. To manage the diversity of \ac{IoT} devices and applications, gateways are required to connect traditional communication networks with the \ac{IoT} domain. However, these gateways are typically centralized, making them impractical for use in the \ac{MANET} environments often encountered during large-scale disasters. Thus, in \cite{mouradian2018nfv}, a distributed \ac{NFV} and \ac{SDN}-based \ac{IoT} gateway architecture \textcolor{black}{was presented}. \textcolor{black}{A prototype is built in which SDN and NFV enable on-the-fly chaining of gateways in the IoT domain.}

\ac{AI}, particularly \ac{ML} and deep learning, has demonstrated significant advantages in risk mitigation, assessing the vulnerability of urban areas to seismic hazards, evaluating site suitability, enhancing early warning systems, and managing natural disasters in the existing literature \cite{allegranti2023use,abdalzaher2021deep,abdalzaher2023seismic,saad2022machine}. The paper \cite{wang2020edge} \textcolor{black}{focused} on supporting mission-critical applications in emergency scenarios by enhancing network adaptability and intelligence.  \ac{AI} and \ac{ML} algorithms are envisioned to be on the edge of the network to analyze data and make real-time decisions. In addition, it uses cognitive radio technology to dynamically allocate spectrum resources based on network conditions and demand. The applications of digital twins in emergency management of civil infrastructure (EMCI) \textcolor{black}{were} surveyed in \cite{cheng2023review}. \textcolor{black}{Additionally, to explain how the digital twin works for EMCI in the mitigation, preparation, response and recovery phases, a framework is proposed.} The authors \textcolor{black}{of} \cite{fan2021disaster} \textcolor{black}{used} digital twin models to simulate and predict the impacts of disasters on urban infrastructure. \textcolor{black}{Their vision goes beyond fragmented digital modeling and includes four main components, namely multi-model data collection and sensing, data integration and analysis, game-theoretic decision making with multiple actors, and dynamic network analysis. } \textcolor{black}{These models integrate AI and IoT technologies to enable real-time monitoring and support effective decision-making. In addition, a data-driven approach is employed to enhance disaster preparedness and response.}

Social media platforms serve three critical roles in disasters: (i) quickly gathering situational awareness information, (ii) facilitating self-organized peer-to-peer help, and (iii) allowing disaster management agencies to hear from the public \cite{zhang2019social}.
During and after a disaster, social media can be leveraged to pinpoint areas where people need assistance with search and rescue, medical aid, and access to food and water. Additionally, social media serves as a vital tool for disseminating information and coordinating relief efforts for those in need. Moreover, for training and simulation, \ac{AR} and \ac{VR} technologies can be used to train responders in simulated disaster scenarios to improve their readiness for real-world situations \cite{zhu2021virtual}.

\subsection{Technology Enablers for Energy Support in Pre-disaster Scenarios}

Technology enablers for energy support in pre-disaster scenarios refer to the use of various technological solutions to ensure the availability and resilience of energy systems before a disaster occurs. These enablers aim to improve energy infrastructure, increase energy generation and storage capabilities, and strengthen the overall resilience of the energy supply to potential disasters.  \ac{RES} can play a crucial role in advancing the sustainability of terrestrial communication systems. \glspl{BS} in 5G communication systems consume approximately four times more energy than those in 4G systems \cite{li2023carbon}. It is the implementation of technologies such as Massive \ac{MIMO}, with more \ac{RF} chains and electronic components, that results in the rising trend in power consumption \cite{musa2023energy}, even if the energy efficiency also improves since the data capacity increases faster than the power \cite{Bjornson2015a}.  As the demand for energy-intensive communication infrastructure continues to grow, integrating \ac{RES} offers a pathway to significantly reduce these systems' carbon footprint. \textcolor{black}{Reference} \cite{samorzewski2023energy} \textcolor{black}{explored} how incorporating \ac{RES} into 5G cellular systems affected overall power consumption, revealing that annual power savings surpassed 50\% when the system relied solely on the conventional energy grid. \textcolor{black}{Reference} \cite{israr2023emission} \textcolor{black}{investigated} a sustainable energy solution for 5G \glspl{BS}, incorporating \ac{RES} and \ac{BESS} to reduce energy consumption and carbon emissions. Integrating \ac{RES} into terrestrial communication systems supports sustainability goals, enhances energy security, and reduces operational costs over time \cite{shen2024integrated}. Moreover, \ac{RES} contribute to the resilience of communication systems by providing a decentralized and diversified energy supply, particularly critical in remote or off-grid areas where traditional power sources may be unreliable or unavailable.

By integrating \ac{BESS} with \ac{RES} such as solar or wind, communication infrastructures can significantly reduce their carbon footprint, contributing to environmental sustainability. The study \cite{kuaban2023modelling} \textcolor{black}{examined} a model that evaluated the energy performance of an off-grid, eco-friendly cellular \ac{BS}, incorporating a solar power system, a \ac{BESS} system, and a \ac{HES}. \textcolor{black}{Reference} \cite{wang2024distribution} \textcolor{black}{investigated} the relationship between the power grid and 5G \glspl{BS} when an energy storage system was involved, highlighting that the energy storage system reduced energy consumption during peak times on the grid. \textcolor{black}{Reference} \cite{bahri2023economic} \textcolor{black}{proposed} the integration of diverse energy sources for \glspl{BS}, such as \ac{HES}, solar \ac{PV} installations, and plug-in \glspl{EV}, highlighting their economic and environmental benefits. \ac{BESS} can be strategically deployed to optimize energy use, balance load demand, and minimize reliance on non-\ac{RES} \cite{zhang2023optimal}. This promotes energy efficiency and enhances the resilience and reliability of communication systems, ensuring they remain operational under various conditions. As the demand for uninterrupted communication grows, the role of \ac{BESS} in supporting sustainable and resilient terrestrial communication networks becomes increasingly vital.

Smart grids enable the integration of multiple distributed energy sources and digital operations using communication technologies \cite{saleh20235g}. They use advanced sensors, automated controls, and real-time data analytics to optimize electricity generation, distribution, and consumption, contributing to sustainability \cite{qays2023key}. \textcolor{black}{Reference} \cite{israr2023renewable} \textcolor{black}{showed} that traffic offloading to terrestrial \glspl{BS} supported by \ac{RES} could reduce grid energy usage and lower investment and operational costs. Meanwhile, the study \cite{wang2023cost} \textcolor{black}{suggested} that 5G \glspl{BS} supported by \ac{RES} and batteries could operate cost-effectively under normal conditions and provide resiliency during disasters, facilitating the recovery of the distribution system. Supporting terrestrial communication systems with the smart grid concept can provide a more resilient, efficient, and environmentally friendly energy infrastructure \cite{taveras2023implications}.

Energy-efficient communication technologies are essential for ensuring the sustainability of terrestrial communication systems. By incorporating low-power IoT communication devices and energy-efficient network equipment, the overall energy consumption of communication networks can be reduced \cite{rahmani2023next}. Moreover, energy savings can be achieved by switching off terrestrial \glspl{BS} based on network traffic, implementing resource management schemes that optimize energy consumption, and applying communication technologies such as D2D and NOMA \cite{larsen2023toward, salamatmoghadasi2023energy, wang2023energy}. \textcolor{black}{Reference} \cite{deevela2024review} \textcolor{black}{investigated} the integration of energy-efficient systems such as diesel generators, solar \ac{PV} panels, wind turbines, fuel cells, and micro-turbines into terrestrial \glspl{BS} and emphasized the reduction of carbon emissions. The study \cite{jin2023survey} \textcolor{black}{examined} energy efficiency methods in \ac{UAV} communication and highlighted battery swapping, solar energy harvesting, \ac{RF} energy harvesting, and laser beam energy harvesting methods to meet the changing needs of \glspl{UAV}. Through energy-efficient power supply tools, network equipment, and communication technologies, communication systems can remain functional during power shortages or outages, thus contributing to terrestrial communication infrastructure's overall sustainability and reliability.

In summary, the above technology enablers can work together to improve the resilience and reliability of energy systems in pre-disaster scenarios and to ensure the availability of critical energy services during and after a disaster.

\subsection{Lessons Learnt and Recommendations}

In the literature, the terms \enquote{in disaster} and \enquote{post disaster} are often used interchangeably, \textcolor{black}{while the “pre-disaster” phase is frequently underrepresented or not clearly distinguished based on the technologies discussed. It is important to emphasize that this distinction should be considered during the development and classification of relevant technologies.} In pre-disaster scenarios, there are already energy producers that promote sustainability, although for earthquake scenarios, there are not many specific solutions available.  Integrating cutting-edge technologies like \ac{GIS}, remote sensing, \ac{IoT}, and \ac{AI} into disaster management plans is crucial for effective disaster management. These technologies offer real-time data, enhance understanding of current situations, and improve early warning systems, proving their essential value in disaster preparedness and response.

Ad hoc networks and UAVs are crucial for pre-disaster communication planning. They can create flexible, quickly deployable communication channels, ensuring a continuous flow of information among emergency responders, local communities, and other stakeholders, even in the most challenging environments. \ac{IoT} and \ac{WSN}s can be used to plan for and warn about disasters by providing real-time data and analysis that can help us understand the situation, identify hazards, and create early warning systems, such as for critical infrastructure like gas and oil refineries. Although \ac{IoT} and \ac{WSN}s hold great promise for disaster management, the variety of \ac{IoT} devices and the need for centralized gateways present challenges, particularly in large-scale disasters. Managing these diverse systems requires innovative solutions, such as decentralized \ac{NFV} and \ac{SDN}-based \ac{IoT} gateway architectures.

Social media and \ac{AI} are becoming more important in predicting, responding to, and managing disasters. Social media helps with early warnings and coordinating relief efforts because it's real-time, and \ac{AI} helps make better decisions by analyzing data in real-time. The deployment of \ac{RES} and \ac{BESS} in communication infrastructures highlights the significance of energy resilience. As communication systems become more energy-intensive, integrating renewable energy with BESS not only reduces the carbon footprint but also ensures the reliability of these systems during disasters.

We summarize our recommendations as follows. Firstly, to overcome the challenges of the variety of \ac{IoT} devices, we should work together to promote compatibility and standardization among communication protocols and data formats. This will make it easier to integrate \ac{IoT} devices into disaster management systems. Secondly, \ac{AI} and \ac{ML} technologies should be further developed to aid real-time decision-making in disaster situations. This involves enhancing cognitive radio technologies for dynamic spectrum allocation and utilizing AI at the network edge to improve adaptability and response times. Thirdly, social media platforms should be integrated into disaster management strategies to enhance early warning systems and coordinate relief efforts. Training responders and communities on effective social media use during disasters can significantly improve results. Fourth, since \ac{RES} and \ac{BESS} have the potential to enhance the resilience of communication systems, we recommend using them more in areas that are often affected by disasters and remote areas. This would involve investing in sustainable energy infrastructure and making sure these systems are strong enough to work in extreme conditions. Moreover, ongoing research and development should focus on improving the integration of various technological enablers, particularly in the context of energy-efficient communication technologies and smart grids. \textcolor{black}{To this end, standardized and interoperable communication interfaces must be developed to ensure seamless integration between communication infrastructure and energy management systems, especially in heterogeneous environments. Additionally, real-time coordination between distributed energy resources and communication networks remains a key challenge, calling for adaptive and energy-aware network protocols that can dynamically respond to fluctuating energy availability and infrastructure conditions. Moreover, joint optimization frameworks that simultaneously consider energy constraints, network topology, and service quality are needed to maximize operational efficiency in post-disaster environments. The integration of renewable-powered microgrids with communication layers should also be explored, particularly in terms of maintaining connectivity when traditional power supplies are disrupted. Further investigation is required into the scalability, latency, and resilience of AI-driven power management strategies for autonomous and self-healing communication networks operating under disaster-induced stress conditions.} This will ensure that communication infrastructures are both sustainable and resilient and capable of meeting the demands of future disaster scenarios. 

Although this study primarily focuses on enhancing the resilience of communication infrastructures through the integration of advanced technologies such as HAPS, UAVs, and RES, the resilience of power system components themselves—and their tight interdependence with ICT systems—remains equally critical. Recent studies (e.g., \cite{ghasemi2023robustness}) have emphasized that failures in either the communication or power network can propagate in a cascading fashion due to tightly coupled interdependencies, often leading to large-scale outages. For instance, ICT components such as base stations, edge devices, and network controllers are heavily reliant on an uninterrupted power supply, while modern smart grids, in turn, depend on ICT layers for control, monitoring, and demand-response actions. This mutual dependency creates a bi-directional vulnerability pathway, particularly during disaster scenarios. Therefore, future research should incorporate co-simulation and co-optimization models that evaluate both networks within a unified framework. Specifically, disaster-resilient designs should consider factors such as the availability and distribution of backup energy systems for ICT nodes, optimal placement of control centers, and fault-tolerant communication strategies within power grids. Moreover, the integration of distributed energy resources with intelligent load balancing mechanisms and localized control can enhance autonomy, thereby mitigating the impact of cascading failures. Understanding and addressing these interdependencies is vital for creating a holistic and robust infrastructure capable of surviving and recovering from extreme disaster scenarios.

Finally, to make pre-disaster communication and early warning systems more effective, public awareness and training should be improved. This includes educating communities on technologies such as mobile alerts and \ac{UAV}s and using \ac{AR} and \ac{VR} technologies to train emergency personnel in realistic disaster response scenarios.

\section{Disaster Response Communication and Energy Planning}
\label{indisaster}

Disaster response communications and energy planning mean developing and implementing a plan to create an effective communication channel with the public during and after a disaster in a sustainable manner. It consists of key elements such as identifying key audiences (e.g., first responders, volunteers, government officials, and the public), developing messages, selecting communication channels, creating a communication protocol, and testing and evaluating the plan. Many research works have been proposed to deploy a temporary communications network in the context of a disaster response \cite{hazra2019novel,teng2020instantaneous,kaisar2021emergency,mondal2021emergency,hazra2021rcpdnpt}. In this section, we examine communication and energy enablers for disaster response phase in detail.

\subsection{Technology Enablers for Communication in Disaster Response Scenarios}

\begin{figure}[htp!]
\centering
\includegraphics[width=\linewidth]{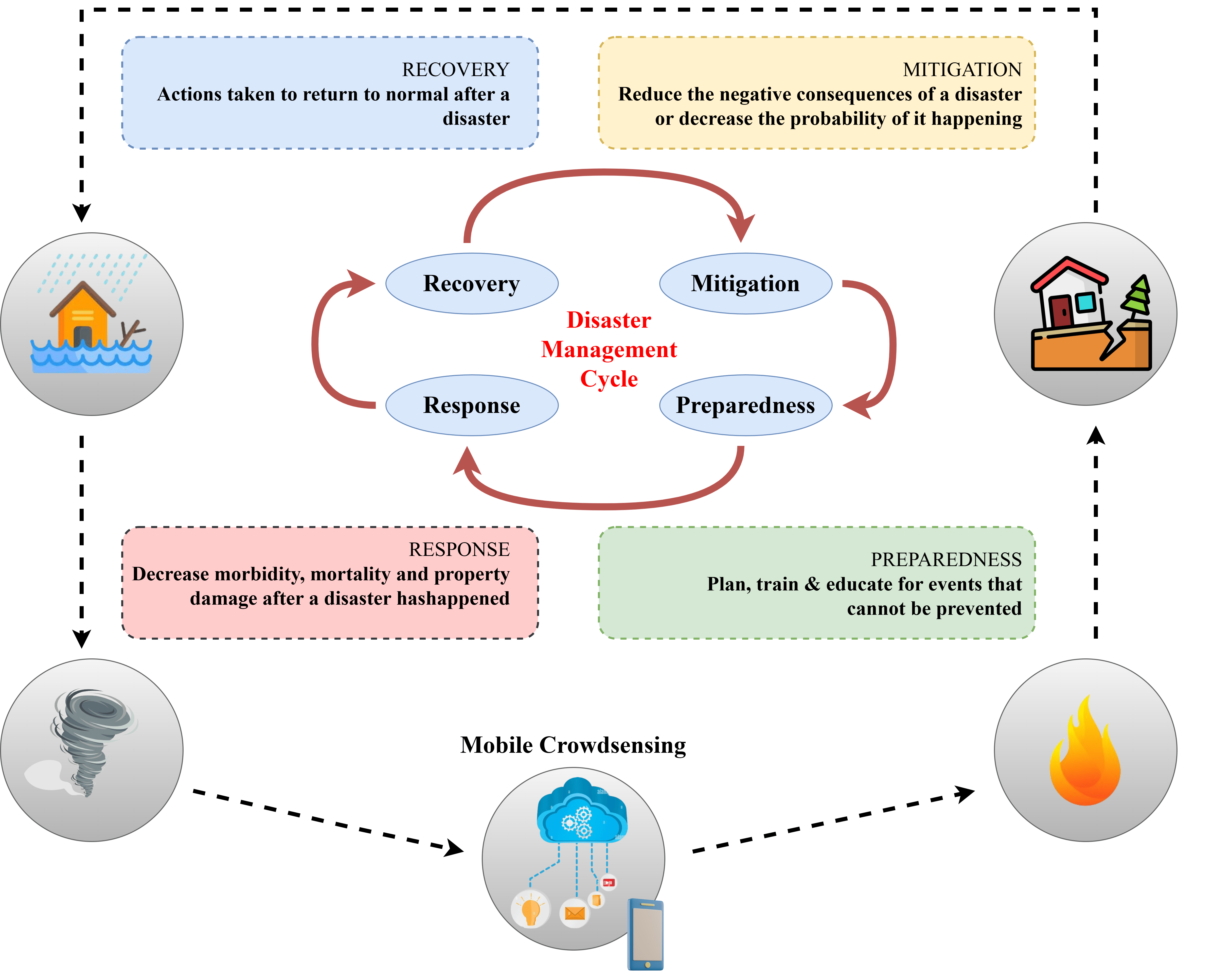}
\caption{Mobile Crowd-sensing-Aided Disaster Management \cite    {cicek2023use}.}
\label{mobile_crowdsourcing}
\vspace{-0.5cm}
\end{figure}

During natural disasters, various stakeholders, including first responders, news agencies, and victims, rely on social media as a primary communication channel to reliably disseminate situational information that reaches a large user base \cite{saroj2020use}.  Officials can use social media platforms to quickly disseminate accurate information, publish detailed instructions or infographics, counter misinformation, and engage with the public.  However, it must also be ensured that the dissemination of information takes place between trusted and reliable nodes in order to guarantee the security of the disseminated information. The authors \textcolor{black}{of} \cite{inal2022use} \textcolor{black}{analyzed} an overview of the characteristics and the potential role of social media (tweets in particular) for risk communication in Turkiye. \textcolor{black}{In particular, relevant five hundred and eighty tweets on disaster topics were discovered, and most of the tweets were related to hydrological and meteorological hazards.}  Georeferencing social media crowd-sourcing data is also crucial for improving situational awareness and requires very low positioning errors.  As given in Fig. \ref{mobile_crowdsourcing}, mobile crowd-sensing applications can engage user communities in emergency response and disaster management for natural hazards.  A review of crowd-sensing through smartphone sensors in disaster incidents \textcolor{black}{was} provided in \cite{cicek2023use}. \textcolor{black}{The authors conclude that crowd-sensing use cases with smartphone sensors is limited for disaster events.}    

Mobile applications can be a powerful tool for disseminating real-time information, evacuation routes and safety guidelines during an earthquake and recovery efforts. By providing users with access to important information, mobile apps can help people make informed decisions and take appropriate actions to stay safe during and after an earthquake. Some examples of successful mobile apps in this respect are the \enquote{MyShake} app\footnote{Online: https://myshake.berkeley.edu/, Available: February 2024.} and ShakeAlert\footnote{Online: https://www.shakealert.org/, Available: February 2024.}. These apps use the sensors in the user's smartphone to detect earthquakes and warn the user at an early stage when an earthquake is detected. In addition to the early warning system, these apps also contain information on what to do in the event of an earthquake, evacuation routes, and a \enquote{Safe Corner} feature that helps users find the safest place to seek shelter in the event of an earthquake.  These apps also include information on earthquake safety, preparedness, and recovery, as well as a map showing the locations of nearby earthquake early warning sensors.

\ac{SDN} technology allows for dynamic and programmable network management, enabling quick adjustments and reconfigurations in response to changing conditions during and after an earthquake. \ac{SDN} can ensure that entities and services such as emergency services, first responders, and critical infrastructure receive the necessary bandwidth and \ac{QoS} guarantees, enhancing their communication capabilities and facilitating effective disaster response. Ogawa et al. \textcolor{black}{proposed} a network operation method using \ac{SDN} to maintain connection quality for disaster areas \cite{ogawa2014network}. \textcolor{black}{They also point out the problems of information infrastructure during disasters and implement the proposed method for maintaining connectivity.}  Web-based offline operation and OpenFlow-based routing control \textcolor{black}{were adopted} in \cite{tabata2016disaster} as a disaster information gathering mechanism. \textcolor{black}{A prototype implementation has shown that the proposed mechanism of dynamically selecting a connection in response to changes in wireless connections based on status and priority achieves a higher information collection rate on the server.}

\textit{Ad hoc} networks and \ac{MANET} can be used for real-time data sharing and coordination between first responders are proposed to improve situational awareness and response capabilities.  \textcolor{black}{Reference} \cite{Ramakrishnan2022} \textcolor{black}{used} \textit{ad hoc} networking techniques to enable dynamic communication between first responders without relying on a fixed infrastructure. \textcolor{black}{The proposed multi-layered ReDiCom (Resilient Disaster Communications) architecture increases reliability and enables efficient data processing in a distributed manner.}  The paper \cite{aliyu2024disaster} \textcolor{black}{extended} the traditional optimized link-state routing (OLSR) protocol of \ac{MANET} with disaster-optimized link-state routing (DS-OLSR) for low-battery devices to extend their battery life. \textcolor{black}{The simulations have shown improvements in terms of energy savings and packets delivery for both sparse and dense network scenarios.} \textcolor{black}{Reference} \cite{zeng2024study} \textcolor{black}{proposed} a deployment strategy for air-to-ground \textit{ad hoc} network nodes in areas without public network access using a Deep Deterministic Policy Gradient (DDPG) method to support spatio-temporal dynamic deployment in disaster response. \textcolor{black}{The results show that when setting up networks for emergencies, higher communication reliability and efficiency is achieved at low cost.}

\ac{D2D} communication is used to enable direct connections between devices, bypassing the damaged infrastructure. The paper \cite{rawat2015towards} \textcolor{black}{leveraged} 5G technology for high-speed and reliable communication in disaster scenarios. It \textcolor{black}{utilized} \ac{D2D} communication to establish direct links between devices, bypassing damaged infrastructure, and \textcolor{black}{focused} on enhancing network resilience through advanced communication protocols and technologies. \textcolor{black}{Reference} \cite{AliD2D2018} \textcolor{black}{used} \ac{D2D} communication to enable direct connections between devices in disaster areas. Power transfer techniques \textcolor{black}{were implemented} to share battery life between devices to extend their operating time. They also \textcolor{black}{used} clustering techniques to organize devices into groups to improve communication efficiency and resource management.

Cellular \ac{LPWAN} technologies such as \ac{LoRa} can also be utilized in disaster response communication planning in the context of earthquakes to provide reliable and efficient communication capabilities \cite{rastogi2022measuring, boccadoro2019quakesense}.  In \cite{poke2023evaluation}, a long-range, low-power mesh network infrastructure \textcolor{black}{was} proposed for disaster response. \textcolor{black}{However, it was found that the implementation of LoRa Mesh Networks had high error rates, although power consumption provided better results.} \ac{LPWAN} technologies have excellent coverage and can reach far distances, making them suitable for disaster areas where traditional communication infrastructure may be damaged or non-functional \cite{rastogi2022measuring}.  They can provide connectivity to a large geographical area, enabling communication between responders, command centers, and affected individuals across a wide range.  They are designed to be power-efficient, allowing devices to operate on battery power for an extended period. This feature is crucial in disaster scenarios where access to electricity may be limited. 

\glspl{UAV} usage is extremely popular for in-disaster scenarios. The authors of \cite{WuFSObasedDrone2020} \textcolor{black}{used} \ac{FSO} technology for high-speed line-of-sight communication links between drones and ground stations. \textcolor{black}{They used} drone-based networks to extend the range of cellular networks in disaster areas and implemented adaptive beam steering to maintain reliable \ac{FSO} links despite drone movements and changes in the environment. \textcolor{black}{Reference} \cite{Kaleem2019UAV} \textcolor{black}{integrated} edge computing capabilities into \glspl{UAV} to process data locally, reduce latency, and improve the responsiveness of disaster management applications. The focus is on ensuring low-latency communication for critical applications in disaster scenarios. For more resilient networks, the authors also \textcolor{black}{proposed} an architecture that used \glspl{UAV} to maintain communication links even when traditional infrastructure was damaged. \textcolor{black}{Reference} \cite{Feng2020SWIPT} \textcolor{black}{used} \ac{SWIPT} technology to transmit both energy and information wirelessly to support IoT devices in disaster areas. \glspl{UAV} deployed \textcolor{black}{were used} to facilitate \ac{SWIPT} and extend communication coverage in remote or damaged areas.

The authors \textcolor{black}{of} \cite{feng2020noma} \textcolor{black}{used} \glspl{UAV} as aerial \glspl{BS} to ensure communication coverage in disaster areas where the infrastructure on the ground is compromised.  \ac{NOMA} \textcolor{black}{was} used for efficient spectrum utilization and allows multiple users to share the same frequency band by superimposing their signals and distinguishing them at the receiver based on their power. Algorithms \textcolor{black}{were} also proposed to optimize energy and resource allocation between users to improve connectivity and network performance. In \cite{wang2023secure}, a secure and efficient information-sharing system called RescueChain \textcolor{black}{was} proposed for UAV-based disaster rescue. Simulations \textcolor{black}{showed} an accelerated consensus process, improved offloading efficiency, reduced energy consumption, and improved user payoffs in finding optimal payment and resource strategies for UAVs and vehicles. The paper \cite{akter2023task} \textcolor{black}{focused} on the problems of task offloading, power consumption, and allocation of computational resources in a multi-layer MEC-enabled UAV network. The approach also \textcolor{black}{considered} the CPU and GPU requirements of tasks, the capacity of the devices (i.e., computational resources, power, and energy), and the constraints on the types of tasks that a UAV could perform. \textcolor{black}{Reference} \cite{Zhao2019UAV} \textcolor{black}{used} UAVs as aerial \glspl{BS} to ensure communication coverage in disaster areas. It also \textcolor{black}{developed} algorithms to optimize UAV deployment and resource allocation. The proposed approach \textcolor{black}{ensured} the seamless integration of UAV-based networks with ground-based infrastructure. \textcolor{black}{Reference} \cite{Sherman2021UAVcharge} \textcolor{black}{used} UAVs to provide temporary cellular coverage in areas where infrastructure was damaged or destroyed. In addition, the UAVs \textcolor{black}{were equipped} with solar panels and batteries to extend their operational life and reduce dependence on conventional energy sources. Network optimization algorithms \textcolor{black}{were also developed} to optimize the placement and operation of the UAVs for maximum coverage and efficiency.  The real-time prediction capability of a digital twin to optimize the offloading decision under uncertainty  in \ac{UAV}-Assisted \ac{MEC}
Emergency Networks \textcolor{black}{was} studied in \cite{wang2023digital}. \textcolor{black}{The simulation results confirmed the superiority of the proposed online matching algorithm under uncertainty to achieve the optimal stability performance.}  The authors of \cite{2018SaxenaD2D} \textcolor{black}{developed} strategies to ensure the survivability and continuity of networks during disasters by utilizing the proximity of devices.  \textcolor{black}{With the proposed D2D-based redistribution for emergency situations, connectivity losses due to power constraints are handled more efficiently and all connected devices have a longer total usage time.}

\ac{HAPS} is another popular disaster response technology in the literature \cite{kement2023sustaining, karaman2024enhancing}. \textcolor{black}{Reference}\cite{kement2023sustaining} \textcolor{black}{used} super-macro \glspl{BS} to supplement terrestrial networks and cope with dynamic traffic demands. They \textcolor{black}{utilized} the large coverage area of \ac{HAPS} to reduce the need for dense terrestrial \glspl{BS}. The main focus is on the energy efficiency of \ac{HAPS}, which uses solar energy and advanced battery systems to minimize carbon emissions. Authors in \cite{karaman2024enhancing} \textcolor{black}{used} \ac{HAPS} to provide connectivity for large-scale coverage areas in disaster regions. \textcolor{black}{With regard to the recent earthquakes in Turkiye, two methods for network management were proposed.} \textcolor{black}{Reference} \cite{liu2021intelligent} \textcolor{black}{used} \ac{RIS} technology to improve signal strength and coverage in communication networks. It \textcolor{black}{explored} the potential of 6G communication technologies in disaster scenarios. Sustainable energy solutions, such as solar power, to support the communication infrastructure \textcolor{black}{were} also incorporated.

\begin{figure*}[htp!]
\centering
\includegraphics[width=0.790\linewidth]{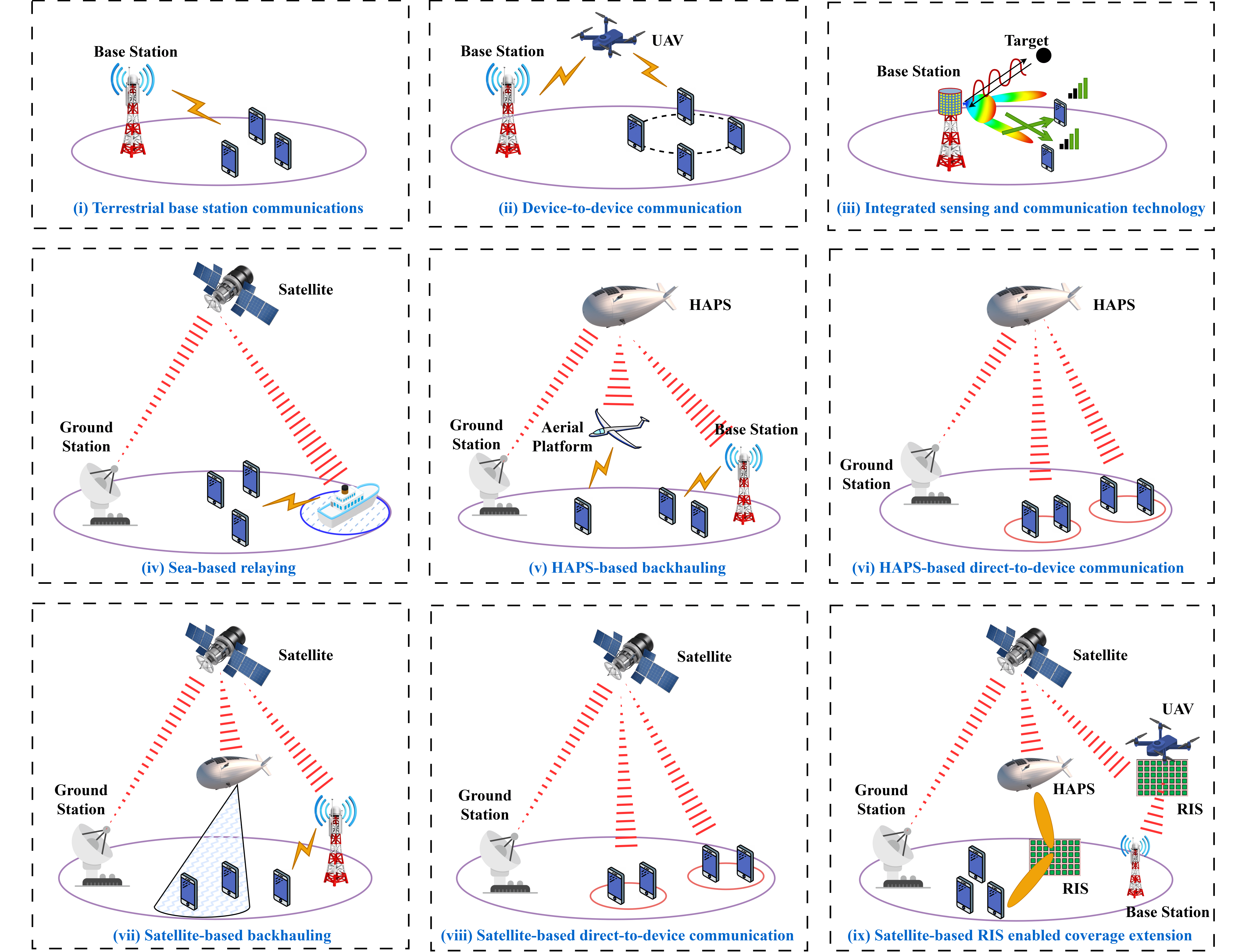}
\caption{Possible use cases of communication technology enablers in a disaster.}
\label{CommEnablersFig}
\vspace{-0.3cm}
\end{figure*}

{Blockchain-based \ac{IoT}}  can be used to enhance the performance of disaster management practices by providing a secure, decentralized, and transparent platform for managing and sharing critical information and resources during and after a disaster \cite{kaur2022biot}. They can be used to securely and transparently share critical information among stakeholders, such as emergency responders, local communities, and government agencies. This can help ensure that all stakeholders have access to the same information and can work together more effectively.    {\ac{AI}}, on the other hand, can be used as a decision-making tool for crisis management and resource allocation. \ac{AI}-driven decision support systems help responders make critical decisions, such as resource allocation and prioritization \cite{sahoh2023role}. For emergent communication, the paper \cite{chafii2023emergent} \textcolor{black}{applied} multi-agent reinforcement learning to enable adaptive and cooperative communication strategies between network nodes. The vision of the paper \textcolor{black}{focused} on the development of communication protocols that could dynamically adapt to changing network conditions and requirements. For emergency communication and information dissemination, chatbots equipped with \ac{NLP} capabilities can deliver real-time information and instructions to the affected population through various communication channels \cite{androutsopoulou2019transforming}.

\textcolor{black}{Figure} \ref{CommEnablersFig} illustrates potential usage scenarios of communication technologies during disasters. 
\textcolor{black}{Figure} \ref{CommEnablersFig}(i) shows the terrestrial BS-supported communication system. \textcolor{black}{Figure} \ref{CommEnablersFig}(ii) presents the \ac{D2D} architecture. \textcolor{black}{Figure} \ref{CommEnablersFig}(iii) depicts the ISAC technology. If the earthquake occurs in a coastal settlement, as shown in Fig. \ref{CommEnablersFig}(iv), sea-based relaying is an option for meeting the communication needs of the region. \textcolor{black}{Figure} \ref{CommEnablersFig}(v) illustrates the scenario where a backhaul link is provided to isolated terrestrial BSs and aerial platforms through HAPS \cite{kurt2021vision}. \textcolor{black}{Figure} \ref{CommEnablersFig}(vi) shows the scenario of direct transmission to ground users in the RF band via HAPS \cite{alfattani2022beyond, alfattani2023resource}. \textcolor{black}{Figure} \ref{CommEnablersFig}(vii) presents the scenario where a backhaul link is provided to isolated terrestrial BSs and aerial platforms via satellite. In contrast, Fig. \ref{CommEnablersFig}(viii) depicts satellite-based direct-to-device communication \cite{bakhsh2024multi}. Being only 20 km away from the ground, HAPS systems benefit from favorable channel conditions and geostationary positions and can provide lower latency than satellite nodes. Moreover, HAPS can cover large areas and be quickly repositioned to offer communication services, especially in regions with damaged or nonexistent terrestrial infrastructure. Although satellite-based direct-to-device communication is becoming feasible, it provides coverage for a short period in the area affected by an earthquake. Moreover, satellite-based direct-to-device communication offers much lower data speeds (enabling SMS sending) than HAPS. It is insufficient to meet the huge communication capacity required for large-scale disasters like earthquakes in Turkiye. For instance, accessing video may be difficult through satellites, while it would be possible via HAPS \cite{kurt2021vision}. In scenarios where terrestrial communication collapses during disasters, deploying \ac{RIS} on terrestrial and aerial platforms can offer significant advantages for dynamic coverage extension \cite{kisseleff2021reconfigurable, chen2022performance}. \textcolor{black}{Figure} \ref{CommEnablersFig}(ix) presents the scenario for satellite-assisted \ac{RIS}-enabled dynamic coverage extension \cite{cao2023ris}.

\subsection{Technology Enablers for Energy Support in Disaster Response Scenarios}

\begin{figure*}[htp!]
\centering
\includegraphics[width=0.790\linewidth]{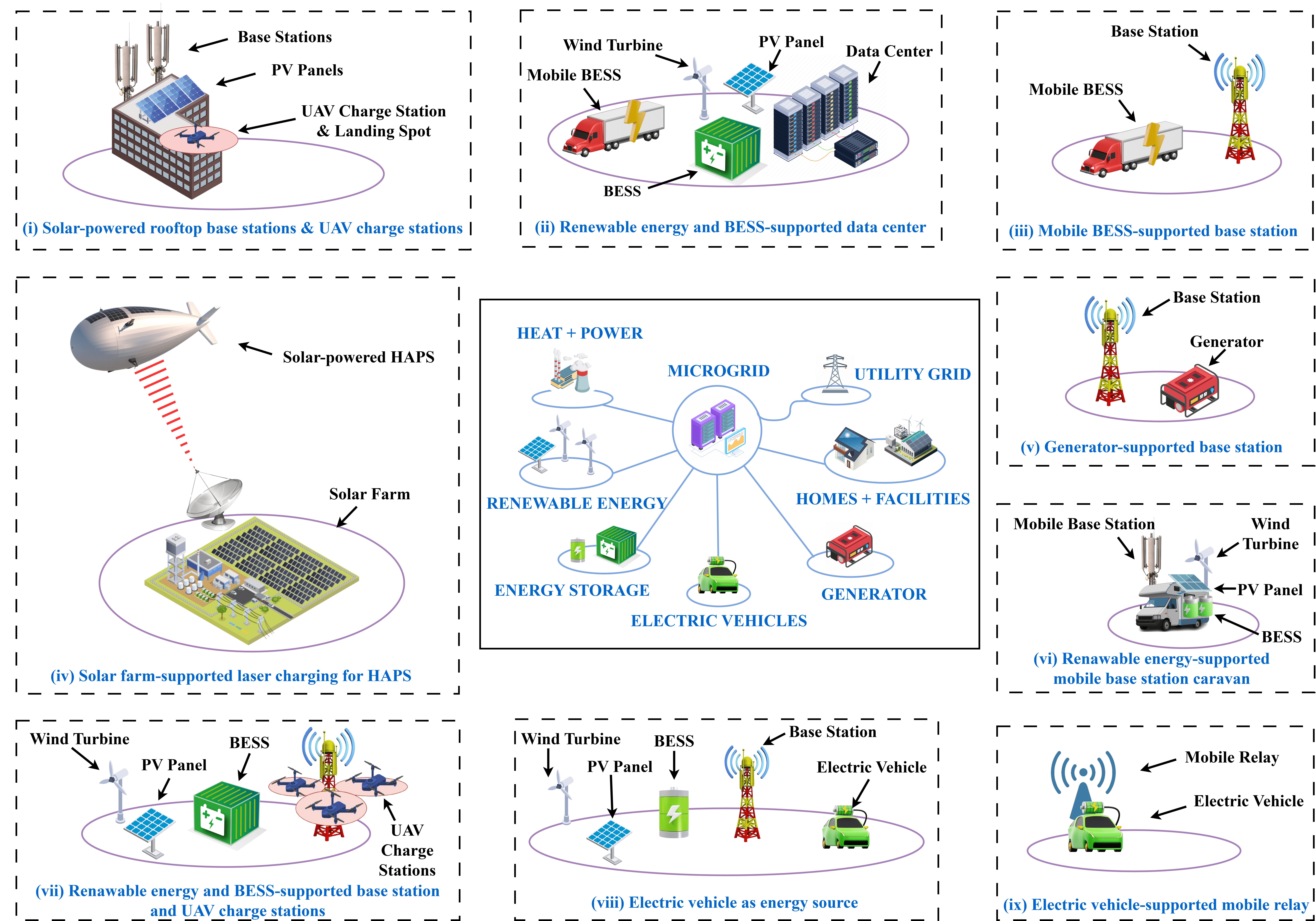}
\caption{Possible use cases of energy technology enablers in a disaster.}
\label{EnergyEnablersFig}
\vspace{-0.3cm}
\end{figure*}


Integrating microgrid components such as \ac{RES}, \ac{BESS}, \glspl{EV}, and generators into critical communication components makes communication systems resilient to energy disruptions during disasters \cite{yadav2020microgrid, rajabzadeh2022enhance}. \textcolor{black}{Figure} \ref{EnergyEnablersFig} illustrates potential usage scenarios of energy-enabling technologies in communication systems during disasters. Installing small-scale solar power plants on building rooftops is becoming increasingly common. The authors \textcolor{black}{of} \cite{galvan2020networked} \textcolor{black}{examined} rooftop solar \ac{PV} and \ac{BESS} to improve power distribution system resilience to natural disasters. The authors \textcolor{black}{of} \cite{jahid2020techno}  \textcolor{black}{explored} the feasibility of optimal power supply solutions for LTE \glspl{BS}, focusing on standalone solar \ac{PV}, hybrid \ac{PV}/wind turbine, hybrid \ac{PV}/diesel generator, and hybrid \ac{PV}/electric grid systems. In \cite{qin2022drone}, the concept of renewable energy-supported charging stations for \glspl{UAV} was explored, showing improved network performance. \textcolor{black}{Reference} \cite{vittal2022smart} \textcolor{black}{implemented} microgrid technology to provide a localized, resilient power supply in emergencies. \ac{RES} such as solar and wind with BESS \textcolor{black}{were integrated} to ensure continuous power availability. \glspl{EV} \textcolor{black}{were} also used as mobile power sources and communication relays in disaster areas.  Since \glspl{BS} are often mounted on rooftops, using energy generated by \ac{PV} panels for \glspl{BS} can be beneficial during disasters. Furthermore, \ac{UAV} charging stations can be installed on building rooftops at specific locations during disasters to better manage \ac{UAV} charging operations. \textcolor{black}{Figure} \ref{EnergyEnablersFig}(i) presents the use case of solar-powered rooftop \glspl{BS} and landing spots.

For critical communication points such as data centers, integrating \ac{RES} and \ac{BESS} can minimize the impact of potential power outages during disasters. If needed, deploying mobile \ac{BESS} can ensure continuous charging and energy supply. Ensuring the power supply for data centers is crucial to maintaining uninterrupted \textcolor{black}{I}nternet and communication services \cite{ahmed2022impacts}. The authors \textcolor{black}{of} \cite{guo2021integrated} \textcolor{black}{studied} the co-planning problem of networked Internet data centers and \ac{BESS} in a smart grid system. \textcolor{black}{A comprehensive case analysis was conducted to demonstrate the effectiveness and appropriateness of the integrated planning methodology using the Multi-Objective Natural Aggregation Algorithm. } \textcolor{black}{Reference} \cite{nazemi2021uncertainty} \textcolor{black}{proposed} a novel restoration mechanism to enhance the resiliency of the distribution grid by deploying mobile energy storage systems. \textcolor{black}{To show the effectiveness, scalability and resiliency of the proposed framework, studies were performed on IEEE 33-node and 123-node test systems.} In light of these studies, \textcolor{black}{Figure} \ref{EnergyEnablersFig}(ii) presents the use case of data centers supported by \ac{RES} and \ac{BESS}.

Increasing the number of tower-type \glspl{BS} can help prevent physical damage due to their greater resilience to disasters' destructive effects. Additionally, supporting tower-type \glspl{BS} with \ac{RES} and \ac{BESS} can enable long-term operation even during potential power outages. \textcolor{black}{Figure} \ref{EnergyEnablersFig}(vii) presents the use case of installing \ac{UAV} charging stations on tower-type \glspl{BS} supported by \ac{RES} and \ac{BESS}. \textcolor{black}{Reference} \cite{yang2020environmental} examine the feasibility of the secondary use of \ac{EV} lithium-ion batteries in \glspl{BS}. Furthermore, \glspl{EV} can be used to feed the battery groups in \glspl{BS}. \textcolor{black}{Figure} \ref{EnergyEnablersFig}(iii) shows the charging of the tower-type \ac{BS}'s energy storage system via a mobile \ac{BESS}, and Fig. \ref{EnergyEnablersFig}(viii) shows the charging of the tower-type \ac{BS}'s energy storage system via an \ac{EV}. Due to \glspl{EV}' mobility advantage and significant energy capacity, Fig. \ref{EnergyEnablersFig}(ix) presents the use case where a mobile relay is mounted on an \ac{EV}. \textcolor{black}{Figure} \ref{EnergyEnablersFig}(v) shows the use case of meeting the energy needs of a \ac{BS} via a generator. HAPS, which have coverage area, substantial channel conditions advantages, and the ability to provide backhauling to small and isolated \glspl{BS} and direct communication with ground users, are expected to meet their energy needs with onboard \ac{PV} modules and battery systems. However, in cases where HAPS requires charging, laser charging technology is among the proposals in the literature \cite{kurt2021vision}. \textcolor{black}{Figure} \ref{EnergyEnablersFig}(iv) presents solar farm-supported laser charging for HAPS as one of the possible use cases.

\subsection{Lessons Learnt and Recommendations}

Several important findings have also emerged from the analysis of the technological enablers for communication and energy support in disaster response scenarios that can guide future disaster management strategies and provide valuable insights into the effectiveness of communication metrics, communication technologies, and energy sources. First, when it comes to prioritizing redundancy and diversity in disaster communication, it is critical to explore how diverse communication channels at physical layer, such as terrestrial networks, satellite links and aerial platforms (e.g. UAVs, \ac{HAPS}) can be integrated with application layer services like social media and mobile applications.  The aim is to ensure a seamless data flow between these communication infrastructures and the user-facing platforms. Satellite or UAV-based networks, for example, can maintain the connection if the terrestrial infrastructure fails. This ensures that important information — such as warnings, real-time updates and safety instructions — can continue to be disseminated via social media platforms and mobile apps. By enabling this interaction between the underlying communication technologies and the user-level applications, we can improve both the robustness of the communication system and the speed at which important information reaches the affected population. On the other hand, there are some case studies where different communication systems have played a crucial role in maintaining connectivity during earthquakes \cite{yulianto2020communication, lam2021network} but an integrated approach with energy and communication systems is still in its infancy.  However, ensuring the security and reliability of these platforms is crucial.  Future efforts should focus on developing secure, scalable, and easily accessible platforms that can be rapidly deployed in disaster scenarios to support first responders, officials, and the affected public. Second, in terms of earthquake response, real-time data analysis is important for decision-making. In this regard, communication and energy metrics such as the battery limitations, throughput, latency and reliability of data transmission can contribute to informed decision-making by authorities and emergency responders.

Thirdly, the use of mesh networks for local communication can be improved, and their focus on improving the resilience of the communication infrastructure could be further exploited.  In the event of damage to the traditional communication infrastructure, mesh networks can provide a means of localized communication. There are several cases where 	\textit{ad hoc} networks formed by devices have enabled effective communication between community members and emergency responders \cite{carreras2022communication}. In the event of natural disasters such as earthquakes, for example, traditional communication systems may fail, leaving people unable to communicate. In such situations, mesh networks can be formed with devices such as smartphones or laptops that allow people to communicate with each other and with the emergency services.  Fourth, UAVs and HAPS are used for aerial surveillance and as communication relays and can play an important role in supporting real-time communication and data collection in hard-to-reach areas or in areas where traditional communication infrastructure has been damaged.  However, the success of these technologies depends on adequate operator training, regulatory compliance and the development of robust algorithms for optimal use and allocation of resources. Future research should investigate the integration of UAV-based networks with ground infrastructure and the development of energy-efficient charging solutions, such as solar-powered charging stations.

Fifth, the introduction of microgrid technology and \ac{RES} is critical to ensuring a stable power supply for communications infrastructure during disasters. The use of solar panels, wind turbines, and battery storage systems has ensured continuous power supply to critical components such as \glspl{BS} and data centers. However, the feasibility and efficiency of these energy solutions depend on the availability of resources and the specific disaster context. To improve energy resilience, future initiatives should focus on the joint planning of renewable energy infrastructures with communication networks, as well as the development of mobile energy solutions such as \glspl{EV} and mobile battery systems to support field operations. Finally, while new technologies such as AI, blockchain, and \ac{XR} have shown promise in other fields, their adoption in disaster management is still in its infancy. These technologies offer opportunities to improve decision-making, secure information sharing, and real-time situational awareness. For example, by overlaying digital information with the real world, \ac{XR} can provide emergency responders with critical information in real-time, enabling them to make more informed decisions.   In addition, \ac{XR} can help responders navigate through damaged buildings and other difficult environments.  However, challenges in terms of scalability, cost, and integration into existing systems need to be addressed. Future research should aim to further develop these technologies and ensure that they are adaptable, secure and meet the specific requirements of disaster scenarios.

\section{Post-Disaster Communication and Energy Planning, Rescue and Evacuation}
\label{afterdisaster}

After a disaster, clear communication and efficient energy management are vital for keeping people safe and helping communities recover. This part of our discussion focuses on the technology that makes these things possible. By looking at new tools and methods, we will show how technology can improve search and rescue, make evacuations smoother, and help us be better prepared for future disasters. The next subsections will go into more detail about these technologies, along with important things we've learned and suggestions for doing things the right way.

\subsection{Technology Enablers for Communication in Post-Disaster Scenarios}

\begin{figure}[htp!]
\centering
\includegraphics[width=\linewidth]{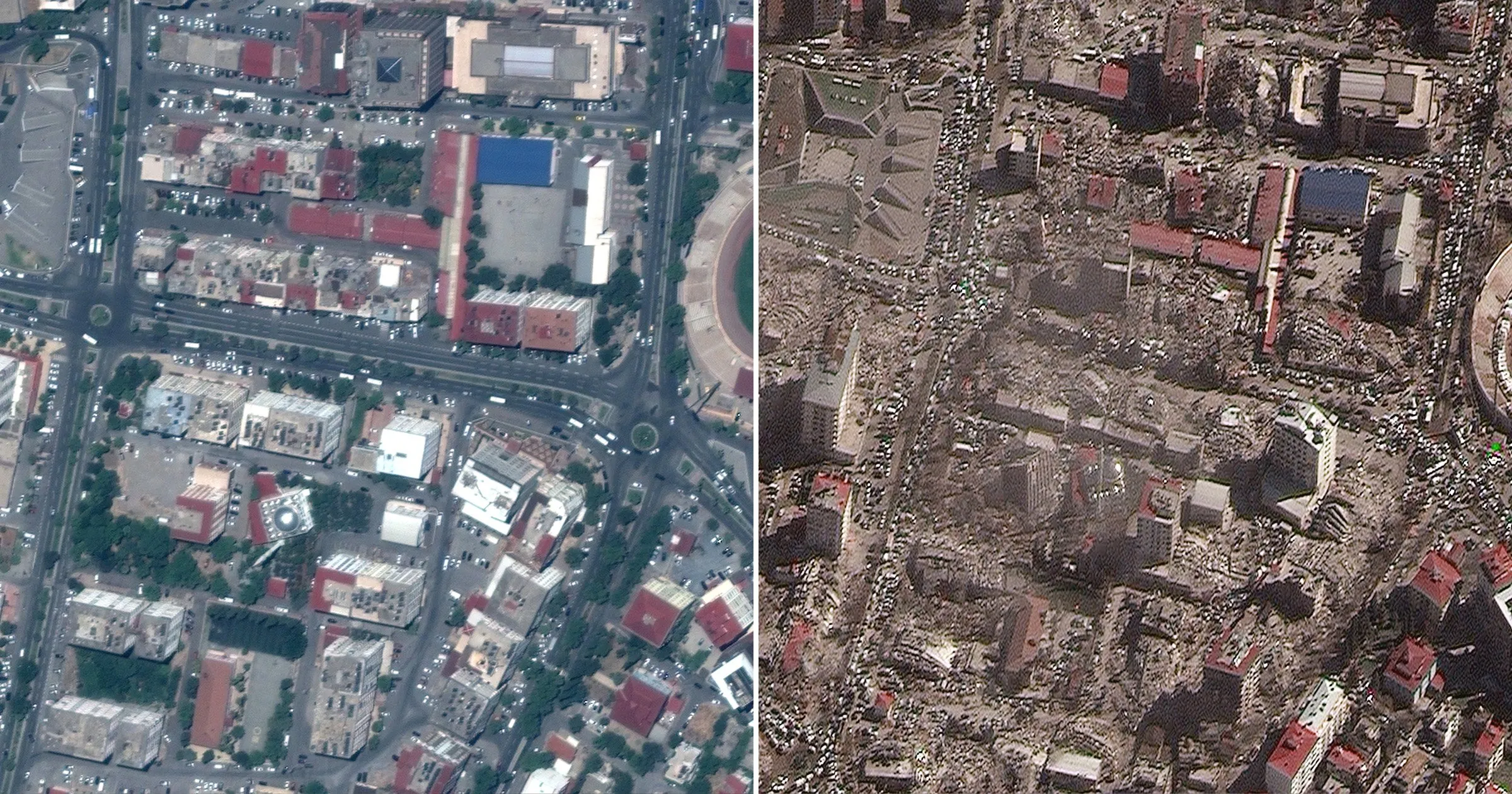}
\caption{The devastating scale of the Turkiye earthquakes before and after satellite pictures.}
\label{disaster_struck_BS}
\vspace{-0.1cm}
\end{figure}

The ground infrastructure is very susceptible to disruption, as shown in Fig. \ref{disaster_struck_BS}, which occurs after disasters and often leads to communication failures. In such situations, alternative solutions beyond the existing architecture are necessary to ensure communication needs are met. Numerous mobile and service providers are putting resources into creating and executing new network architectures and strategies to improve the effectiveness of existing wireless communication systems. Many technical solutions have been proposed as potential candidates for supporting data transmission in disaster-affected regions. In the following, to prevent communication disruptions in a post-disaster case, the solutions proposed in the literature will be detailed.

Satellite networks can provide reliable communications in areas where terrestrial infrastructure has been damaged or destroyed and enable communications over long distances and in remote locations. While the primary role of satellites in post-disaster situations is to provide backhaul support to aerial nodes, an intriguing area of research is the development of low-latency LEO satellites that can also offer direct access to user devices \cite{2022Matracia}. This might be a reality if the 6G technology supports satellite BSs and these satellites use similar frequency bands, so devices can connect to them without modification, but with lower rates due to the low-gain antennas. Many studies in the past decade \textcolor{black}{explored} the use of satellites in disaster scenarios \cite{2017Wang,takahashi2013disaster,casoni2015integration,volk2021emergency,dai2020capacity, zhou2021integrated}.In \cite{2017Wang}, the hybrid Satellite-Aerial-Terrestrial Networks in emergency scenarios \textcolor{black}{were studied}. The wireless technologies used in this hybrid scenario \textcolor{black}{were summarized}, and some challenges like interoperability, QoS assurance, and security \textcolor{black}{were discussed}. \textcolor{black}{Reference} \cite{volk2021emergency} introduced three innovative architectures, backed by experimental data, utilizing satellite communication (SATCOM) for public protection and disaster relief (PPDR). In \cite{zhou2021integrated}, a post-disaster technology \textcolor{black}{was proposed} that integrated satellite-based and ground-based emergency networks. The nodes of the network, in particular the mesh \ac{AP} nodes,  and the portable satellite stations, have a portable design that can be easily transported by the network operator and deployed at the appropriate location. 

Deploying aerial networks is another alternative for restoring network connectivity quickly, as the recovery of terrestrial networks in disaster-affected areas often takes considerable time. \ac{UAV}s can adjust their positions and access hard-to-reach areas, making them ideal for use in disaster-stricken regions \cite{erdelj2017help} \cite{Zhao2019UAV}\cite{zhu2023cooperative}. In \cite{erdelj2017help}, it \textcolor{black}{was discussed} how \ac{UAV}s could be utilized for various tasks such as assessing the extent of damage, identifying affected areas, and aiding in search and rescue missions during natural disasters. Additionally, the importance of integrating \ac{UAV}s with \ac{WSN}s to enhance data collection and communication capabilities in disaster-stricken areas \textcolor{black}{was emphasized}. In \cite{Zhao2019UAV}, \ac{UAV} trajectory and scheduling optimization \textcolor{black}{was}
studied for emergency networks in disaster. In \cite{zhu2023cooperative}, a cooperative trajectory planning approach to deliver emergency communications quickly and efficiently \textcolor{black}{was} presented by considering a time-constrained disaster-affected area. In \cite{Panda2019UAV}, \ac{UAV}-assisted \ac{Wi-Fi} based emergency network model \textcolor{black}{was designed} for post-disaster management. Considering post-disaster scenarios, a mesh communication architecture for drone swarms that can maintain intra-network connectivity between drones \textcolor{black}{was} proposed in \cite{kurt2021distributed}.  In \cite{liu2023construction}, an algorithm based on particle swarm optimization (PSO) \textcolor{black}{was proposed} for constructing a \ac{FANET} that can provide communication coverage for as many users as possible on the ground and can send information back to the rescue center as quickly as possible in a post-disaster scenario is proposed. \textcolor{black}{Reference} \cite{matracia2023comparing} used stochastic geometry tools to estimate both the average and local coverage probability of a wireless network using Aerial Reconfigurable Intelligent Surfaces (ARIS) in post-disaster scenarios. The paper \cite{Matracia2023SG} used \ac{UAV}s to restore cellular connectivity in disaster-affected areas. Strategies \textcolor{black}{were} developed to improve the resilience and reliability of \ac{UAV}-aided networks. In addition, new methods \textcolor{black}{were} proposed to mitigate interference in dense \ac{UAV} deployments. The authors \textcolor{black}{of} \cite{Niu2021EEmax} developed techniques to maximize the energy efficiency of \ac{UAV}-assisted emergency communication networks. Optimization algorithms have been used to determine the best deployment and operational strategies for \ac{UAV}s. The paper also incorporated \ac{RES} to power the \ac{UAV}s and reduce overall energy requirements. The authors \textcolor{black}{of} \cite{wang2022task} addressed the problem of limited battery and computing resources with an \ac{UAV} system based on fog computing. A stable matching algorithm \textcolor{black}{was} proposed to avoid transmission competitions and enable cooperation among \glspl{UAV} and \glspl{UGV}. A Ubiquity Network (UbiQNet) architecture, where drones can form a mesh network to allow users to communicate their situation and location to emergency responders, \textcolor{black}{was} studied in \cite{ganesh2021architecture}. Although most of the studies focused on \ac{UAV}s as aerial networks for disaster management, only some studies focused on other aerial network components like tethered balloons \cite{arimura2014new,nakajima2015balloon,alsamhi2018disaster,gomez2013realistic}, and \ac{HAPS} \cite{karaman2024enhancing}.

Mesh networks can be established using devices such as smartphones or portable routers in a post-disaster scenario. These networks can provide local communications capabilities even when traditional infrastructure is disrupted. In \cite{abdel2020efficient}, two lightweight and fast authentication mechanisms \textcolor{black}{were} proposed that address the physical limitations of mm-wave communication in wireless mesh networks for post-disaster scenarios. The performance of IEEE 802.11s (the reference standard for wireless mesh networks (WMNs)) is evaluated in \cite{erturk2019ieee} for post-disaster communications.  \textcolor{black}{Reference} \cite{gupta2021optimal} \textcolor{black}{provided} algorithms for the optimal placement of UAVs forming Aerial Mesh Networks (AMNs) to support communication in a post-disaster scenario.

\ac{D2D} technology is also an effective communication enabler in post-disaster scenarios where traditional communication infrastructure is damaged or overloaded \cite{AliD2D2018,Deepak2019Over,deepak2019robust,Zhao2019UAV,hossain2020smartdr,Ramakrishnan2022,zhou2019drone,thomas2019finder}.
The authors \textcolor{black}{of} \cite{AliD2D2018} combined clustering technology with \ac{D2D} communication within a cellular network, ensuring communication services could continue even if parts of the cellular infrastructure were compromised. \textcolor{black}{Reference} \cite{deepak2019robust} evaluated the benefits of communication using \ac{D2D}  technologies and cellular systems operating in underlay mode to recover communication networks in the post-disaster scenario. Coverage probability \textcolor{black}{was} improved in the disaster \textcolor{black}{striken} area. In \cite{zhou2019drone}, two important technologies \ac{UAV} and \ac{D2D} for post-disaster scenarios \textcolor{black}{were} combined. This study explored drone-initiated, D2D-aided multihop multicast networks for the rapid dissemination of emergency alert messages in public safety scenarios. In \cite{thomas2019finder}, a framework named Finding Isolated Nodes Using D2D for Emergency Response (FINDER) \textcolor{black}{was} developed to identify and reconnect isolated mobile nodes in disaster areas, aiming to prevent asset damage and loss of life.

In addition to early warning and situational awareness, \ac{IoT} networks can also be used to support post-disaster recovery efforts \cite{kamruzzaman2017study,liu2018transceiver,2021Ali}. For example, \ac{IoT}-based sensors can be used to monitor the status of critical infrastructure, such as power grids and transportation networks, and to provide real-time data on the progress of recovery efforts. In the literature, \ac{IoT} technology \textcolor{black}{was} used for different natural disasters. In \cite{bushnaq2021role}, the coexistence of \ac{UAV} and \ac{IoT} technologies \textcolor{black}{were} used for wildfire detection. In \cite{hasan2021search}, a flood disaster \textcolor{black}{was} considered, and an IoT-aided flood management framework utilizing water, ground, and air networks \textcolor{black}{were} proposed.

\ac{ML} techniques \textcolor{black}{were} proved valuable in post-disaster and crisis management, e.g. recovering and consolidating information, searching and rescuing with limited human interaction, and post-disaster assessment \cite{chamola2020disaster,sreelakshmi2022machine}. \ac{ML} algorithms can be used for different aims. For example, they can optimize the distribution of resources like food, water, medical supplies, and personnel by analyzing the needs of affected areas, predicting demand, and identifying the most efficient delivery routes \cite{chamola2020disaster}. They can also analyze satellite images, drone footage, and other visual data to automatically assess the extent of damage to infrastructure, buildings, and natural landscapes\cite{alidoost2017application,cooner2016detection,sheykhmousa2019post}. \ac{ML} algorithms can also process vast amounts of social media data to identify real-time information on the needs of survivors, locate people in need of rescue, and track the spread of information and misinformation during the disaster recovery process \cite{jamali2019social,sadhukhan2018producing}.

\subsection{Technology Enablers for Energy Support in Post-Disaster Scenarios}

Technology enablers for energy support in post-disaster scenarios are essential for creating a resilient and sustainable energy infrastructure that can withstand the challenges posed by natural disasters and emergencies while ensuring the availability of power for critical functions.  For sustainable reconstruction, the use of \ac{RES} and microgrid systems can ensure a reliable and sustainable energy supply during recovery and reconstruction \cite{alobaidi2022distribution}.

Communication networks might experience prolonged outages after disasters or grid failures due to the unavailability of conventional power sources. \ac{RES} are among the candidates for the recovery and resilience of power grid and terrestrial communication systems. \textcolor{black}{Reference} \cite{galvan2020networked} \textcolor{black}{showed} the potential of rooftop solar \ac{PV} and \ac{BESS} to enhance distribution grid resilience against disasters. \textcolor{black}{The authors assessed system performance under moderate and heavy damage scenarios using resilience metrics in two case studies.} \textcolor{black}{Reference} \cite{deevela2024review} \textcolor{black}{presented} solutions that integrate \ac{RES}, such as diesel generators, solar \ac{PV} panels, wind turbines, fuel cells, and micro-turbines, into terrestrial \glspl{BS} to make them more resilient against energy-related problems in post-disaster scenarios. \textcolor{black}{Reference} \cite{faraci2023green} \textcolor{black}{examined} how renewable energy-supported charging stations met the charging needs of \glspl{UAV} and enabled them to operate within a flying ad-hoc network structure in disaster recovery scenarios. The deployment of renewable energy-supported systems not only aids in disaster recovery efforts but also enhances the long-term resilience and sustainability of terrestrial communication networks.

\ac{BESS} can provide a resilient and immediate energy source, ensuring the continuous operation of critical communication infrastructure in post-disaster situations. The study \cite{dugan2021application} \textcolor{black}{investigated} mobile energy storage systems and \glspl{EV} for their use as energy sources during power outages in disaster scenarios. \textcolor{black}{The study highlighted that enhancing power grid resilience with mobile energy resources requires jointly modeling transport and grid constraints, along with cost–benefit considerations.} \textcolor{black}{Reference} \cite{saboori2023enhancing} \textcolor{black}{examined} how truck-mounted \ac{BESS} could enhance the resilience of distribution systems and facilitate recovery. \textcolor{black}{The proposed model, tested on a 33-bus grid and 15-node transport network, demonstrated improved resilience and utilization of renewable resources.} \textcolor{black}{Reference} \cite{ferraro2020uninterruptible} \textcolor{black}{explored} using \ac{BESS} at \glspl{BS} as a resilient power supply. In post-disaster scenarios, \ac{BESS} can be strategically deployed with mobile \glspl{BS} or other essential network components to ensure that even remote areas can regain connectivity quickly, thus aiding in disaster response and recovery efforts.

Smart grids are pivotal in recovering terrestrial communication systems in a disaster. By dynamically reallocating energy resources, smart grids can prioritize power delivery to critical communication infrastructure, such as \glspl{BS} and data centers, thereby minimizing downtime. Smart grids can seamlessly integrate \ac{RES} and battery storage systems, providing diversified and resilient energy supplies. \textcolor{black}{Reference} \cite{hossain2020smartdr} \textcolor{black}{developed} smart distribution networks to enhance the resilience and recovery capabilities of power grids post-disaster. Advanced data analytics \textcolor{black}{were used} to assess damage, predict recovery needs, and optimize resource allocation. \ac{RES} and energy storage systems \textcolor{black}{were also incorporated} for sustainable recovery.

In post-disaster environments, where the energy infrastructure is often compromised, using low-power IoT devices, energy-efficient network equipment, and optimized communication methodologies becomes paramount. \glspl{UAV} offer the advantage of rapid deployment for disaster management and search and rescue operations \cite{saif2024multi, saif2023skyward}. The studies in \cite{ariman2023energy, qu2023environmentally} \textcolor{black}{examined} the energy-efficient coordination of multi-drones for disaster response. \textcolor{black}{The results showed that energy-efficient optimization models enhance coverage, network connectivity, and power efficiency performance.} The study \cite{dabiri2024enabling} \textcolor{black}{investigated} the use of \glspl{UAV} for backhaul links in post-disaster scenarios and addressed their charging needs through mobile distributed charge stations. \textcolor{black}{The authors also proposed algorithms to minimize outage probability and evaluated their performance in two different scenarios.} \textcolor{black}{Reference} \cite{atat2023efficient}, considering the charging requirements of \glspl{UAV}, \textcolor{black}{proposed} a \ac{UAV}-based solution for detecting and recovering damaged transmission lines after a disaster. Moreover, energy harvesting is crucial for meeting the energy needs of aerial communication platforms, which can be quickly deployed and play an essential role in addressing communication requirements in post-disaster \cite{khalid2023computational, ghosh2023uav}.

\subsection{Lessons Learnt and Recommendations}

Disasters often cause traditional communication infrastructure to fail, emphasizing the need for alternative and resilient communication solutions. Ground-based networks are especially vulnerable to damage, requiring the quick deployment of alternative communication strategies. Satellite networks, especially those in low Earth orbit, have become a crucial technology for keeping communications going in disaster-stricken areas. They can provide backhaul support and direct user access over long distances if the users are in line-of-sight, which is extremely helpful when ground-based networks are damaged. However, challenges like latency, interoperability, and security in combined satellite-aerial-terrestrial networks still need to be studied further. Drones have been shown to be effective in restoring network connectivity and providing vital data in difficult-to-reach areas. Research has demonstrated their versatility in assessing damage, searching for and rescuing people, and improving communication capabilities in disaster-stricken regions. However, optimizing drone deployment strategies and energy efficiency remains a key area for ongoing research.

Mesh networks, created by devices like smartphones and portable routers, offer a decentralized and reliable communication structure when traditional infrastructure is not available. Additionally, \ac{D2D} communication has shown potential in maintaining connectivity even when parts of the cellular network are damaged. Combining drones and \ac{D2D} technologies has further improved the speed and reliability of emergency communications.  \ac{IoT} and \ac{ML} are valuable tools for disaster management. \ac{IoT} devices can monitor critical infrastructure and provide real-time data, while \ac{ML} algorithms can optimize resource distribution, assess damage, and process large amounts of social media data to support rescue efforts.

After disasters, \ac{RES} like solar panels and wind turbines, along with battery storage systems, are essential for keeping communication networks running. These energy solutions help with immediate disaster recovery and also contribute to the long-term resilience of the energy infrastructure. Smart grids are crucial for dynamically distributing energy resources to critical communication infrastructure during disasters. They allow for the smooth integration of \ac{RES} and storage systems, making power grids more resilient and able to recover quickly. Additionally, energy-efficient drone coordination and energy-harvesting techniques are vital for maintaining communication in disaster-affected areas.

We summarize our recommendations for post-disaster cases as follows. First, we recommend further developing systems that combine satellite, aerial, and ground-based networks to make communication systems more resilient after disasters. This will create more reliable and flexible communication solutions for emergencies. Addressing challenges related to interoperability and security in hybrid communication networks is crucial. Future research should focus on developing standardized protocols and security measures that can seamlessly integrate different communication technologies in disaster situations. Second, further research on optimizing drone flight paths, scheduling, and energy management will improve their effectiveness in disaster response. Using \ac{RES} and energy-efficient technologies in drone operations can extend their operational capabilities during prolonged disaster recovery periods. Third, it will be good to encourage the creation of community-based mesh networks in areas prone to disasters. These networks, built using affordable devices like smartphones and portable routers, can be the first line of communication when traditional infrastructure fails. Fourth, it must be improved the development of \ac{D2D} communication protocols that can work independently of centralized networks. These protocols should be optimized for energy efficiency and low-latency communication to ensure reliable connectivity during disasters. Fifth, it is important to ensure that \ac{IoT} and \ac{ML} technologies are seamlessly integrated with existing emergency management systems. This involves creating interoperable platforms that can aggregate and analyze data from various sources and provide a unified view for decision-makers. Sixth, policies and infrastructure investments should prioritize the integration of \ac{RES} into communication networks. This will not only support immediate disaster response but also contribute to building a more resilient and sustainable energy infrastructure for the future. Finally, governments and utilities should invest in smart grid technologies that can dynamically allocate energy resources during disasters. Smart grids, in combination with renewable energy and storage systems, will be crucial in minimizing communication network downtime and ensuring a continuous power supply during emergencies.

These recommendations aim to provide a comprehensive approach to improving disaster resilience and ensuring the continuity of communication and energy networks in emergencies. By addressing both technological and organizational challenges, these recommendations can help build more robust systems capable of withstanding and quickly recovering from disaster impacts.

\section{Existing Vendor Products, Services, \textcolor{black}{Solutions}}
\label{vendors}

In the context of emergency communications, there have been different solutions from numerous vendors, and their products play crucial roles in shaping emergency response strategies. Recovering from a disaster entails numerous challenges, posing potential obstacles to the swift restoration of essential services and effective communication. The nature of disasters, whether natural or human-made, varies across geographies, influencing their impact. Consequently, both local and large-scale companies have developed diverse recovery solutions tailored to maintain communication continuity. These entities must be well-prepared for scenarios wherein entire infrastructures may be destroyed, the networks may be congested, and the power supply may be disrupted, leaving telecommunication towers without electricity. This preparation involves considering data consistency, ensuring the integrity of intricate services, and standing ready to address the aftermath of complete infrastructure destruction. In this section, we briefly explore a compilation of current solutions dedicated to supporting mobile infrastructure, enhancing network resilience, and facilitating vital communication in disaster-affected regions. Our examination primarily delves into the services and products offered by well-known companies that have PPDR products and services. PPDR agencies have generally relied on narrowband radio systems such as Terrestrial Trunked Radio (TETRA) and APCO P25 for critical voice communications. However, with the growing need for real-time data and video, advanced broadband solutions like LTE-Advanced and 5G are becoming essential. These newer technologies offer ultra-reliable, low-latency communications, which are crucial for handling the high data rates required by video and situational awareness applications. The evolution of these technologies is also driven by international efforts to harmonize spectrum usage, particularly in the $700$ MHz and $800$ MHz bands, as outlined by ITU’s Resolution 646 and related reports \cite{series2017use}. This shift towards integrated broadband and narrowband systems will allow PPDR networks to ensure resilience, reliability, and efficiency in life-saving missions. 

There is an organization called TETRA and Critical Communications Association (TCCA) that plays a significant role in the development and promotion of critical communication technologies, particularly for PPDR. The main focus of TCCA is to support the development and standardization of technologies like TETRA and other critical communications solutions, including LTE and 5G, for mission-critical communications. TCCA collaborates with key organizations responsible for developing open standards in critical mobile communications, including ETSI and 3GPP. Its members are involved in a worldwide range of activities, from designing, manufacturing, and building to implementing, utilizing, analyzing, promoting, and deploying critical communication solutions.

\textcolor{black}{Ericsson, a prominent member of TCCA, has established a strong presence in the public safety domain through its dedicated disaster response initiatives. The company actively supports humanitarian efforts by providing wireless Internet access to relief workers via Wi-Fi-enabled devices, leveraging its global infrastructure and technical expertise~\cite{mohan2020review}. In collaboration with China Mobile, Ericsson has piloted 5G-based disaster management solutions, demonstrating the potential of next-generation connectivity in emergency scenarios~\cite{ericsson2022china}. Its enterprise business unit, Cradlepoint, focuses on fail-over use cases, enabling rescue operations and sustaining vital services through rapidly deployable communication nodes. The solution portfolio of Ericsson includes advanced edge computing capabilities via MEC, which significantly enhances efficiency by enabling localized, low-latency services and optimizing spectrum use. With the reservation of a $2 \times 5$ MHz segment in Band 68 for PPDR agencies in Europe, Ericsson has showcased how cross-vendor \ac{MCX} deployments in this band can support mission-critical services such as push-to-talk, video streaming, and control room integration~\cite{ericsson2023band68}. The Tactical Humanitarian Operations Response (THOR), jointly developed with Verizon, exemplifies the commitment of Ericsson to resilient design through a mobile 5G ultra-wideband and MEC-based platform intended for rapid response~\cite{blackwell2023ad}. Demonstrations at major events, including CCW 2023, underline the efficacy of Ericsson’s ecosystem in supporting real-time communication, situational awareness, and scalable response coverage across local and national levels.}

Huawei is another key member of the PPDR and ETSI TC-RT organizations under the International Telecommunication Union (ITU-R).  Addressing various applications like major event security or emergency rescue, it has developed the eLTE Mission Critical Communication System (MCCS) solution. This solution empowers governments to establish a comprehensive national critical communication network (NCCN) as a shared strategic infrastructure. \textcolor{black}{The NCCN concept includes multi-mode support, remote coverage capabilities, and fallback options using hybrid networks. Huawei explicitly frames their solution as essential for national security and large-scale disaster response, reinforcing the resilience to disasters. In addition, the system of Huawei is designed to provide situational awareness, reduce incident response times, and allow hybrid PPDR deployment (dedicated + public networks). This flexibility supports a wide range of PPDR use cases with real-world mission-critical requirements while enhancing its efficacy. Huawei emphasizes low power optimization and spectrum efficiency through TDD synchronization, adaptive antennas, and advanced frame ratios (e.g., 8:2 DL/UL). Their focus on minimizing interference and maximizing spectrum utilization suggests efficient operation during both routine and emergency situations.}
To ensure effective emergency response during disasters, offering ubiquitous coverage, and restoring communication systems, three distinct deployment modes are provided for NCCN. This allows each country to choose the most suitable deployment mode based on its specific requirements. For detailed information, one can refer to the NCCN eLTE7.0 brochure \cite{huaweiNCCN}.

Nokia, as one of the important players in the telecommunication world and an active member of TCCA, has made important contributions to PPDR services. It provides mission-critical communication services designed for emergency responders and public safety organizations. Utilizing its advanced 4G LTE and 5G networks, Nokia ensures \ac{URLLC}, high data throughput, and nationwide coverage. These networks are optimized for PPDR use cases like real-time video streaming, situational awareness, and command and control functions. Nokia integrates Mission Critical Push-to-Talk (MCPTT), Mission Critical Video (MCVideo), and Mission Critical Data (MCData) into its solutions, supporting seamless communication even in the most demanding situations. Additionally, the platform is built with robust security measures, such as encryption and resilient network slicing, ensuring the confidentiality and availability of critical communications during disaster recovery and emergency operations \cite{nokia}. \textcolor{black}{The system of Nokia has TDD-optimized AirScale RAN and IP/MPLS backhaul to enhance network efficiency, offering scalable deployment options through hybrid public-private configurations.} \textcolor{black}{Its platform also supports legacy interoperability with systems like TETRA and P25, while enhancing operational efficacy with IoT sensor fusion, location tracking, and bio-sign monitoring capabilities for real-time situational awareness.} \textcolor{black}{In terms of resilience, the architecture includes geo-redundant cores, mission-critical transport paths, and edge failover mechanisms to ensure service continuity even under adverse conditions.}

ZTE, another global provider of telecommunications equipment and solutions, has introduced an important product known as the All-in-One Nomadic 5G solution for disaster relief following the $6.8$ magnitude earthquake in Luding County, Sichuan, on the Richter scale. It provides for the evolving needs of rapid response in disaster zones and beyond. ZTE's All-in-One Nomadic 5G integrates essential components of a 5G network into a single, portable unit, facilitating swift deployment and connectivity in emergencies. It supports high data rates up to $10$ Gbps, low latency, and multiple access points, while its compact design ensures a compact design for quick transport and setup, making it ideal for remote or temporary locations where rapid network establishment is critical. With these features tailored for rapid response scenarios, such as disaster zones, the solution ensures seamless communication and data transmission, enabling efficient coordination of relief efforts and support services \cite{albanese2021sardo}. \textcolor{black}{To enhance deployment flexibility and operational independence, 5G CampSite product of ZTE integrates edge computing, backhaul, and radio access functions into a unified plug-and-play system.} \textcolor{black}{Its smart O\&M tools enable deployment in approximately 20 minutes, while the built-in NodeEngine and SmartEdge Gateway support localized real-time data processing and private 5G applications.} \textcolor{black}{The system is further optimized through energy-efficient features such as AAU hibernation and intelligent energy scheduling, contributing to reduced operational overhead.} \textcolor{black}{In terms of resilience, the solution includes support for 4G/5G dual-mode access, autonomous local traffic offloading, and fallback connectivity options via satellite or public networks—ensuring continuity of communication in disconnected or infrastructure-damaged environments.} Power is the most crucial requirement in disaster relief communications. Recognizing this critical need, BlackStarTech offers the Broadband Beacon system, a solution designed to address emergency communication challenges seamlessly. It is a portable 4G/5G Node that can be rapidly deployable on private LTE networks designed for emergency and mobile operations. It supports various network connections, including satellite, \ac{Wi-Fi}, and Ethernet, and can be operational within 10 minutes. Powered by a portable battery, the system ensures long-term communication without reliance on external power sources \cite{blackstar}.

Motorola Solutions stands as a prominent figure, offering a broad spectrum of solutions finely tuned for public safety. They focus on creating resilient, mission-critical mobile networks designed for emergency communications and disaster recovery. Motorola provides solutions with high-power, high-tower solutions like deployable \ac{CoW} and simplex operations, ensuring robust, off-network communication through direct or repeater modes. Their system operates across harmonized frequency bands, including $380-400$ MHz for narrowband and $703-862$ MHz for broadband LTE, supporting cross-border and multi-agency collaboration. These features enable reliable, fast deployment and secure communication essential for real-time, mission-critical operations \cite{motorola_ppdr_solutions}. Among their offerings are P25 digital radio systems, ASTRO 25 land mobile radio systems, and Emergency CallWorks, an innovative 911 call-handling solution engineered to meet the demands of next-generation emergency response scenarios \cite{motorola}. 

Cisco Systems, using its expertise in networking technologies, provides sophisticated emergency communications solutions. Notably, Cisco Emergency Responder automates emergency response processes within IP telephony networks, while Cisco Kinetic for Cities serves as a versatile platform initially tailored for smart city initiatives but adaptable for emergency response scenarios as well \cite{cisco}. \textcolor{black}{Cisco Emergency Responder (CER) enhances operational efficiency by automatically detecting the location of IP phones based on IP subnets and switch port mappings, enabling seamless provisioning and accurate location assignment without manual intervention. This integration with Cisco Unified Communications Manager (CUCM) facilitates plug-and-play deployment across enterprise telephony systems.} \textcolor{black}{In terms of efficacy, CER supports accurate emergency call routing to Public Safety Answering Points (PSAPs) by attaching Automatic Location Information (ALI) and Emergency Location Identification Numbers (ELIN) to calls. It also provides in-building Emergency Response Location (ERL) mapping and detailed logging of emergency communications to aid accountability and coordination.} \textcolor{black}{To ensure resilience, the system supports failover mechanisms using Computer Telephony Integration (CTI) route points and Default ERL configurations, maintaining emergency call routing capabilities even during network or CUCM outages. Additionally, fallback testing functions allow administrators to validate communication paths under failure conditions, ensuring continuity during critical events \cite{ciscoCER2023}.}

AT\&T, a major telecommunications company, delivers diverse emergency communication solutions to address various needs. Their flagship offering, AT\&T FirstNet, stands out as a dedicated communications platform designed specifically for first responders. Moreover, AT\&T Wireless Emergency Alerts ensure the timely dissemination of critical information by delivering emergency notifications directly to mobile devices \cite{firstnet}. 

Sierra Wireless, specializing in IoT solutions, offers products purpose-built for emergency communications. Their AirLink routers and gateways are engineered to provide reliable connectivity. Additionally, their Emergency Vehicle Gateway (EVG) enables first responders to establish mobile \ac{Wi-Fi} hotspots in emergency vehicles, facilitating seamless communication and data exchange in the field \cite{SierraWireless2024}. \textcolor{black}{The AirLink platform supports dual 5G radios and Wi-Fi 6, equipped with cognitive wireless steering across cellular, Ethernet, and wireless interfaces, enabling adaptive, high-speed communication under dynamic conditions.} \textcolor{black}{Vehicle Area Networks (VANs) established through AirLink devices integrate multiple systems such as tablets, body-worn cameras, and radios to maintain operational awareness and improve coordination.} \textcolor{black}{Support for Computer-Aided Dispatch (CAD) and Automatic Vehicle Location (AVL) enables real-time tracking of personnel and vehicles, while electronic patient care record (ePCR) access allows first responders to exchange medical data securely with hospitals and control centers.} \textcolor{black}{The AirLink Management Service (ALMS) further improves operational efficiency by offering over-the-air (OTA) firmware updates, remote configuration, and fleet diagnostics.} \textcolor{black}{In terms of resilience, the system supports multi-network failover (including FirstNet, Verizon Frontline, and the Emergency Services Network in the UK), meets IP64 and MIL-STD-810G ruggedization standards, and offers fallback alerting mechanisms to ensure communication reliability during infrastructure outages.} 

Tait Communications, recognized as a global leader in critical communications solutions, offers a comprehensive suite of products tailored to meet the unique challenges of emergency response. Their portfolio includes Project 25 (P25) and Digital Mobile Radio (DMR) digital radios, renowned for their reliability and interoperability in mission-critical environments. Furthermore, their Unified Vehicle platform integrates voice, data, and video communications within emergency vehicles \cite{tait}. \textcolor{black}{Unified Critical Communications (UCC) architecture of Tait enables seamless roaming and bearer selection across multiple network technologies, including Land Mobile Radio (LMR), Long-Term Evolution (LTE), satellite, and Wi-Fi, based on parameters such as power consumption, cost, and signal availability.} \textcolor{black}{This dynamic multi-bearer capability enhances operational efficiency by intelligently utilizing existing infrastructure while reducing deployment complexity.} \textcolor{black}{For efficacy, the platform supports real-time integration of Computer-Aided Dispatch (CAD), Automatic Vehicle Location (AVL), video surveillance, and sensor data, enabling rapid situational awareness and response coordination.} \textcolor{black}{It has been deployed in high-demand environments such as mining operations, public transit systems, and emergency services, where uninterrupted communication is essential.} \textcolor{black}{In terms of resilience, Tait’s solutions feature autonomous fallback modes, including simplex operation and site trunking, allowing users to maintain direct communication even when core network infrastructure fails.} \textcolor{black}{Open standard support ensures interoperability and prevents vendor lock-in, promoting sustainable and fault-tolerant deployments across agencies.}

\textcolor{black}{Finally, Airbus is also actively contributing to PPDR communications with its advanced 5G slicing capabilities \cite{apostolakis2024network}, \cite{zayas2024ppdrslicing}. One notable solution is the Airbus M6 application, which supports secure real-time video streaming for mission-critical users. Airbus collaborated with the University of Málaga to test PPDR, specific 5G network slices using the Victoria Networks testbed. The setup involved separate slices for commercial and PPDR users, each allocated half of the available bandwidth. Round-trip time measurements showed significant latency advantages for the PPDR slice (as low as 11–16 ms), compared to the commercial slice (latency up to 100 ms) under congestion. This demonstration highlights the capabilities of Airbus in delivering ultra-reliable, prioritized communication services using network slicing—crucial for maintaining QoS in emergency response scenarios.}

To provide a clearer comparison of the technical characteristics of the existing real-life solutions, Table~\ref{tab:industry_comparison} summarizes key attributes across latency, deployment speed, interoperability, energy efficiency, with their extra features. This comparative view highlights the trade-offs and deployment contexts for each solution.

\begin{table*}[htp!]
\scriptsize
\centering
\caption{\textcolor{black}{Comparison of Industry Solutions for Disaster Communication}}
\begin{tabular}{|p{1.8cm}|p{1.6cm}|p{1.2cm}|p{3.2cm}|p{2.5cm}|p{2.5cm}|p{2.5cm}|}
\hline
\rowcolor{gray!25}
\textbf{Vendor / Solution} & \textbf{Network Type} & \textbf{Latency (ms)} & \textbf{Special Features} & \textbf{Efficiency} & \textbf{Efficacy} & \textbf{Resilience} \\
\hline
Ericsson (Cradlepoint, THOR) & 5G + MEC & $\sim$5 & THOR platform, 5G UWB, MEC; Band 68 PPDR ecosystem & Up to 150 hrs autonomous runtime; edge-local processing & Deployed in nationwide disaster relief trials (e.g., CCW); \ac{MCX}-ready & Dual-mode failover, MEC fallback; Band 68 coverage \\
\hline
Huawei (eLTE MCCS, NCCN) & eLTE / LTE & 20–30 & Three deployment modes, national NCCN & TDD sync, adaptive antennas; optimized DL/UL ratios (e.g., 8:2) & Situational awareness and hybrid PPDR deployment & Multi-mode, hybrid fallback; remote deployment ready \\
\hline
Nokia (\ac{MCX} over LTE/5G) & 4G/5G LTE & 10–20 & MCPTT, MCVideo, encryption, slicing & TDD-optimized AirScale RAN; Supports hybrid and scalable deployment & Supports real-time \ac{MCX}; integrates legacy systems, such as TETRA, P25; enhances situational awareness with IoT, video, and sensor fusion & Geo-redundant core; mission-critical transport; supports fallback over legacy networks \\
\hline
ZTE (Nomadic 5G) & 5G & $\sim$10 & Portable, 10 Gbps, multiple access points &Plug-and-play deployment; integrated edge-RAN-backhaul; AAU hibernation and energy policies; 20-minute setup & Supports low-latency video and real-time data relay; enhances situational awareness between field teams and command centers & Disaster-proven; 4G/5G dual-mode; local offloading; satellite and fallback backhaul; edge autonomy \\
\hline
Motorola Solutions & P25 / LTE & 100 (P25), 50 (LTE) & CoW, P25, ASTRO 25, harmonized spectrum & Field-deployable towers; simplex options & Proven in mission-critical voice/data deployments & Cross-border/interagency interoperability; high survivability \\
\hline
Cisco Systems & IP / LTE & $\sim$50 & Emergency Responder, Kinetic for Cities & Automated emergency routing and location assignment via IP subnet/SNMP; plug-and-play with CUCM integration & Provides ALI/ELIN for accurate PSAP routing; logs emergency calls; supports building-level ERL mapping & Failover via CTI and Default ERLs; maintains PSAP routing during CUCM/server outages; supports fallback testing \\
\hline
AT\&T FirstNet & LTE (Band 14) & 30–50 & Public safety band, emergency alerts & Field-optimized; access to priority LTE spectrum & Used by $>5,000$ agencies in U.S. & Dedicated Band 14; hardened LTE core \\
\hline
Sierra Wireless (AirLink, EVG) & LTE / Wi-Fi & 30–50 & Mobile Wi-Fi hotspots, vehicle integration & Vehicle-powered, low infrastructure dependency; Dual 5G + Wi-Fi 6; cognitive wireless WAN steering; OTA management via ALMS &  Enables VANs and real-time ePCRs; CAD/AVL support; used by $50\%$ of U.S. state police and top EMS fleets & Multi-network failover (e.g., FirstNet, ESN); IP64 ruggedness; fallback alerts and end-to-end cloud security \\
\hline
Tait Communications & DMR / P25 / Unified Vehicle & $\sim$100 & Unified voice/data/video, vehicle integration & Dynamic bearer selection (LMR, LTE, etc.) based on power/cost/availability; minimizes infrastructure overhead  & Supports real-time voice/data/video; CAD/AVL integration; proven in mining, transit, and emergency fleets & Multi-bearer fallback (LMR/LTE/simplex); site autonomy; no single point of failure; open standard-based \\
\hline
Airbus (M6 app, Victoria Networks) & 5G + Slicing & 11–16 (PPDR slice) & 5G slicing, priority traffic, M6 real-time video streaming & Dedicated network slices with resource guarantees & Validated at CCW and EU demo trials under congestion & Priority fallback, network slice survivability \\
\hline
\end{tabular}
\label{tab:industry_comparison}
\end{table*}

\textcolor{black}{While the reviewed products and services demonstrate considerable progress, several key limitations remain that hinder their full effectiveness in disaster scenarios. These limitations can be listed as follows.
\begin{itemize} 
\item \textit{Lack of multi-network interoperability:} Current solutions often operate in silos and have limited integration with terrestrial, aerial, and satellite systems. Therefore, the systems are prone to some vulnerabilities that arise during handover or fallback operations under degraded network conditions.
\item \textit{Insufficient power autonomy and deployment readiness:} Many systems, including mobile base stations and drone-supported relays, face challenges in terms of battery life, setup time or reliance on external infrastructure - critical constraints in power-outage scenarios. 
\item \textit{Limited satellite communication integration:} Despite the value of non-terrestrial communications in disaster-affected or remote areas, few commercial offerings provide native support or dual-mode capabilities to ensure coverage when terrestrial networks are unavailable.
\item \textit{Lack of standardization and certification:} The lack of a common certification framework for PPDR-specific compliance and interoperability across vendors creates uncertainty, especially for multi-agency and cross-border operations.
\item \textit{Inadequate AI/ML assurance and governance:} While AI/ML is increasingly being used for automation and situational awareness, many solutions lack standardized assurance mechanisms, explainability and ethical safeguards— - important requirements in time-sensitive, high- stakes environments.
\end{itemize} 
Addressing these gaps would significantly enhance the practical value and scalability of PPDR technologies in diverse disaster contexts.}

\section{Standardization and Projects}
\label{standadization}

Standardization efforts in PPDR communications are critical in ensuring emergency services operate seamlessly and effectively across different regions and jurisdictions. These efforts aim to create uniform protocols and technologies that enhance the interoperability, reliability, and efficiency of communication systems used during disasters. The advent of 5G technology has provided a significant acceleration to these standardization initiatives, offering advanced features such as high-speed connectivity, low latency, and network slicing, which are essential for modern PPDR operations.

While the aforementioned vendors and their products in the previous section represent significant advancements in emergency communications, there are still areas for further exploration and improvement. One aspect to consider is the interoperability of different systems and technologies, ensuring seamless communication between various agencies and organizations involved in disaster response efforts. Standardization efforts, such as those promoted by organizations like ITU \cite{itu_r_m2377_2_2023}, TCCA \cite{tcca_5g_critical_2021}, \cite{tcca_mc_broadband_2022}, ETSI \cite{etsi_tr_102445_2008}, and 3GPP \cite{etsi_tr_136762} play a crucial role in facilitating interoperability and should be further emphasized. Moreover, given the dynamic nature of disasters and emergencies, continuous innovation is essential. Vendors should focus on developing adaptive and resilient solutions that can quickly adapt to changing conditions and emerging threats. This may involve leveraging emerging technologies such as artificial intelligence, machine learning, and edge computing to enhance situational awareness, automate response processes, and optimize resource allocation. Furthermore, there is a need for comprehensive testing and validation of emergency communication systems under realistic disaster scenarios. This will help identify potential weaknesses and areas for improvement, ensuring that these systems perform effectively when they are needed most. Collaborative efforts between vendors, emergency responders, government agencies, and research institutions can facilitate such testing and validation processes. Finally, greater attention should be paid to the ethical and social implications of emergency communication technologies. This includes considerations such as privacy, data security, and the equitable distribution of resources. Vendors should actively engage with stakeholders to address these concerns and ensure that their products and solutions uphold ethical standards and promote social equity in disaster response efforts. The ongoing advancements in 5G technology have opened new avenues for enhancing PPDR communications. Projects like 5G-EPICENTRE, Fuge-5G, PPDR-5G, Respond-A, and 5G Safety are at the forefront of leveraging the capabilities of 5G to improve the efficiency, reliability, and interoperability of emergency response systems. These recent projects have important contributions to the standardization of PPDR systems, and they are summarized in Table \ref{PPDRprojects}.

\begin{table*}[htp!]
\scriptsize
\centering 
\caption{Some recent 5G-based PPDR projects}
\begin{tabular}{|>{\raggedright\arraybackslash}p{2cm}|>{\raggedright\arraybackslash}p{4cm}|>{\raggedright\arraybackslash}p{4cm}|>{\raggedright\arraybackslash}p{4cm}|>{\raggedright\arraybackslash}p{2cm}|>{\raggedright\arraybackslash}p{2cm}|}
\hline
\rowcolor{gray!25}
\textbf{Project Name} & \textbf{Project Objective} & \textbf{Technologies/Tools Used} & \textbf{Impact} & \textbf{Status} \\ \hline
5G-EPICENTRE & Develop a holistic ecosystem for next-gen PPDR communication networks using 5G technologies. & 5G-enabled PPDR solutions, network resilience, mission-critical communications. & Improve PPDR operations efficiency, enhancing public safety and security. & Ongoing \\ \hline
Fuge-5G & Evaluate pilots interconnecting Non-Public and Public 5G Networks, exploring 5GC deployments on various clouds. & 5G networks, network slicing, edge computing, IoT devices, AI and ML. & Better emergency communication, situational awareness, network resilience, and cost savings. & Ongoing \\ \hline
PPDR-5G & Develop and validate 5G solutions to enhance PPDR capabilities, leveraging 5G's features for improved systems. & 5G networks, network slicing, edge computing, IoT devices, AI and ML, AR. & Faster response times, reliable networks, better resource allocation, increased safety, and improved collaboration. & Ongoing \\ \hline
Respond-A & Develop advanced communication and information systems for emergency response and public safety operations. & 5G networks, IoT devices, AI, big data analytics, real-time communication platforms. & Improve speed, accuracy, and coordination of emergency responses, saving lives and reducing the impact of disasters. & Ongoing \\ \hline
5G Safety & Evaluate 5G technology for PPDR, using 5G features to improve communication systems for emergency services. & Mobile apps, multimedia capabilities, backend servers, PPDR-IoT devices, drones, body-worn cameras, Pro-M’s “ProPhone” MCPTT app. & Faster, coordinated emergency responses, reliable networks, efficient resource use, increased safety, and strengthened collaboration. & Completed (2018-2021) \\ \hline
\textcolor{black}{RECODIS} &  Resilience of communication networks under disaster-induced failures & Optical networks, physical-layer security, supervised and unsupervised machine learning, anomaly detection. & Enhanced robustness of critical infrastructures and faster recovery from large-scale network disruptions. & Completed (2016–2020) \\ \hline 
\end{tabular}
\label{PPDRprojects}
\end{table*}

The 5G-EPICENTRE project aims to provide an open, federated, end-to-end 5G experimentation platform designed to lower barriers for European SMEs in the public safety market. This platform will allow experimentation with services such as \ac{MCX} communications, data, and video using a cloud-native, microservices-based architecture. Key technical features include network slicing, which allows PPDR users to have dedicated network resources, and \ac{URLLC} for real-time high-definition video and other critical services. The platform integrates the latest 5G technologies like Multi-access Edge Computing (MEC), allowing rapid information access and service reliability, crucial for public safety applications. Additionally, 5G-EPICENTRE contributes to aligning PPDR services with ongoing standardization efforts, especially the \ac{MCX} standards, to ensure interoperability and secure communication. The project also supports ITU-defined service types such as \ac{eMBB}, \ac{mMTC}, and \ac{URLLC}, advancing the state of 5G technologies for public safety.

The FUDGE-5G project focuses on enabling private 5G networks by developing a service-based architecture and cloud-native principles to meet the specific needs of various sectors. These private networks are critical for ensuring high performance, security, and flexibility in use cases such as public safety, industry 4.0, media delivery, and virtual offices. The project integrates SDN and NFV for network orchestration, providing unified access to multiple technologies (5G NR, \ac{Wi-Fi}, Ethernet). One of the main innovations is the seamless connectivity across 5GC and vertical applications, ensuring dynamic resource allocation and enhanced security features for mission-critical services. Key use cases include remote media production, industrial automation, and mission-critical PPDR, where the project’s architecture ensures real-time communication and secure, isolated tactical networks. FUDGE-5G, with its pilots interconnecting public and non-public 5G networks, explores multi-vendor 5G core deployments on public and private clouds, thus setting precedents for future standards.

The Project PPDR-5G network will enhance emergency response, coordination, and threat mitigation, particularly during natural disasters or terrorism. It will deploy a disaster-resilient 5G mobile network along the Hungary-Ukraine border, providing secure and real-time communication for police, border guards, and ambulances. The network will include several 5G gNBs and a 5G standalone (SA) core network, ensuring private cloud-native security. With downlink/uplink speeds of at least 3 Mbps/2 Mbps at the cell edge, low latency under 5 ms, and MIMO technology, it will support up to 500 devices and enable services like telemedicine, real-time video transmission, and enhanced border protection. The focus of Project Respond-A is on real-time situational awareness, incident management, and multi-agency coordination through advanced communication systems exemplifying the integration of cutting-edge technologies like IoT, AI, and big data analytics.  The 5G Safety project, having demonstrated the use of mobile applications, PPDR-IoT devices, and other broadband-intensive applications, provides valuable insights and frameworks that can be standardized and replicated across different regions. Collectively, these initiatives not only enhance the operational capabilities of emergency services but also drive the development of global standards for disaster communication and PPDR systems, ensuring that best practices and technological innovations are widely adopted.

\textcolor{black}{Another relevant initiative is COST Action RECODIS ~\cite{7550596}, which addressed the resilience of communication networks under disaster-induced failures. These failures, resulting from natural disasters, weather events, technical faults, or malicious attacks, can severely disrupt services in critical infrastructures. The project focused on both preventive measures and responsive mechanisms to maintain service continuity during large-scale disruptions.}

Last but not least, a recently started COST Action titled "AlertHub: Warning Communication Knowledge Network" is another important project in enhancing the public protection and disaster relief systems by focusing on effective warning communication \cite{cost_alerthub}. It aims to address the challenges in warning communication arising from climate change-related disasters, enhancing the effectiveness of disaster management and safeguarding communities.

\textcolor{black}{Despite significant progress in the standardization of 5G-based PPDR communications, several crucial aspects remain open for harmonization, both within the EU and globally. One pressing area is the implementation and regulation of Quality of Service, Priority, and Pre-emption (QPP) mechanisms. While QPP is already defined in 3GPP standards, its rollout remains inconsistent due to varying interpretations of EU net neutrality regulations (Regulation (EU) 2015/2120), which may hinder prioritization for critical services. National exemptions, as adopted by countries like Finland, France, and Belgium, provide partial solutions, but a unified legal framework at the EU level is needed. Moreover, the realization of the EU Critical Communication System by 2030 aims to ensure cross-border interoperability across the Schengen area. This calls for a harmonized QPP policy and technical standard across member states. }

\textcolor{black}{Globally, several challenges mirror those in the EU. The lack of standardized \ac{D2D} communication in LTE is being addressed through emerging 5G NR sidelink technologies, which have shown promise in pilot deployments. However, their commercial maturity and chipset support remain limited. In addition, spectrum harmonization, especially for Band 68 (698–703 / 753–758 MHz) and Band n79 (4.9 GHz), is vital for ensuring seamless interoperability across borders, as these bands are increasingly allocated for PPDR in both Europe and the Asia-Pacific regions.} \textcolor{black}{Another essential requirement is the development of sustainable certification and conformance testing frameworks for \ac{MCX} services, especially for vendors and network operators. TCCA is advancing such processes, and government procurement models are encouraged to include mandatory certification criteria. Furthermore, as demonstrated in public safety deployments across Korea, the U.S., Japan, and Scandinavia, integration of 5G features such as network slicing, MEC, and NR sidelink must be standardized in terms of performance, security, and resilience metrics to support next-generation PPDR applications.}

\textcolor{black}{Lastly, ethical and societal concerns such as privacy-preserving data sharing, lawful surveillance, and equitable access must be addressed within standardized governance frameworks. These must span not only technical interoperability but also accountability in AI/ML-driven decision-making for mission-critical systems. Addressing these regulatory, technical, and societal gaps will be key to achieving a truly global, resilient, and future-proof PPDR communication infrastructure.}

\section{Case Study: Turkiye Earthquakes}
\label{case_study}

On February 6, 2023, devastating earthquakes with magnitudes of 7.7 and 7.6 hit Turkiye and Syria\footnote{Since we do not have reliable data for Syria, we will focus on the effects and actions in Turkiye.}, and the response of mobile network operators in Turkiye became critical post-disaster. The earthquakes, which led to considerable destruction and loss of life, severely impacted the telecommunications infrastructure. The earthquakes centered in Kahramanmaraş, Turkiye, affected 11 provinces, covering an area of $115,000$ $\textrm{km}^{2}$. According to the Turkish Strategy and Budget Office (TSBO), over half a million buildings were severely damaged in the earthquakes, resulting in more than 53,000 death tolls \cite{TRdisasterreport}. The earthquakes caused substantial damage to the telecommunications infrastructure.  More than 12 million mobile subscribers existed in the earthquake-affected region, accounting for approximately 14\% of the country. The total damage to the telecommunications sector in public and private sectors is at least 185 million USD \cite{TRdisasterreport}. In this section, we explore the extent to which the earthquake affected the telecommunication infrastructure and how the industry addressed these unprecedented challenges. \textcolor{black}{In light of the lessons learned from the Turkiye earthquake use case, the current wireless infrastructure should be further improved to make it more robust against natural disturbances. To this end, this section also discusses what problems wireless infrastructure have in case of other natural hazards (e.g., floods and wildfires) and what the potential solutions are.}

\subsection{Reasons for Network Failure}
Immediately after the earthquakes, numerous regions experienced disruptions in mobile network services. In some locations, the network failure was exacerbated by the enormity of the disaster, which overwhelmed the existing infrastructure and emergency response capabilities. Below, we summarize the primary reasons for network failure in the area.
\subsubsection{Physical Damage} The earthquake caused extensive damage to over 500,000 buildings, many of which housed critical telecommunications infrastructure. The Association of Mobile Telecommunications Operators (m-TOD) in Turkiye reported that 2,451 out of 8,900 \glspl{BS} of three operators in the 11 provinces affected by the earthquake were offline due to severe damage \cite{mTODData}. As detailed in \cite{turkcellData, vodafoneData}, it is clear that most BS installations in the region (nearly two-thirds) were mounted on rooftops. In Turkiye, the backhaul connections of BSs are fiber-connected at levels of 35\%, while the remaining ones have radio link connectivity\cite{turkcellData}. As a result of this, the collapse of these buildings during the earthquake damaged these stations and disrupted the radio link connections essential for communication between them. This resulted in significant initial disruptions in mobile network services across the impacted areas. Right after the earthquake, a large part of the mobile network was down, with estimates indicating that up to 60\% of the network in the hardest-hit regions was initially non-operational. Some provinces had severe cases, such as Kahramanmaraş Province, which experienced a dramatic 94\% drop in \textcolor{black}{I}nternet traffic following the second earthquake \cite{cloudfareStatistics}. 
\subsubsection{Power Outages} The earthquake inflicted widespread harm on the power grid, resulting in extensive power outages. This greatly impacted the functioning of \glspl{BS} and other essential infrastructure. Although most of the tower-type stations remained intact, many \glspl{BS} did not have sufficient backup power, causing prolonged outages. The reliance on generators, which were often inadequate or quickly ran out of fuel, underscored the need for more robust power backup solutions. Generators have an endurance of 3-4 hours, after which they need to be refueled. Due to harsh climate and field conditions and transportation challenges, the necessary fuel for these generators couldn't be supplied continuously \cite{turkcellData, vodafoneData}. These power outages worsened the situation over time. While some initial steps were taken to restore communication, network disruptions worsened post-disaster due to power issues, with mobile network connectivity dropping to almost zero after 24 hours \cite{cloudfareStatistics}.
\subsubsection{Network Congestion} As communication standards have advanced (2G, 3G, 4G) and the population density in residential areas has increased, particularly in urban areas, there is a need for shorter distances between \glspl{BS} and more frequent \ac{BS} installations. Due to the earthquake-affected area's current urban development, this necessitates rooftop \ac{BS} installations. For instance, according to data from one of the MNOs in Turkiye, 93\% of the \glspl{BS} in the region are LTE-compatible \cite{turkcellData}. Although mobile \glspl{BS} were sent to the area by telecom operators after the earthquake, the capacity of mobile \glspl{BS} is lower than that of fixed \glspl{BS} \cite{turktelekomData}. Besides, the sudden spike in call and data traffic following the earthquake led to substantial network congestion \cite{cloudfareStatistics}. According to network traffic reports from one of the MNOs in Turkiye, data traffic surged by 260\% and voice traffic by 9150\% within the first half an hour after the earthquake, compared to the previous day \cite{yagan2024fast}. The high volume of communication attempts overwhelmed the already compromised infrastructure, causing dropped calls, slow data speeds, and overall network instability. Additionally, not only BSs but also some backhaul fiberoptic links and data centers suffered damage. This exacerbated network congestion issues due to a lack of reliable routing and data balancing algorithms. Consequently, some regions experienced outages despite the RAN over the air (e.g., BS to UE) still being operational. Implementing a distributed network with smaller, more numerous cell sites could have mitigated the impact of any single point of failure. This approach would enhance overall network resilience and redundancy.
\subsection{Response by Network Operators} The response by network operators in Turkiye to the 2023 earthquakes showcased both the strengths and weaknesses of the current telecommunications infrastructure. The actions taken by the MNOs are detailed as follows. \subsubsection{ Restoring Communication} All three \glspl{MNO} deployed mobile \glspl{BS} and \ac{CoW} to offer temporary network coverage in the impacted regions. As reported by mToD in Turkiye, the initial step involved sending a total of 190 mobile \glspl{BS} to the area. These units were pivotal in re-establishing communication lines, particularly in areas with compromised permanent infrastructure. However, transporting the \ac{CoW} to the regions faced delays due to the magnitude of the crisis, damaged roads, traffic congestion, and regulations, as the trucks weren't initially categorized as first responders and were not allowed to enter the areas immediately. Mounted on the Communication Disaster Plan, operators sent more than 2,200 personnel, nearly 500 mobile, caravan, and trailer-type \glspl{BS}, approximately 3,500 generators, and 8 emergency communication vehicles to the affected regions instantly following the earthquake. In addition, UAVs paired with the mobile \glspl{BS} were deployed. Furthermore, TÜRKSAT set up VSAT satellite terminals and \ac{Wi-Fi} access points at 163 sites \cite{btkData}. Due to the physical damage to the main transportation routes in the region caused by the earthquake, challenging weather conditions, and heavy traffic trying to access the area, it took more than 36 hours for the mobile \glspl{BS} and equipment to reach the earthquake-affected region \cite{turkcellData, vodafoneData}. In the first 48 hours, close to 40\% of the damaged network capacity was restored. By the end of the first week, about 80\% of the network in the affected areas was operational again \cite{cloudfareStatistics}. Full restoration, however, took several weeks due to the intricate and time-consuming repairs to the physical infrastructure. Also, according to TSBO data, around 8.4 million USD has been spent by operators to sustain the infrastructure at the first step \cite{TRdisasterreport}.   \subsubsection{ Free Services} Network operators offered free call, SMS, and data packages to enable affected residents to communicate with their families and emergency services \cite{mTODBilgilendirme}. This step was critical in maintaining social connectivity and support during the disaster. \subsubsection{ Coordination with Authorities} Mobile operators collaborated closely with government agencies to focus restoration efforts on critical areas, such as hospitals and emergency shelters, ensuring effective healthcare for those affected \cite{turktelekomData}.

\subsection{\textcolor{black}{Lessons Learnt and} Evaluations}

The implementation of temporary BSs and the use of satellite communication significantly reduced the impact of network outages. The coordinated efforts between mobile network operators, governmental bodies, and international partners led to a more structured and efficient response. However, the initial phase proved ineffective, highlighting the need to enhance telecommunication infrastructure and organizational strategies to mitigate the effects of such disasters. Below, we outline the key areas for improvement based on the Turkiye earthquakes to ensure reliable communication services during future events.  By addressing these points, \glspl{MNO} can better their resilience and response efficiency in subsequent disasters, ultimately providing improved connectivity and support for affected communities.
\subsubsection{Disaster-Resilient Infrastructure} The earthquake underscored the necessity for infrastructure capable of withstanding natural disasters. Increasing redundancy in network design with more robust backhaul connections can lessen the effects of future incidents. Constructing cell towers with materials and designs that can endure seismic activity is essential, involving retrofitting existing structures and ensuring new ones meet higher standards. Establishing a more distributed network with smaller, numerous cell sites can minimize single-point failures, thereby enhancing overall network resilience and redundancy. In \cite{karaman2024enhancing}, the authors have addressed the Turkiye earthquakes and proposed a HAPS-based solution on the RAN side, demonstrating that even a single HAPS can largely meet the communication needs of the region. Expanding the use of NTN communication (e.g., satellite or HAPS) as a backhaul and/or RAN option can ensure reliable connectivity when terrestrial networks fail. In \cite{yagan2024fast}, the authors proposed integrating terrestrial and non-terrestrial nodes using emerging 6G technologies, such as cell-free MIMO, \ac{RIS}, and joint communication and sensing technologies for rapid network recovery. Integrating NTN links into the standard network framework ensures seamless fail-over during emergencies. Developing portable and quickly deployable network solutions for activation during disaster scenarios is crucial. \subsubsection{Reliable Power Solutions for Infrastructure} BSs should be fitted with batteries providing extended backup power and include \ac{RES} like solar panels. This ensures continuous operation during prolonged power outages. Additionally, creating a more efficient logistics plan to deploy mobile generators can maintain network functions during disasters.
\subsubsection{Pre-Disaster Planning and Coordination} Conducting regular disaster preparedness exercises and establishing clear emergency protocols with telecom operators and governmental agencies can enhance coordination and response times in future disasters. This also includes preparation and training before disaster strikes. Though the response was relatively swift, more comprehensive pre-disaster training and simulations could enhance the efficiency and effectiveness of restoration efforts.
\subsubsection{Community Engagement} Promoting community awareness and preparedness programs ensures the public understands what to expect and how to stay connected during emergencies. Clear communication about network restoration schedules and available services can reduce uncertainty and anxiety.

\textcolor{black}
{\subsection{Communication Challenges Across Evolving Disaster Scenarios: Floods, Wildfires, Hurricanes, and Industrial Accidents}}

\textcolor{black}{In the above use case scenario, we focused on the Turkiye earthquakes since we had reliable data from the region to make proper assessments. Nevertheless, the current wireless infrastructure faces similar problems in all natural events. While earthquakes lead to immediate infrastructure collapse, disasters such as floods and wildfires develop over time, causing progressive and varying disruptions to communication networks. In such cases, the affected areas expand dynamically, requiring flexible and adaptable network architectures.}

\begin{table}[!t]
\footnotesize
    \caption{Comparison of Disaster Types: Impacts on Infrastructure, Challenges and Solutions}
    \centering
    \begin{tabular}{|p{1.4cm}|p{1.8cm}|p{2cm}|p{2cm}|}
\hline
\rowcolor{gray!25}
\textbf{Disaster Type} & \textbf{Infrastructure Impact} & \textbf{Key Challenges} & \textbf{Suggested Solutions} \\
\hline
Earthquake & Immediate destruction & Base station failures, power outages & Emergency cellular networks (e.g., portable base stations, LEO satellites) \\
\hline
Flood & Gradual but widespread damage & Fiber-optic failures, network congestion & Floating base stations, UAV-assisted relays, satellite backup \\
\hline
Wildfire & Expanding disaster zones & Signal interference, relocation of responders & UAV-based relays, AI-driven network repositioning, mesh networking \\
\hline
Hurricane & Wind \& flood-related destruction & Wide-area outages, tower collapses, line-of-sight interruptions & HAPs, wind-resistant base stations, backup satellite networks \\
\hline
Industrial Accident & Restricted access zones & Hazard exposure, radiation shielding, remote operation needs & Autonomous robotic relays, radiation-hardened devices, secure long-range communication \\
\hline
\end{tabular}
    \label{tab:disaster_types}
\end{table}

\textcolor{black}{On the one hand, flooding often disrupts both wired and wireless networks, as rising water damages underground fiber- optic cables, cellular towers and electrical grids. In addition, network congestion occurs during emergency operations, especially in urban and densely populated areas. For example, the recent floods in Pakistan in 2022 affected more than 33 million people and caused severe network disruptions. In Sindh and Balochistan, 4G networks were largely down due to submerged infrastructure. The terrestrial infrastructure could not cope with these challenges, so the emergency teams deployed Starlink satellite terminals. In addition, network restoration was carried out in some regions using UAVs to set up temporary coverage zones in remote, flooded areas. Although the above measures have attempted to restore communications quickly, further technical solutions should be developed to deal with floods. Some of the recommended measures are as follows. Floating base stations can be deployed in flooded areas to restore connectivity before the water recedes. AI-based congestion control can play a crucial role in dynamic resource allocation to manage peaks in traffic. Satellite-based networks would be a key player, e.g. the integration of LEO satellites for instant coverage.}

\textcolor{black}{On the other hand, wildfires pose unique challenges due to their high-speed spread, unpredictable movement, and smoke interference, which can degrade wireless signals. The need for mobile and self-adapting communication networks is critical in these scenarios. For instance, the recent Hawaii Wildfires on Maui Island in 2023 were one of the deadliest wildfires and caused mass cellular network failures, with over 1000 cell sites damaged. AI-driven predictive fire modeling was used to optimize evacuation routes and preemptively reposition network infrastructure. Besides, drone-based mobile relays provided emergency connectivity in hard-to-reach areas. Based on the lessons from this catastrophe, some of the recommended actions to better cope with network failures are given as follows. UAV-assisted mobile base stations have a crucial role in rapid deployment in fire-affected zones. Mesh networking can ensure decentralized communication when the infrastructure is destroyed. Moreover, AI-driven early warning systems are vital in optimizing network placement and resource allocation.}

\textcolor{black}{In addition to floods and wildfires, hurricanes pose unique challenges due to their wide-area impact and high wind speeds, often causing both physical destruction of infrastructure and service disruptions due to wind-induced outages and flooding. For example, during Hurricane Maria in 2017, more than 95 \% of cell sites in Puerto Rico were rendered inoperative, severely hindering emergency response and coordination efforts. Recommended approaches include wind-resistant \ac{BS} designs, satellite-based backup systems, and \ac{HAPS} to maintain line-of-sight connectivity in severely impacted regions.}

\textcolor{black}{Industrial accidents, such as chemical plant explosions or nuclear incidents, differ in nature by introducing hazardous zones where human access is severely restricted. Communication must therefore rely on robotic or autonomous relays, radiation-hardened devices, and secure, encrypted communication protocols. The Fukushima Daiichi disaster in 2011 highlighted the need for long-range wireless solutions that could operate in high-radiation zones without endangering human operators.}

\textcolor{black}{Based on the aforementioned discussions, Table \ref{tab:disaster_types} summarizes the differences in communication challenges across disaster types and some of the recommended solutions.}

\section{Open Issues, Challenges and Future Directions}
\label{issues}

\subsection{Open Issues and Challenges  in Realizing Resilient and Sustainable Infrastructures}

Existing and emerging technologies in communication networks have greatly improved disaster management efforts, but they also come with several limitations, open issues, and challenges that need to be addressed despite the research efforts. Some of the key limitations of these technologies:

\begin{itemize}[leftmargin=*]
    \item \textit{Data Privacy and Security:} Increased reliance on data transmission and storage can lead to data breaches or unauthorized access, which can compromise sensitive information related to disaster management. In addition, the use of emerging technologies for disaster management can raise concerns about government surveillance and the erosion of privacy rights.
    \item \textit{Sustainability:} New green communication metrics that take into account energy efficiency, carbon footprint, and environmental sustainability of networks and the services provided have recently been defined by IETF Internet-Draft in \cite{Clemm2023}. These metrics are concerned with carbon footprint at various levels of network communication. According to \cite{Clemm2023},  green metrics are categorized into four main levels: (i) At the device/equipment level, the energy consumption of a device as a whole or components (e.g., line cards, individual ports) is measured. (ii) At the flow level, metrics related to the aggregate energy consumption of packets across the flow are investigated. (iii) At the path level, metrics related to energy consumption of paths and path segment selection are studied.  (iv) At the network level, metrics representing a global view of the network as a system for aggregated sustainability are investigated. Energy consumption may differ in each technological domain of mobile networks. According to sources \cite{koltagoing,alliance2021network}, the \ac{RAN} accounts for 73\% of the energy consumption in the mobile network, followed by the core network at 13\%, datacenters at 9\%, and other operations at 5\%.  The RAN energy consumption encompasses \glspl{BS} and all the associated infrastructure, such as inverters, rectifiers, repeaters, and \ac{MBH} transport. As per sources \cite{koltagoing,alliance2021network}, the energy consumed by a \ac{BS} is distributed among air conditioning (40\%), radio processing (40\%), power (7\%), baseband processing (6.5\%), and main control (3.5\%). The authors \textcolor{black}{of} \cite{larsen2023towards} analyzed and compared the potential of current and future energy consumption minimization techniques and provide guidelines for energy-efficient future mobile networks.
    \item \textit{Resiliency and Robustness:} Communication failures caused by disasters have proven to be much more dynamic and extensive than traditional random failures. These incidents often result in so-called \enquote{regional failures}, which indicate the concurrent breakdown of network elements in certain geographical areas \cite{2021Ali}, \cite{7550596}. Even taking into account the additional costs of resilience and extended coverage, a study by \cite{Bauer2003ACA} examines the use of commercial cellular networks and devices for mission-critical high-speed broadband communications. The results show that the PPDR communications delivery model with commercial LTE operation is the most cost-effective option from a purely financial perspective. Nevertheless, it was found that the biggest challenge in implementing this approach is not the technological aspect of building a robust network, but rather the regulatory, legal, and contractual framework. Regarding the technological aspect, studies are currently being carried out to create disaster-resilient networks \cite{mauthe2016disaster}. Self-organizing networks (SONs) have been studied intensively as they are one of the emerging areas in beyond 5G systems. Integrating self-configuration, self-optimization, and self-healing functions in network structures is increasingly recognized as a crucial aspect of ensuring resilient communication. Self-optimization has been investigated for small cells, LTE, \ac{D2D}, and 5G networks \cite{FOURATI2021108435}, \cite{fi14030095}, \cite{3GPP22}. 
    \item \textit{Infrastructure Vulnerability:} Communication infrastructure, such as cell towers and data centers, can be vulnerable to natural disasters like  earthquakes which can disrupt communication networks when they are needed most \textcolor{black}{\cite{preventionweb2023}}.
    \item \textit{Limited Coverage in Rural and Remote Areas:} Emerging technologies like \textcolor{black}{Beyond 5G, 6G} and advanced satellite communication may not be readily available in rural or remote disaster-prone areas, leaving these regions with limited connectivity \textcolor{black}{\cite{hamza2021locating}}. Moreover, the users might not have access to device equipment that can connect to the latest network technologies.
    \item \textit{Interoperability:} In disaster management, different agencies and organizations may use different communication systems and technologies that are not always interoperable. This can hinder the seamless exchange of information during a crisis \textcolor{black}{\cite{mani2023navigating}}. 
    \item \textit{Cost and Accessibility:} Implementing and maintaining advanced communication technologies and infrastructure can be expensive, making them less accessible for lower-income communities or countries. In particular, the fundamental component of resilience is redundancy, which creates additional costs since extra equipment is deployed even if it is not needed under pre-disaster conditions. 
    \item \textit{Regulatory and Legal Barriers:} \textcolor{black}{Integrated and heterogeneous} communication networks may be subject to regulatory hurdles \textcolor{black}{(e.g. requiring thorough examination)}, limiting the availability of certain technologies \textcolor{black}{\cite{kagai2024rapidly}}.
    \item \textit{Resource Constraints:} Some technologies, especially emerging ones, require substantial resources for research, development, and deployment, which may not be available to all regions or organizations \textcolor{black}{\cite{sumbal2025managing}}.
    \item \textit{Human Error and the need for zero-touch automation:} Human error can still play a significant role in communication network failures, from misconfiguration to mishandling of equipment. Therefore, it is important to transition to a zero-touch network and infrastructure management systems \textcolor{black}{\cite{liyanage2022survey}}.
    \item \textit{Misinformation and rumors:} Quickly combating misinformation and disinformation is essential for disaster communications \textcolor{black}{\cite{omar2024disaster}}. Monitoring social media platforms and other channels enables the identification and correction of false information, helping to prevent the spread of rumors and confusion. Data collection from social media for ML-based rescue coordination can also risk building on misconceptions and rumors, which degrade the rescue operations.
    \item \textcolor{black}{\textit{Cybersecurity:} A critical yet often overlooked component of resilient communication infrastructure in disaster scenarios, especially in man-made crises like terrorism or cyberattacks. During such events, emergency networks face increased vulnerability to threats like spoofing, denial-of-service attacks, and misinformation, which can severely disrupt response efforts \cite{spellman2023emergency}. Emerging solutions such as zero-trust architectures, blockchain, quantum-safe encryption, and AI-driven threat detection can offer promising defenses to ensure secure, reliable, and tamper-resistant communication. However, to build truly resilient systems, cybersecurity must be integrated from the outset, with disaster frameworks incorporating secure protocols, trusted identity management, and cross-agency coordination for operating in hostile digital environments.}
\end{itemize}

\textcolor{black}{These open issues can also be examined under different disaster phases—pre - disaster, in- disaster and post - disaster— each of which presents different but sometimes overlapping challenges:}

\textcolor{black}{\textit{Pre-Disaster Phase:}  Open challenges include ensuring data privacy and security in risk assessments and early warning systems, developing a sustainable, energy-efficient communications infrastructure, improving resilience and robustness to disasters, and addressing regulatory and legal barriers that can delay the introduction of technologies. The lack of interoperability between different systems and limited coverage in remote areas are also critical issues at this phase.}

\textcolor{black}{\textit{In-Disaster Phase:} The biggest challenges revolve around real time response and maintaining operational continuity. Cybersecurity threats, misinformation and rumors, and human error in crisis coordination can disrupt disaster response efforts. In addition, the vulnerability of infrastructure and limited network coverage in affected areas pose significant barriers to effective communication. The need for cost-effective and rapidly deployable solutions is also a pressing issue at this phase.}

\textcolor{black}{\textit{Post-Disaster Phase:} Recovery and rebuilding efforts pose long-term challenges. Data privacy and security concerns remain, especially when dealing with sensitive information of the affected population. Restoring resilient infrastructure, ensuring interoperability for long-term recovery and overcoming resource constraints are critical. Cybersecurity remains a risk, as weakened systems may be more vulnerable to attacks during the rebuilding phase.}

Disaster management professionals and policymakers need to consider these challenges when developing and implementing disaster preparedness, response, and recovery communication strategies. They also need to develop contingency plans in the event that these technologies are not available or reliable due to the above limitations. Although the technologies discussed in this paper are powerful tools, their effectiveness also depends on factors such as the policy framework, community engagement, and the capacity of local institutions. In addition, ethical considerations and privacy concerns must be taken into account when using these technologies in disaster response.

\subsection{Future Directions}

\textcolor{black}{In recent years, the increasing frequency and severity of natural disasters worldwide have highlighted critical gaps in communications infrastructure, particularly when connectivity is needed most. Notable examples include the 7.2 magnitude Haiti Earthquake in 2021, Hurricane Ian that struck the United States and Cuba in 2022, the Hawaii Wildfires on the island of Maui in Hawaii in 2023 and the devastating Kahramanmaraş Earthquakes in Turkiye during the same year. These events underline the urgent need for resilient and rapidly deployable communication solutions tailored to disaster scenarios.}

Future directions in earthquake disaster response and communication technologies should focus on integrating advanced technologies and enhancing the resilience and adaptability of communication networks while  \textcolor{black}{maintaining scalability and cost-effectiveness. Improving system resilience requires both technological innovation and cross-layer optimization — from the physical communication infrastructure to network management and application layer intelligence.} \textcolor{black}{First, there is an increasing need to co-design communication and energy systems that are customised for each phase of disaster response (pre-, in-, and post- disaster). Research should focus on the development of modular and rapidly deployable systems, such as UAV-based edge networks and microgrids powered by renewable energy, which can self-configure according to the evolving needs of a particular disaster scenario. Energy-efficient communication protocols, adaptive power management for off-grid devices and robust network topologies are essential to ensure long-term operation with minimal reliance on infrastructure.} \textcolor{black}{Second, the integration of AI at multiple layers of the communication stack offers substantial opportunities to enhance system resilience. Specifically: (i) At the network layer, reinforcement learning (RL) and multi-agent reinforcement learning (MARL) can be used for dynamic spectrum access, topology reconfiguration, routing decisions, and autonomous UAV swarm coordination to ensure robust connectivity even under infrastructure disruption. (ii) At the data processing layer, supervised and unsupervised learning algorithms can support real-time damage assessment by analyzing satellite imagery and drone footage, enabling faster and more accurate situational awareness. (iii) At the application layer, natural language processing (NLP) models and large language models (LLMs) can assist in parsing social media content for extracting real-time distress signals, identifying misinformation, and improving communication between responders and the public. (iv) Generative AI models, such as transformers, can simulate potential disaster propagation paths, predict demand surges in aid resources, and generate proactive response plans by learning from historical disaster data.}

\textcolor{black}{These AI capabilities can also be embedded at the edge using edge intelligence, allowing real-time, low-latency decision-making close to the data source, which is particularly important in disconnected or degraded environments. Third, future systems should incorporate resilient-by-design cybersecurity mechanisms. With the rise of hybrid physical- cyber disasters, the integration of AI-based anomaly detection and blockchain for secure and traceable data exchange is critical to protecting the integrity and availability of communications.} \textcolor{black}{Fourth, emerging technologies such as quantum communication, quantum key distribution (QKD) and quantum computing should be explored for ultra-secure high-throughput communication and predictive simulation in disaster management. While these technologies are still in the early stages, they have the potential to redefine secure, large-scale coordination between agencies in high-stakes environments. Finally, the development of standardized testbeds and digital twins for disaster scenarios will be crucial for benchmarking new systems under realistic conditions. These platforms should integrate simulated physical environments, virtual communication networks and real-time AI agents to evaluate performance, resilience and adaptability before actual deployment.}

\section{Conclusions}
\label{conclusion}

In this paper, we have explored the critical role of sustainable and resilient communications infrastructure in disaster response and management scenarios. Through a comprehensive analysis of new communication technologies, energy solutions, and their integration across pre-disaster, in-disaster, and post-disaster phases, we have emphasized the importance of a multi-layered approach that combines space-based, air-based, sea-based, and ground-based networks. This integrated system ensures continuous connectivity, even in the most challenging conditions where traditional infrastructure may be compromised. 

Our study emphasizes the need for sustainability, redundancy, and diversity in communication channels, including the use of satellites, UAVs, HAPS, and mesh networks, to improve the robustness of communication networks during disasters. We also highlighted the importance of real-time data analysis and energy management, which are critical for informed decision-making and sustainable operations in disaster zones.  The case study of the Turkiye earthquakes further illustrated the practical application of these technologies and demonstrated how integrated communication and energy systems can significantly improve disaster response and recovery efforts. However, despite advances in communications technologies, challenges remain around energy efficiency, regulatory compliance, and the need for further research into cultural factors and new technologies such as quantum communications. Looking to the future, it is clear that continued innovation and interdisciplinary collaboration will be critical to addressing these challenges and realizing the full potential of sustainable and resilient communications infrastructures. By addressing the open issues and using the lessons learned,  more adaptable and resilient systems can be built that can better respond to the increasing frequency and severity of natural disasters.


\bibliographystyle{IEEEtran}
\bibliography{biblio_Rev2_clean}  

\vfill

\end{document}